\documentclass[12pt]{article}
\pdfoutput=1
\usepackage{booktabs}
\usepackage{tabulary}
\usepackage{tabularx}
\usepackage{multirow}
\usepackage{rotating}
\usepackage{epsfig}
\usepackage{psfrag}
\usepackage{latexsym}
\usepackage{indentfirst}
\usepackage{fancyhdr}
\usepackage{dsfont}
\usepackage{amsmath}
\usepackage{amssymb}
\usepackage{amsfonts}
\usepackage{mathrsfs}
\usepackage{amsthm}
\usepackage{pifont}
\usepackage{dsfont}
\usepackage{array}
\usepackage{chngpage}
\usepackage{longtable}
\usepackage{cite}
\usepackage{bbold}
\usepackage{color}
\usepackage{braket}
\usepackage{colordvi}
\usepackage{fancybox}
\usepackage[footnotesize]{caption2}
\usepackage{graphicx}
\usepackage[center,footnotesize,hang]{subfigure}
\usepackage{bbm}
\usepackage{url}
\usepackage[colorlinks, linkcolor=red, anchorcolor=black, citecolor=green]{hyperref}
\usepackage{harpoon}
\usepackage{textcomp}
\usepackage{enumitem}
\usepackage{mathrsfs}
\usepackage{lscape}
\usepackage{makecell}
\usepackage{pdflscape}
\usepackage{adjustbox}
\usepackage{hhline}
\allowdisplaybreaks

\newcommand{\PreserveBackslash}[1]{\let\temp=\\#1\let\\=\temp}
\newcolumntype{C}[1]{>{\PreserveBackslash\centering}p{#1}}
\newcolumntype{R}[1]{>{\PreserveBackslash\raggedleft}p{#1}}
\newcolumntype{L}[1]{>{\PreserveBackslash\raggedright}p{#1}}
\newcommand{\bq}{\begin{eqnarray}}
\newcommand{\nq}{\end{eqnarray}}

\newcommand{\cgone}[2]{
\begin{center}
  \begin{tabular}{|p{#1}|}
    \hline
    #2 \\
    \hline
\end{tabular}
\end{center}
}

\newcommand{\cgtwo}[4]{
\begin{center}
  \begin{tabular}{|p{#1}|p{#2}|}
    \hline
    #3 & #4 \\
    \hline
\end{tabular}
\end{center}
}

\begin{document}
\title{
\begin{flushright}
\hfill\mbox{\small USTC-ICTS/PCFT-20-36} \\[5mm]
\begin{minipage}{0.2\linewidth}
\normalsize
\end{minipage}
\end{flushright}
{\Large \bf Fermion Masses and Mixing from Double Cover and Metaplectic Cover of $A_5$ Modular Group \\[2mm]}}
\date{}

\author{
Chang-Yuan~Yao$^{a}$\footnote{E-mail: {\tt
yaocy@nankai.edu.cn}},~~ Xiang-Gan~Liu$^{b,c}$\footnote{E-mail: {\tt
hepliuxg@mail.ustc.edu.cn}},~~ Gui-Jun~Ding$^{b,c}$\footnote{E-mail: {\tt
dinggj@ustc.edu.cn}}
\\*[20pt]
\centerline{
\begin{minipage}{\linewidth}
\begin{center}
$^a${\it \small
School of Physics, Nankai University, Tianjin 300071, China}\\[2mm]
$^b${\it \small Peng Huanwu Center for Fundamental Theory, Hefei, Anhui 230026, China} \\[2mm]
$^c${\it \small
Interdisciplinary Center for Theoretical Study and  Department of Modern Physics,\\
University of Science and Technology of China, Hefei, Anhui 230026, China}\\
\end{center}
\end{minipage}}
\\[10mm]}
\maketitle
\thispagestyle{empty}

\begin{abstract}

We perform a comprehensive study of the homogeneous finite modular group $A'_5$ which is the double covering of $A_5$. The integral weight and level 5 modular forms have been constructed up to weight 6 and they are decomposed into the irreducible representations of $A'_5$. Then we perform a systematical analysis of the $A'_5$ modular models for lepton masses and mixing. The phenomenologically viable models with minimal number of free parameters and the results of fit are presented. We find out 15 models with 9 real free parameters which can accommodate the experimental data of lepton sector. After including generalized CP symmetry, 9 viable models with 7 free parameters are found out. We apply $A'_5$ modular symmetry to the quark sector, and a quark-lepton unification model is given. The framework of modular invariance is extended to include the rational weight modular forms of level 5. The ring of modular forms at level 5 can be generated by two algebraically independent weight $1/5$ modular forms denoted by $F_1(\tau)$ and $F_2(\tau)$. We give the expressions of the rational weight modular forms of level 5 up to weight $3$ and arrange them into the irreducible multiplets of finite metaplectic group $\widetilde{\Gamma}_5\cong A'_5\times Z_5$. A neutrino mass model with $\widetilde{\Gamma}_5$ modular symmetry is presented, and the phenomenological predictions of the model are analyzed numerically.

\end{abstract}
\newpage

\section{\label{sec:intro} Introduction}

The quark masses and charged lepton masses as well as the neutrino mass squared differences have been precisely measured, yet we still don't know the exact value of the lightest neutrino mass while it is constrained by the cosmology data and the direct neutrino mass measurement experiments through beta decay. From the smallest neutrino mass of order of a fraction of an eV to the top quark mass being 173 GeV, the range of fermion masses span over at least 12 orders of magnitudes. In addition, the quark and charged lepton masses exhibit a hierarchical pattern:
$m_d/m_b\geq m_e/m_\tau\gg m_u/m_t$,
$m_s/m_b\simeq m_{\mu}/m_\tau\gg m_c/m_t$. Moreover, the quark and lepton mixing angles also span several orders of magnitudes, the quark mixing angles are small, while the lepton sector has two large mixing angles $\theta_{12}$, $\theta_{23}$ and one small mixing angle $\theta_{13}$ which is of the same order of magnitude as the quark Cabibbo mixing angle~\cite{Zyla:2020zbs}. It is one of the greatest challenges in particle physics to understand the observed flavor structure of quarks and leptons from first principle in terms of few underlying parameters. In the Standard Model, the fermion masses and mixing matrices arise from the Yukawa interactions with Higgs. The flavor dependence of the Yukawa couplings is the origin of the fermion masses and the mixing matrices.

Several different scenarios have been proposed to understand quark and lepton
masses and mixing angles as well as CP phases.
In view that the symmetry principles have proven very successful in physics,
one resorts to flavor symmetry to constrain the Yukawa interaction so that the fermion mass spectra and mixing matrices could be explained. In the last twenty years, the non-abelian discrete flavor symmetry has been widely studied, and it is found particularly suitable to reproduce the large lepton mixing angles, see~\cite{Altarelli:2010gt,Ishimori:2010au,King:2013eh,King:2014nza,King:2015aea,King:2017guk,Feruglio:2019ktm} for review. The flavor symmetry should be broken in a non-trivial way in discrete flavor symmetry models, and the vacuum expectation values (VEVs) of the flavons are frequently required to be oriented along certain directions in generation space. It is technically quite tricky to dynamically realize the desired vacuum alignment, thus the traditional flavor models appear complicated. Although the flavor symmetry can constrain the mixing angles and CP violation phases, one generally can not make quantitative predictions for fermion masses whose observed values are usually reproduced by tuning the available parameters.

Recently modular invariance as flavor symmetry has been proposed to address the flavor puzzle~\cite{Feruglio:2017spp}. In this new framework, the Yukawa couplings transform nontrivially under the modular symmetry and they are modular forms which are holomorphic function of the complex modulus $\tau$. In the simplest modular symmetry model, the vacuum expectation value of $\tau$ is the unique source of flavor symmetry breaking such that the vacuum alignment problem is simplified. Models with modular flavor symmetry can be quite predictive, and the neutrino masses and mixing parameters can be predicted in terms of few input parameters. The superpotential is completely fixed by modular symmetry in the limit of unbroken supersymmetry, however the K\"ahler potential is less constrained. Thus the predictive power of this framework could be reduced~\cite{Chen:2019ewa} by additional parameters in the general K\"ahler potential. This drawback can be overcome by incorporating the traditional flavor symmetry. Both group representations and modular weights of matter fields would be severely constrained in this new scheme, consequently the structure of the K\"ahler potential and superpotential becomes much more restrictive~\cite{Nilles:2020nnc,Nilles:2020kgo,Nilles:2020tdp,Baur:2020jwc,Nilles:2020gvu}.

The inhomogeneous finite modular group $\Gamma_N$ for $N=2$~\cite{Kobayashi:2018vbk,Kobayashi:2018wkl,Kobayashi:2019rzp,Okada:2019xqk}, $N=3$~\cite{Feruglio:2017spp,Criado:2018thu,Kobayashi:2018vbk,Kobayashi:2018scp,deAnda:2018ecu,Okada:2018yrn,Kobayashi:2018wkl,Novichkov:2018yse,Nomura:2019jxj,Okada:2019uoy,Nomura:2019yft,Ding:2019zxk,Okada:2019mjf,Nomura:2019lnr,Kobayashi:2019xvz,Asaka:2019vev,Gui-JunDing:2019wap,Zhang:2019ngf,Nomura:2019xsb,Wang:2019xbo,Kobayashi:2019gtp,King:2020qaj,Ding:2020yen,Okada:2020rjb,Nomura:2020opk}, $N=4$~\cite{Penedo:2018nmg,Novichkov:2018ovf,deMedeirosVarzielas:2019cyj,Kobayashi:2019mna,King:2019vhv,Criado:2019tzk,Wang:2019ovr,Gui-JunDing:2019wap,Wang:2020dbp}, $N=5$~\cite{Novichkov:2018nkm,Ding:2019xna,Criado:2019tzk} and $N=7$~\cite{Ding:2020msi} have been studied to construct models for neutrino masses and mixing, and the weights of the modular forms have to be even integers. The modular forms of general integral weights have been discussed in~\cite{Liu:2019khw}, and the finite modular group is enhanced to the homogeneous finite modular group $\Gamma'_N$ which is the double covering of $\Gamma_N$. Some flavor models based on $\Gamma'_3\cong T'$~\cite{Liu:2019khw,Lu:2019vgm}, $\Gamma'_4\cong S'_4$~\cite{Liu:2020akv,Novichkov:2020eep} and $\Gamma'_5\cong A'_5$~\cite{Wang:2020lxk} have been proposed. Recently the modular invariance framework has been further extended to include rational weight modular forms~\cite{Liu:2020msy}. It was found that the modular group should be extended to its metaplectic cover group,
and the rational weight modular forms can be arranged into irreducible multiplets of the  finite metaplectic group $\widetilde{\Gamma}_N$~\cite{Liu:2020msy}. Furthermore, modular symmetry can be combined with generalized CP symmetry, and the consistency conditions between these two symmetries require $\tau\rightarrow -\tau^{*}$ up to a modular transformation under the action of CP~\cite{Novichkov:2019sqv}.
A comprehensive analysis about flavor symmetry, CP symmetry and modular invariance in string theory has been given in~\cite{Baur:2019kwi,Baur:2019iai}.
From a top-down perspective, typical compactifications of extra dimensions give rise to low energy effective theories depending on several moduli.
The modular invariant supersymmetric theories with single modulus has been extended to automorphic supersymmetric theory where several moduli can occur~\cite{Ding:2020zxw}. As an example, the so-called Siegel forms were investigated in detail, the direct product of multiple modular symmetry~\cite{deMedeirosVarzielas:2019cyj,King:2019vhv} is now embedded in this Siegel modular symmetry as a special case in which the moduli space is factorized into several independent tori.

The phenomenology of modular symmetry has been extensively discussed in the literature. Besides the fermion masses and flavor mixing, the modular symmetry has been applied to dark matter models~\cite{Nomura:2019jxj,Nomura:2019lnr} and leptogenesis~\cite{Asaka:2019vev,Wang:2019ovr,Behera:2020sfe}. The modular symmetry could naturally produce texture zeroes of fermion if odd weight modular forms are taken into account~\cite{Lu:2019vgm}. The modular symmetry has been implemented in $SU(5)$ grand unification theory~\cite{deAnda:2018ecu,Kobayashi:2019rzp}. The dynamics  of  modular  symmetry could potentially be tested at present and future neutrino oscillation experiments~\cite{Ding:2020yen}. There are many papers on inhomogeneous finite modular groups $\Gamma_N$, but the homogeneous finite modular groups $\Gamma'_N$ are less studied. In this work, we shall perform a comprehensive analysis of $\Gamma'_5=A'_5$ which is double covering of $A_5$ group. The traditional flavor symmetry $A'_5$ has been studied in the literature~\cite{Everett:2010rd,Hashimoto:2011tn,Chen:2011dn} and it was spontaneously broken by the VEVs of some flavons. In the present work, no flavon other than the modulus $\tau$ is introduced. We construct the weight 1 modular forms of level 5 by two methods: the first one is from two weight $1/5$ modular forms $F_1(\tau)$ and $F_{2}(\tau)$, and the second one is by the Klein forms. These two methods give the same results for the weight 1 modular forms which can be arranged into a sextet of $A'_5$, and the tensor products of weight 1 modular forms give modular forms of higher weight. According to the representations and weights of matter fields, a systematical classification of lepton model with $A'_5$ modular symmetry is performed. The neutrino masses are assumed to arise from effective Weinberg operator or type-I seesaw mechanism, both models with two right-handed neutrinos and three right-handed neutrinos are considered. We aim to find out all $A'_5$ modular models with minimal number of free parameters in this work. We numerically scan over the parameter space of each model, and find that 25 (including 15 models with 9 free parameters and 10 models with 10 parameters ) models can explain the experimental data in lepton sector, as shown in table~\ref{tab:lepton_models}.
We incorporate in the generalized CP symmetry to further improve the predictive power of the models, all the coupling constants are constrained to be real in our working basis, and we find that 19 out of 25 models can be compatible with experimental data. Furthermore, we extend the $A'_5$ modular symmetry to quark sector, and give a typical model which can explain the experimental data of both quarks and leptons simultaneously.

Rational weight modular forms can be defined at level 5, and they can be arranged into irreducible representations of the 5-fold metaplectic cover group $\widetilde{\Gamma}_5$ which is isomorphic to $A'_5 \times Z_5$. We find that the two weight $1/5$ modular forms $F_1(\tau)$ and $F_2(\tau)$ furnish a doublet $Y^{(\frac{1}{5})}_{\mathbf{2}^4}(\tau)$ of $\widetilde{\Gamma}_5$. All integral weight modular forms of level 5 can be obtained from the tensor products of $Y^{(\frac{1}{5})}_{\mathbf{2}^4}(\tau)$ with the Clebsch-Gordan (CG) coefficient of $\widetilde{\Gamma}_5$, nevertheless it is not necessary to examine the constraints relating redundant higher weight modular multiplets in this manner. Furthermore, we present a concrete model with rational weight modular forms.

The remainder of the paper is organized as follows. In section~\ref{sec:MDF}, we briefly introduce the relevant concepts of modular symmetry and modular groups, the integral weight and level 5 modular forms are constructed and they are decomposed into irreducible multiplets of $A'_5$.
In section~\ref{sec:gCP-A5DC}, we discuss the generalized CP symmetry compatible with $A'_5$ modular symmetry. In section~\ref{sec:lepton-models}, we perform a systematical classification of $A'_5$ modular lepton models, the phenomenologically viable models with minimal number of free parameters
and the numerical results of the fit are presented.
In section~\ref{sec:quark-lepton}, we extend the $A'_5$ modular symmetry to explain the quark masses and CKM mixing matrix, and a quark-lepton unified model is presented. The rational weight modular forms at level 5 are considered in section~\ref{sec:rational-weight-MF}, and they are arranged into irreducible multiplets of $\widetilde{\Gamma}_5$. Then a benchmark lepton model involving rational weight modular forms is presented. Section~\ref{sec:conclusion} is devoted to our conclusion and discussions. The group theory of $A'_5$ and the CG coefficients in our working basis are presented in Appendix~\ref{sec:A5_group_DC}. We give another approach of constructing modular forms of weight 1 and level 5 from Klein forms in Appendix~\ref{app:weight-1-MF}. We prove two identities between theta constants and Klein forms in Appendix~\ref{sec:proof}. The expressions of linearly independent higher integral weight modular forms with $k=4, 5, 6$ and higher rational weight modular forms with $r=6/5, 7/5,\ldots, 14/5,3$ are given in Appendix~\ref{sec:higher-weight-MF}.

\section{\label{sec:MDF}Modular symmetry and modular forms of level $5$}

The two-dimensional special linear group $\Gamma=SL(2,\mathbb{Z})$ over the integers acts on the upper half complex plane by the linear fractional transformation
\begin{equation}
\label{eq:linear-fraction-transformation}\tau\mapsto \gamma\tau=\frac{a\tau+b}{c\tau+d}, ~~~~\gamma=\begin{pmatrix}
a  ~&  b  \\
c  ~&  d
\end{pmatrix}\,,
\end{equation}
where $a$, $b$, $c$, $d$ are integers fulfilling $ad-bc=1$, and $\tau$ is the complex modulus with $\texttt{Im}(\tau)>0$.
If $c\neq0$ then $-d/c$ maps to $\infty$ and $\infty$ maps to $a/c$, and if $c=0$ then $\infty$ maps to $\infty$. Since $\gamma$ and $-\gamma$ induce the same linear fractional transformation, the group of transformations in Eq.~\eqref{eq:linear-fraction-transformation} is isomorphic to the projective special linear group $\overline{\Gamma}=PSL(2, \mathbb{Z})\cong SL(2,\mathbb{Z})/\{I, -I\}$, which is the quotient of  $SL(2, \mathbb{Z})$ by its center $\{I, -I\}$ with $I$ being 2-dimensional identity matrix. $\Gamma$ and $\overline{\Gamma}$ are usually called homogeneous and inhomogeneous modular groups respectively. The group $\Gamma$ has infinite elements and it can be generated by two matrices
\begin{equation}
S=\begin{pmatrix}
0 ~& 1 \\
-1 ~& 0
\end{pmatrix},  ~~~~  T=\begin{pmatrix}
1 ~& 1 \\
0 ~& 1
\end{pmatrix}\,.
\end{equation}
The matrices $S$ and $T$ are often referred to as modular inversion and  translation respectively,
\begin{equation}
S:~ \tau\mapsto-\frac{1}{\tau}\,,~~~~~T:~\tau\mapsto \tau+1\,.
\end{equation}
We check immediately that in $\Gamma$ we have the relations
\begin{equation}
S^2=-I,~~~~S^4=(ST)^3=I,~~~~~S^2T=TS^2
\end{equation}
and also $(TS)^3=I$ which is equivalent to $(ST)^3=I$. The corresponding relations in $\overline{\Gamma}$ are $S^2=(ST)^3=I$. The   homogeneous modular group $\Gamma$ has a series of infinite normal subgroups $\Gamma(N)$ ($N=1, 2, 3,\ldots$) defined as,
\begin{equation}
\Gamma(N)=\left\{\begin{pmatrix}
a  ~&  b  \\
c  ~& d
\end{pmatrix}\in SL(2, \mathbb{Z}),~~ \begin{pmatrix}
a  ~&  b  \\
c  ~& d
\end{pmatrix}=\begin{pmatrix}
1 ~& 0 \\
0  ~ & 1
\end{pmatrix}~(\texttt{mod}~N)
\right\}\,,
\end{equation}
which is called the principal congruence subgroup of level $N$. We have $\Gamma=\Gamma(1)$ and the element $T^N$ belongs to $\Gamma(N)$, i.e. $T^N\in\Gamma(N)$. For $N=1$ and $N=2$, we can define $\overline{\Gamma}(N)=\Gamma(N)/\{I, -I\}$. When $N>2$, we have $\overline{\Gamma}(N)=\Gamma(N)$ because the element $-I$ doesn't belong to $\Gamma(N)$. The homogeneous finite modular group $\Gamma'_N$ and inhomogeneous finite modular group $\Gamma_N$ can be obtained from the modular groups $\Gamma$ and $\overline{\Gamma}$ as the quotient groups
$\Gamma'_N=\Gamma/\Gamma(N)$ and $\Gamma_N=\overline{\Gamma}/\overline{\Gamma}(N)$. Therefore we have $\Gamma'_2\cong \Gamma_2$, and $\Gamma_N$ is isomorphic to the quotient group $\Gamma'_N/\{I, -I\}$ for $N>2$. The finite modular groups $\Gamma'_N$ and $\Gamma_N$ can also be obtained by imposing an additional condition $T^N=1$, so that their defining relations in terms of the generators $S$ and $T$ are\footnote{The homogeneous finite modular group $\Gamma'_N$ can also be equivalently presented as $S^2=R,~(ST)^3=R^2=T^N=1, ~RT=TR$.}
\begin{eqnarray}
\nonumber&& \Gamma'_N:~S^4=(ST)^3=T^N=1, ~S^2T=TS^2\,,\qquad\quad   N\leq5\,,    \\
&& \Gamma_N:~S^2=(ST)^3=T^N=1\,,\qquad\qquad\qquad\qquad~~ N\leq5\,.
\end{eqnarray}
Note that additional relations are necessary in order to render the groups $\Gamma'_N$ and $\Gamma_N$ finite for $N>5$~\cite{deAdelhartToorop:2011re}. For $N$ taking small values, $\Gamma_N$ and $\Gamma'_N$ are isomorphic to permutation groups and their double covering groups, $\Gamma_2=\Gamma'_2\cong S_3$, $\Gamma_3\cong A_4$, $\Gamma'_3\cong T'$, $\Gamma_4\cong S_4$, $\Gamma'_4\cong S'_4$, $\Gamma_5\cong A_5$ and $\Gamma'_5\cong A'_5$ which will be studied in the present work.

Modular forms of integral weight $k$ and level $N$ are holomorphic functions $f(\tau)$ satisfying the modular transformation property,
\begin{equation}
\label{eq:modular-transfm}f(\gamma\tau)=(c\tau+d)^kf(\tau),~~~\gamma\in\Gamma(N)\,.
\end{equation}
Modular forms of weight $k$ and level $N$ span a linear space of finite dimension.
It is always possible to choose a basis in this space such that a set of modular forms $f_i(\tau)$ are arranged into weight $k$ modular form multiplet $Y^{(k)}_{\mathbf{r}}(\tau)=(f_1(\tau), f_2(\tau), \ldots)^{T}$ in the irreducible representation $\mathbf{r}$ of the homogeneous finite modular group $\Gamma'_N$.
Under the full modular group $\Gamma$, the modular multiplet $Y^{(k)}_{\mathbf{r}}(\tau)$ transforms as~\cite{Liu:2019khw}
\begin{equation}
\label{eq:MF-decomp}Y^{(k)}_{\mathbf{r}}(\tau)\mapsto Y^{(k)}_{\mathbf{r}}(\gamma\tau)=(c\tau+d)^k\rho_{\mathbf{r}}(\gamma)Y^{(k)}_{\mathbf{r}}(\tau)\,,
\end{equation}
where $\gamma$ denotes a representative element in $\Gamma'_N$, $(c\tau+d)^k$ is the automorphy factor of $\gamma$ with modular weight $k$.
If the finite modular group is $\Gamma_N$, the modular weight $k$ should be even. It is sufficient to require that Eq.~\eqref{eq:MF-decomp} holds for $\gamma$ equal to the generators $S$ and $T$,
\begin{equation}
\label{eq:MF-decom-ST}Y^{(k)}_{\mathbf{r}}(-1/\tau)=(-\tau)^k\rho_{\mathbf{r}}(S)Y^{(k)}_{\mathbf{r}}(\tau),~~~~Y^{(k)}_{\mathbf{r}}(\tau+1)=\rho_{\mathbf{r}}(T)Y^{(k)}_{\mathbf{r}}(\tau)\,.
\end{equation}
Applying Eq.~\eqref{eq:MF-decomp} to $\gamma=S^2=R$, we have
\begin{equation}
Y^{(k)}_{\mathbf{r}}(\tau)=(-1)^{k}\rho_{\mathbf{r}}(R)Y^{(k)}_{\mathbf{r}}(\tau)\,,
\end{equation}
which implies
\begin{equation}
\rho_{\mathbf{r}}(R)=(-1)^{k}=\left\{\begin{array}{cc}
1,~&~ \text{even~}k \\
-1,~&~ \text{odd~}k
\end{array}
\right.\,.
\end{equation}
Hence the odd and even weight modular forms furnish irreducible representations of $\Gamma'_N$ in which the element $R=S^2$ is represented by $-1$ and $+1$ respectively.

\subsection{Integral weight modular forms of level 5}

The linear space of weight $k$ and level 5 modular forms has dimension $5k+1$, and it can be explicitly constructed in terms of the Dedekind eta function and Klein forms, as shown in Eq.~\eqref{eq:MF_Gamma5}.
In the following, we proceed to construct integral weight modular forms of level 5 by using weight $1/5$ modular forms. It has been proved that the linear space of modular form of weight $1/5$ and level 5 is spanned by two algebraically independent modular functions $F_1(\tau)$ and $F_2(\tau)$ with~\cite{ibukiyama2000modular}
\begin{equation}
F_1(\tau)= e^{-\frac{\pi i}{10}}\dfrac{\theta_{(\frac{1}{10},\frac{1}{2})}(5\tau)}{\eta(\tau)^{3/5}},~~~F_2(\tau)= e^{-\frac{3\pi i}{10}}\dfrac{\theta_{(\frac{3}{10},\frac{1}{2})}(5\tau)}{\eta(\tau)^{3/5}}\,,
\end{equation}
where $\theta_{(m',m'')}(\tau)$ is the theta constant and $\eta(\tau)$ is the Dedekind eta function, see section~\ref{sec:rational-weight-MF} for more details about the definitions and properties of these two special functions.
The transformation rules of $F_1(\tau)$ and $F_2(\tau)$ under the modular generators $S$ and $T$ are
\begin{eqnarray}
\nonumber&&F_1(\tau)\stackrel{S}{\rightarrow}(-\tau)^{1/5}e^{i\frac{\pi}{10}} \sqrt{\frac{1}{\sqrt{5}\phi}}\,\left[\phi F_1(\tau)+ F_2(\tau)\right],~~~F_1(\tau)\stackrel{T}{\rightarrow}F_1(\tau)\,,\\
&&F_2(\tau)\stackrel{S}{\rightarrow}(-\tau)^{1/5}e^{i\frac{\pi}{10}} \sqrt{\frac{1}{\sqrt{5}\phi}}\,\left[ F_1(\tau)-\phi F_2(\tau)\right],~~~F_2(\tau)\stackrel{T}{\rightarrow}e^{i\frac{2\pi}{5}} F_2(\tau)\,,
\end{eqnarray}
with $\phi=(1+\sqrt{5})/2$.
The modular space of level 5 can be generated by $F_1(\tau)$ and $F_2(\tau)$, and each modular form of integral weight $k$ and level 5 can be written as a polynomial of degree $5k$ in $F_1(\tau)$ and $F_{2}(\tau)$:
\begin{equation}
\sum_{i=0}^{5k} c_i F^i_{1}(\tau)F^{5k-i}_2(\tau)\,.
\end{equation}
Because $F_1(\tau)$ and $F_{2}(\tau)$ are algebraically independent, all terms in above polynomial are linearly independent, and obviously the number of independent terms matches with the correct dimension $5k+1$. Without loss of generality, we can choose a set of basis vectors of the weight 1 modular space as
\begin{equation}
F^5_1(\tau), ~~F^4_1(\tau) F_2(\tau), ~~F^3_1(\tau) F^2_2(\tau), ~~F^2_1(\tau) F^3_2(\tau), ~~F_1(\tau) F^4_2(\tau),~~F^5_2(\tau)\,.
\end{equation}
Solving the decomposition equation of modular forms in Eq.~\eqref{eq:MF-decom-ST}, we find that these six modular functions can be organized into a six dimensional representation $\mathbf{6}$ of $\Gamma'_5=A'_5$. In the representation basis collected in table~\ref{tab:rep-matrices}, the sextet modular form is given by
\begin{equation}
Y^{(1)}_{\mathbf{6}}(\tau)= \begin{pmatrix} F_1^5 + 2 F_2^5  \\  2 F_1^5 - F_2^5 \\ 5 F_1^4 F_2 \\ 5\sqrt{2}F_1^3 F_2^2 \\-5\sqrt{2} F_1^2 F_2^3 \\ 5F_1 F_2^4 \end{pmatrix}\equiv \begin{pmatrix}
Y_1(\tau) \\
Y_2(\tau) \\
Y_3(\tau) \\
Y_4(\tau) \\
Y_5(\tau) \\
Y_6(\tau)
\end{pmatrix}\,.
\end{equation}
The same result is obtained in section~\ref{sec:rational-weight-MF}, where the weight 1 modular forms are constructed from the tensor product of weight $1/5$ modular forms by using the CG coefficient of $\widetilde{\Gamma}_5$. Here $\widetilde{\Gamma}_5$ is the 5-fold metaplectic covering of $A_5$, and more details are explained in section~\ref{sec:rational-weight-MF}. The $q-$expansion of $Y_i(\tau)$ is given by Eq.~\eqref{eq:wt1-MF-1/5}.

The higher weight modular forms can be constructed from tensor products of $Y_{\mathbf{6}}$. For example, there are 11 independent weight 2 modular forms of level 5, which can be decomposed into two triplets and a quintet transforming in the $\mathbf{3}$, $\mathbf{3'}$ and $\mathbf{5}$ irreducible representations of $A'_5$. Without loss of generality we can choose the weight 2 modular forms as
\begin{equation}
  \begin{split}
Y_{\mathbf{3}}^{(2)}&=(Y_{\mathbf{6}}^{(1)}Y_{\mathbf{6}}^{(1)})_{\mathbf{3}_{1, s}}=
\begin{pmatrix}
 -2 \left(Y_1 Y_2+Y_4 Y_5-Y_3 Y_6\right) \\
 \sqrt{2} \left(Y_5^2-2 Y_2 Y_3\right) \\
 -\sqrt{2} \left(Y_4^2+2 Y_1 Y_6\right) \\
\end{pmatrix}\,,\\
Y_{\mathbf{3'}}^{(2)}&=(Y_{\mathbf{6}}^{(1)}Y_{\mathbf{6}}^{(1)})_{\mathbf{3'}_{1, s}}=
\begin{pmatrix}
 2 \left(Y_1 Y_2+Y_3 Y_6\right) \\
 2 Y_5 Y_6-2 Y_2 Y_4 \\
 -2 \left(Y_3 Y_4+Y_1 Y_5\right) \\
\end{pmatrix}\,,\\
Y_{\mathbf{5}}^{(2)}&=(Y_{\mathbf{6}}^{(1)}Y_{\mathbf{6}}^{(1)})_{\mathbf{5}_{1, s}}=
\begin{pmatrix}
 -\sqrt{6} \left(Y_1^2+Y_2^2\right) \\
 2 \left(Y_5^2+Y_1 Y_3+Y_2 Y_3+\sqrt{2} Y_4 Y_6\right) \\
 2 \left(Y_3^2+\sqrt{2} \left(Y_2 Y_4+Y_5 Y_6\right)\right) \\
 2 \left(Y_6^2+\sqrt{2} Y_3 Y_4-\sqrt{2} Y_1 Y_5\right) \\
 2 \left(Y_4^2-\sqrt{2} Y_3 Y_5+\left(Y_2-Y_1\right) Y_6\right) \\
\end{pmatrix}\,.
\end{split}
\end{equation}
We check that the $q$-expansions of $Y^{(2)}_{\mathbf{3}}$, $Y^{(2)}_{\mathbf{3'}}$ and $Y^{(2)}_{\mathbf{5}}$ are identical with those of the weight 2 modular forms of level 5 reached in Refs.~\cite{Novichkov:2018nkm,Ding:2019xna} up to some irrelevant overall constants. From the Kronecker product $\mathbf{6}\otimes \mathbf{6}=\mathbf{1_a}\oplus \mathbf{3_{1,s}}\oplus \mathbf{3_{2,s}}\oplus \mathbf{3'_{1,s}}
\oplus \mathbf{3'_{2,s}}\oplus \mathbf{4_s}\oplus \mathbf{4_a}\oplus \mathbf{5_{1,s}}\oplus \mathbf{5_{2,a}}\oplus \mathbf{5_{3,a}}$, we know that the contraction $Y_{\mathbf{6}}^{(1)}Y_{\mathbf{6}}^{(1)}$ can give rise to two additional weight two triplet modular forms $\left(Y_{\mathbf{6}}^{(1)}Y_{\mathbf{6}}^{(1)}\right)_{\mathbf{3_{2,s}}}$ and $\left(Y_{\mathbf{6}}^{(1)}Y_{\mathbf{6}}^{(1)}\right)_{\mathbf{3'_{2,s}}}$. From the $q$-expansion we see that they are proportional to $Y^{(2)}_{\mathbf{3}}$ and $Y^{(2)}_{\mathbf{3'}}$ respectively:
\begin{equation}
\left(Y_{\mathbf{6}}^{(1)}Y_{\mathbf{6}}^{(1)}\right)_{\mathbf{3_{2,s}}}=-\frac{3}{2}Y^{(2)}_{\mathbf{3}},\quad \left(Y_{\mathbf{6}}^{(1)}Y_{\mathbf{6}}^{(1)}\right)_{\mathbf{3'_{2,s}}}=\frac{3}{4}Y^{(2)}_{\mathbf{3'}}\,,
\end{equation}
which implies
\begin{eqnarray}
\nonumber&&2Y_1^2+3Y_1Y_2-2 Y_2^2-3Y_3Y_6-Y_4Y_5=0\,,\\
\nonumber&&3Y_5^2+4Y_1 Y_3-2Y_2 Y_3-4\sqrt{2} Y_4 Y_6=0\,, \\
\nonumber&&3Y_4^2+4\sqrt{2}Y_3Y_5+2\left(Y_1+2Y_2\right)Y_6=0\,,\\
\nonumber&&2Y_1^2+3Y_1Y_2-2Y_2^2+4Y_4Y_5+7Y_3Y_6=0\,, \\
\nonumber&&2\sqrt{2}Y_3^2+\left(4Y_1-7Y_2\right)Y_4+3Y_5Y_6=0\,,\\
\label{eq:constraint1_wt1MF}&&2\sqrt{2} Y_6^2+\left(7 Y_1+4 Y_2\right)Y_5+3Y_3Y_4=0\,.
\end{eqnarray}
Moreover we find the modular forms $Y_i(\tau)$ ($i=1,\ldots,6$) satisfy the following constraints,
\begin{eqnarray}
\nonumber&&2 Y_5^2+2\left(Y_2-2Y_1\right)Y_3-\sqrt{2}Y_4Y_6=0\,, \\
\nonumber&&\sqrt{2}Y_3^2-\left(3Y_1+Y_2\right)Y_4-Y_5Y_6=0\,, \\
\nonumber&&\sqrt{2}Y_6^2+\left(Y_1-3 Y_2\right)Y_5-Y_3Y_4=0\,,\\
\label{eq:constraint2_wt1MF}&&\sqrt{2}Y_4^2-\sqrt{2}\left(Y_1+2Y_2\right)Y_6+Y_3Y_5=0\,.
\end{eqnarray}
Hence the contraction $\left(Y_{\mathbf{6}}Y_{\mathbf{6}}\right)_{\mathbf{4_{s}}}$ is vanishing, i.e.,
\begin{equation}
\left(Y_{\mathbf{6}}Y_{\mathbf{6}}\right)_{\mathbf{4_{s}}}=0\,.
\end{equation}
\begin{table}[t!]
\centering
\begin{tabular}{|c|c|}
\hline  \hline

Modular weight $k$ & Modular forms $Y^{(k)}_{\mathbf{r}}$ \\ \hline

$k=1$ & $Y^{(1)}_{\mathbf{6}}$\\  \hline

$k=2$ & $Y^{(2)}_{\mathbf{3}}, ~Y^{(2)}_{\mathbf{3'}}, ~Y^{(2)}_{\mathbf{5}}$\\ \hline

$k=3$ & $Y^{(3)}_{\mathbf{4'}},~ Y^{(3)}_{\mathbf{6}I},~Y^{(3)}_{\mathbf{6}II}$\\ \hline

$k=4$ & $Y^{(4)}_{\mathbf{1}},~Y^{(4)}_{\mathbf{3}},~ Y^{(4)}_{\mathbf{3'}}, ~Y^{(4)}_{\mathbf{4}}, ~Y^{(4)}_{\mathbf{5}I}, ~Y^{(4)}_{\mathbf{5}II}$\\ \hline

$k=5$ & $Y^{(5)}_{\mathbf{2}},~ Y^{(5)}_{\mathbf{2'}}, ~Y^{(5)}_{\mathbf{4'}},~Y^{(5)}_{\mathbf{6}I},~Y^{(5)}_{\mathbf{6}II},~Y^{(5)}_{\mathbf{6}III}$\\ \hline

$k=6$ & $Y^{(6)}_{\mathbf{1}},~Y^{(6)}_{\mathbf{3}I},~Y^{(6)}_{\mathbf{3}II},~Y^{(6)}_{\mathbf{3'}I},~Y^{(6)}_{\mathbf{3'}II},~Y^{(6)}_{\mathbf{4}I},~Y^{(6)}_{\mathbf{4}II},~Y^{(6)}_{\mathbf{5}I},~Y^{(6)}_{\mathbf{5}II}$\\ \hline \hline
\end{tabular}
\caption{\label{Tab:Level5_MF}Summary of integral weight modular forms of level $N=5$ up to weight 6, the subscript $\mathbf{r}$ denote the transformation property under homogeneous finite modular group $SL(2, Z_5)\cong A'_5$. }
\end{table}
The linear space of modular forms of weight $k=3$ and level 5 has dimension $5k+1=16$, and they can be decomposed into a quartet and two sextets transforming as $\mathbf{4'}$ and $\mathbf{6}$ under $A'_5$,
\begin{equation}
  \begin{split}
Y_{\mathbf{4'}}^{(3)}&=(Y_{\mathbf{6}}^{(1)}Y_{\mathbf{3'}}^{(2)})_{\mathbf{4'}}=
\begin{pmatrix}
 -\sqrt{6} Y_3 Y_{\mathbf{3}',1}^{(2)}-\sqrt{3} Y_6 Y_{\mathbf{3}',2}^{(2)}+\sqrt{6} Y_5 Y_{\mathbf{3}',3}^{(2)} \\
 -2 Y_4 Y_{\mathbf{3}',1}^{(2)}+Y_1 Y_{\mathbf{3}',2}^{(2)}-3 Y_2 Y_{\mathbf{3}',2}^{(2)}+Y_6 Y_{\mathbf{3}',3}^{(2)} \\
 -2 Y_5 Y_{\mathbf{3}',1}^{(2)}-Y_3 Y_{\mathbf{3}',2}^{(2)}+\left(3 Y_1+Y_2\right) Y_{\mathbf{3}',3}^{(2)} \\
 -\sqrt{6} Y_6 Y_{\mathbf{3}',1}^{(2)}-\sqrt{6} Y_4 Y_{\mathbf{3}',2}^{(2)}+\sqrt{3} Y_3 Y_{\mathbf{3}',3}^{(2)} \\
\end{pmatrix}\,, \\
Y_{\mathbf{6}I}^{(3)}&=(Y_{\mathbf{6}}^{(1)}Y_{\mathbf{3}}^{(2)})_{\mathbf{6}_{1}}=
\begin{pmatrix}
 -Y_1 Y_{\mathbf{3},1}^{(2)}-\sqrt{2} Y_3 Y_{\mathbf{3},3}^{(2)} \\
 Y_2 Y_{\mathbf{3},1}^{(2)}+\sqrt{2} Y_6 Y_{\mathbf{3},2}^{(2)} \\
 Y_3 Y_{\mathbf{3},1}^{(2)}-\sqrt{2} Y_1 Y_{\mathbf{3},2}^{(2)} \\
 \sqrt{2} Y_5 Y_{\mathbf{3},3}^{(2)}-Y_4 Y_{\mathbf{3},1}^{(2)} \\
 Y_5 Y_{\mathbf{3},1}^{(2)}+\sqrt{2} Y_4 Y_{\mathbf{3},2}^{(2)} \\
 \sqrt{2} Y_2 Y_{\mathbf{3},3}^{(2)}-Y_6 Y_{\mathbf{3},1}^{(2)} \\
\end{pmatrix} \,, \\
Y_{\mathbf{6}II}^{(3)}&=(Y_{\mathbf{6}}^{(1)}Y_{\mathbf{5}}^{(2)})_{\mathbf{6}_{3}}=
\begin{pmatrix}
 \sqrt{2} Y_1 Y_{\mathbf{5},1}^{(2)}+\sqrt{3} \left(Y_6 Y_{\mathbf{5},2}^{(2)}-\sqrt{2} Y_5 Y_{\mathbf{5},3}^{(2)}-\sqrt{2} Y_4 Y_{\mathbf{5},4}^{(2)}+Y_3 Y_{\mathbf{5},5}^{(2)}\right) \\
 \sqrt{2} Y_2 Y_{\mathbf{5},1}^{(2)}+\sqrt{3} \left(Y_6 Y_{\mathbf{5},2}^{(2)}-\sqrt{2} Y_5 Y_{\mathbf{5},3}^{(2)}+\sqrt{2} Y_4 Y_{\mathbf{5},4}^{(2)}-Y_3 Y_{\mathbf{5},5}^{(2)}\right) \\
 \sqrt{3} \left(Y_1 Y_{\mathbf{5},2}^{(2)}-Y_2 Y_{\mathbf{5},2}^{(2)}+\sqrt{2} Y_4 Y_{\mathbf{5},5}^{(2)}\right)-2 \sqrt{2} Y_3 Y_{\mathbf{5},1}^{(2)} \\
 \sqrt{2} Y_4 Y_{\mathbf{5},1}^{(2)}+\sqrt{6} \left(Y_3 Y_{\mathbf{5},2}^{(2)}+\left(Y_2-Y_1\right) Y_{\mathbf{5},3}^{(2)}\right) \\
 \sqrt{2} Y_5 Y_{\mathbf{5},1}^{(2)}-\sqrt{6} \left(Y_1 Y_{\mathbf{5},4}^{(2)}+Y_2 Y_{\mathbf{5},4}^{(2)}-Y_6 Y_{\mathbf{5},5}^{(2)}\right) \\
 \sqrt{3} \left(\sqrt{2} Y_5 Y_{\mathbf{5},2}^{(2)}+\left(Y_1+Y_2\right) Y_{\mathbf{5},5}^{(2)}\right)-2 \sqrt{2} Y_6 Y_{\mathbf{5},1}^{(2)} \\
\end{pmatrix}\,,
  \end{split}
\end{equation}
where $Y_{\mathbf{3}}^{(2)}\equiv (Y_{\mathbf{3}, 1}^{(2)}, Y_{\mathbf{3}, 2}^{(2)}, Y_{\mathbf{3}, 3}^{(2)})^{T}$, $Y_{\mathbf{3}'}^{(2)}\equiv (Y_{\mathbf{3}', 1}^{(2)}, Y_{\mathbf{3}', 2}^{(2)}, Y_{\mathbf{3}', 3}^{(2)})^{T}$ and $Y_{\mathbf{5}}^{(2)}\equiv (Y_{\mathbf{5}, 1}^{(2)}, Y_{\mathbf{5}, 2}^{(2)}, Y_{\mathbf{5}, 3}^{(2)}, Y_{\mathbf{5}, 4}^{(2)}, Y_{\mathbf{5}, 5}^{(2)})^{T}$ are denoted, and similar notations are adopted for the modular forms in this work. The explicit expressions of the higher weight modular forms are given in Appendix~\ref{sec:higher-weight-MF}. The structure of modular forms at level 5 is summarized in table~\ref{Tab:Level5_MF} up to weight 6.

\section{\label{sec:gCP-A5DC}Combining generalized CP with $A'_5$ modular symmetry}

It has been established that the modular symmetry group $SL(2,\mathbb{Z})$ can be consistently combined with the generalized CP symmetry such that $SL(2,\mathbb{Z})$ is enhanced to $GL(2,\mathbb{Z})$~\cite{Novichkov:2019sqv}, and the CP transformation is represented by the matrix
\begin{equation}
\mathscr{C}=\begin{pmatrix}
1 ~&  0 \\
0  ~& -1
\end{pmatrix}\,.
\end{equation}
The CP transformation has to act on the complex modulus $\tau$ as~\cite{Novichkov:2019sqv,Baur:2019kwi,Acharya:1995ag,Dent:2001cc,Giedt:2002ns}
\begin{equation}
\tau\stackrel{\mathcal{CP}}{\longrightarrow}-\tau^{*}\,.
\end{equation}
If we perform a CP transformation, followed by a modular transformation and subsequently an inverse CP transformation, we end up with
\begin{equation}
\label{eq:consistency-chain}\tau\stackrel{\mathcal{CP}}{\longrightarrow}-\tau^{*}\stackrel{\gamma}{\longrightarrow} -\frac{a\tau^{*}+b}{c\tau^{*}+d}\stackrel{\mathcal{CP}^{-1}}{\longrightarrow}\frac{a\tau-b}{-c\tau+d}=u(\gamma)\tau\,,
\end{equation}
where
\begin{equation}
u(\gamma)=\begin{pmatrix}
a ~& -b  \\
-c  ~&  d
\end{pmatrix}=\mathscr{C}\gamma\mathscr{C}^{-1}\in SL(2,\mathbb{Z})\,.
\end{equation}
Obviously we have $u(S)=S^{-1}$ and $u(T)=T^{-1}$. Therefore the generalized CP transformation corresponds to an automorphism $u(\gamma)$ of the modular group. A second CP transformation $CP_2$ for the automorphism $u(S)=-S^{-1}$ and $u(T)=-T^{-1}$ can possibly be defined if the level $N$ is even~\cite{Novichkov:2020eep}. Obviously $CP_2$ is not allowed in the context of $N=5$ which is an odd number. For a generic chiral supermultiplets $\varphi$, its transfromation properties are characterized by $k$ and $\rho_{\mathbf{r}}$, where $-k$ is the modular weight of $\varphi$ and $\rho_{\mathbf{r}}$ is an irreducible representation of the homogeneous finite modular group $\Gamma'_N$. Under a modular transformation $\gamma$, the multiplet $\varphi(x)$ transforms according to
\begin{equation}
\varphi(x)\stackrel{\gamma}{\longrightarrow} (c\tau+d)^{-k}\rho_{\mathbf{r}}(\gamma)\varphi(x)\,.
\end{equation}
The action of generalized CP symmetry on the multiplet $\varphi$ is
\begin{equation}
\varphi(x)\stackrel{\mathcal{CP}}{\longrightarrow} X_{\mathbf{r}}\overline{\varphi}(x_{\mathcal{P}})\,,
\end{equation}
where $x_{\mathcal{P}}=(t, -\vec{x})$, a bar denotes the hermitian conjugate superfield, and $X_{\mathbf{r}}$ is a unitary matrix that acts on flavor space. Now we consider the same transformation chain in Eq.~\eqref{eq:consistency-chain},
\begin{equation}
\label{eq:varphi-chain}\varphi(x)\stackrel{\mathcal{CP}}{\longrightarrow} X_{\mathbf{r}}\overline{\varphi}(x_{\mathcal{P}})
\stackrel{\gamma}{\longrightarrow}(c\tau^{*}+d)^{-k} X_{\mathbf{r}}\rho^{*}_{\mathbf{r}}(\gamma)\overline{\varphi}(x_{\mathcal{P}})\stackrel{\mathcal{CP}^{-1}}{\longrightarrow}
(-c\tau+d)^{-k} X_{\mathbf{r}}\rho^{*}_{\mathbf{r}}(\gamma)X^{-1}_{\mathbf{r}}\varphi(x)\,.
\end{equation}
Consistency between the modular and CP trnasformations requires that the resulting transformation should be equivalent to the modular transformation $u(\gamma)$, i.e., $\mathcal{CP}\circ \gamma\circ\mathcal{CP}^{-1}=u(\gamma)$.
One can read out the transformation of $\varphi(x)$ under $u(\gamma)$ as,
\begin{equation}
\label{eq:ugamma-act}\varphi(x)\stackrel{u(\gamma)}{\longrightarrow}(-c\tau+d)^{-k}\rho_{\mathbf{r}}(u(\gamma))\varphi(x)\,.
\end{equation}
Comparing Eq.~\eqref{eq:varphi-chain} with Eq.~\eqref{eq:ugamma-act}, we reach the consistency condition~\cite{Novichkov:2019sqv},
\begin{equation}
\label{eq:consistency-cond}X_{\mathbf{r}}\rho^{*}_{\mathbf{r}}(\gamma)X^{-1}_{\mathbf{r}}=\rho_{\mathbf{r}}(u(\gamma))\,.
\end{equation}
For each irreducible representation $\mathbf{r}$ of $\Gamma'_N$, the above consistency condition fixes the generalized CP transformation $X_{\mathbf{r}}$ up to an overall phase. It is sufficient to require that Eq.~\eqref{eq:consistency-cond} holds for $\gamma$ equal to $S$ and $T$,
\begin{equation}
X_{\mathbf{r}}\rho^{*}_{\mathbf{r}}(S)X^{-1}_{\mathbf{r}}=\rho^{\dagger}_{\mathbf{r}}(S),\qquad X_{\mathbf{r}}\rho^{*}_{\mathbf{r}}(T)X^{-1}_{\mathbf{r}}=\rho^{\dagger}_{\mathbf{r}}(T)\,.
\end{equation}
In our chosen representation basis in table~\ref{tab:rep-matrices} at level $N=5$, both generators $S$ and $T$ are represented by symmetric matrices in all irreducible representations of $A'_5$. Therefore the CP transformation $X_{\mathbf{r}}$ is exactly the canonical CP,
\begin{equation}
\label{eq:Xr_S4p}X_{\mathbf{r}}=1\,.
\end{equation}
Under the action of CP, the modular forms of weight 1 and level 5 transform as
\begin{equation}
Y^{(1)}_{\mathbf{6}}(\tau)\stackrel{\mathcal{CP}}{\longrightarrow} Y^{(1)}_{\mathbf{6}}(-\tau^{*})=[Y^{(1)}_{\mathbf{6}}(\tau)]^{*}\,.
\end{equation}
All the CG coefficients in our working basis are real, consequently the identity $Y^{(k)}_{\mathbf{r}}(\tau)\stackrel{\mathcal{CP}}{\longmapsto} Y^{(k)}_{\mathbf{r}}(-\tau^{*})=[Y^{(k)}_{\mathbf{r}}(\tau)]^{*}$ is satisfied for any integral weight modular forms $Y^{(k)}_{\mathbf{r}}(\tau)$ at level 5, and CP invariance would enforce all coupling constants to be real. Once CP invariance is incorporated in, the symmetry of the theory would be enhanced. As a result, the homogeneous finite modular group $A'_5$ would be enlarged to the CP extended finite modular group $\Gamma^{*}_5$ defined by~\cite{Nilles:2020nnc}
\begin{equation}
\Gamma^{*}_5: S^4=T^5=(ST)^3=\mathscr{C}^2=1,\quad S^2T=TS^2\,,~~\mathscr{C}S\mathscr{C}^{-1}=S^{-1}\,,~~\mathscr{C}T\mathscr{C}^{-1}=T^{-1}\,.
\end{equation}

\section{\label{sec:lepton-models}Systematical classification of minimal lepton models based on $A'_5$ modular symmetry}

In this section,  we shall perform a systematical classification of all minimal lepton mass models with the $A'_5$ modular symmetry. We shall adopt the bottom-up approach of modular invariance in~\cite{Feruglio:2017spp}. We let $\Phi_I$ to denote a set of chiral superfields, it transforms under the modular transformation in the following way,
\begin{equation}
\label{eq:modularTrs_Phi}
\tau\to \gamma\tau=\frac{a\tau+b}{c\tau+d}\,,\qquad
\Phi_I\to (c\tau+d)^{-k_I}\rho_I(\gamma)\Phi_I\,,~~\gamma=\begin{pmatrix}
a   &  b  \\
c   &  d
\end{pmatrix}\in SL(2,\mathbb{Z})\,,
\end{equation}
where $-k_I$ is the modular weight of $\Phi_I$, and $\rho_I(\gamma)$ is the unitary representation of the representative element $\gamma$ in $\Gamma_N$.
There are no restrictions on the possible value of $k_I$ since the supermultiplets $\Phi_I$ are not modular forms. In the present work, the  K\"ahler potential is taken to be the minimal form~\cite{Feruglio:2017spp},
\begin{equation}
\mathcal{K}(\Phi_I, \overline{\Phi}_I; \tau,\bar{\tau}) =-h\Lambda^2 \log(-i\tau+i\bar\tau)+ \sum_I (-i\tau+i\bar\tau)^{-k_I} |\Phi_I|^2~~~,
\end{equation}
which gives rise to the kinetic terms for both the scalar and fermionic components of the supermultiplets $\Phi_I$ and the modulus superfield $\tau$. The superpotential $\mathcal{W}(\Phi_I,\tau)$ can be expanded in power series of the involved supermultiplets $\Phi_I$,
\begin{equation}
\mathcal{W}(\Phi_I,\tau) =\sum_n Y_{I_1...I_n}(\tau)~ \Phi_{I_1}... \Phi_{I_n}\,,
\end{equation}
where $Y_{I_1...I_n}$ is a modular multiplet of weight $k_Y$ and it transforms in the presentation $\rho_{Y}$ of $\Gamma'_{N}$,
\begin{equation}
\tau\to \gamma\tau =\dfrac{a\tau+b}{c\tau+d}\,,\qquad Y(\tau)\to Y(\gamma\tau)=(c\tau+d)^{k_Y}\rho_{Y}(\gamma)Y(\tau)\,.
\end{equation}
Modular invariance entails that each term of the superpotential should be invariant under the action of $\Gamma'_N$ and its modular weight should be vanishing, i.e.,
\begin{equation}
k_Y=k_{I_1}+...+k_{I_n},~\quad~ \rho_Y\otimes\rho_{I_1}\otimes\ldots\otimes\rho_{I_n}\ni\mathbf{1}\,.
\end{equation}
In the following, we shall perform a comprehensive and systematical study of
possible quark and lepton models with the $A_5'$ modular symmetry. We shall utilize the advantage of the modular symmetry and no flavon fields other than $\tau$ is introduced. We assign the Higgs doublets $H_u$ and $H_d$ to two singlets $\mathbf{1}$ and their modular weights $k_{H_u, H_d}$ are vanishing.

\begin{table}[!ht]
\centering
\begin{tabular}{|c|c|c|c|c|c|c|c|c|c|c|}
\hline\hline
 & $C^i$ & $C^{ii}$ & $C^{iii}$ & $C^{iv}$ & $C^{v}$ & $C^{vi}$ & $C^{vii}$ & $C^{viii}$ & $C^{ix}$ & $C^{x}$ \\ \hline
 $\rho_L$ & $\mathbf{3}$ & $\mathbf{3'}$ & $\mathbf{3}$ & $\mathbf{3}$ & $\mathbf{3'}$ & $\mathbf{3'}$ & $\mathbf{3}$ & $\mathbf{3}$ & $\mathbf{3'}$ & $\mathbf{3'}$ \\ \hline
 $\rho_{E^c_i}$ & $\mathbf{1}\oplus\mathbf{1}\oplus\mathbf{1}$ & $\mathbf{1}\oplus\mathbf{1}\oplus\mathbf{1}$ & $\mathbf{2}\oplus\mathbf{1}$ & $\mathbf{2'}\oplus\mathbf{1}$ & $\mathbf{2}\oplus\mathbf{1}$ & $\mathbf{2'}\oplus\mathbf{1}$ & $\mathbf{3}$ & $\mathbf{3'}$ & $\mathbf{3}$ & $\mathbf{3'}$\\ \hline
  \# of models & $1$ & $1$ & $6$ & $8$ & $8$ & $6$ & $2$ & $3$ & $3$ & $2$\\
 (total number) & $(1)$ & $(1)$ & $(6)$ & $(9)$ & $(9)$ & $(6)$ & $(3)$ & $(3)$ & $(3)$ & $(3)$ \\
\hline\hline
\end{tabular}
\caption{\label{tab:lepton-rep}The number (in bracket) of possible charged lepton models for different representation assignment of $L$ and $E^{c}_{1,2,3}$ under the finite modular group $A'_5$ up to weight 6 modular forms. We don't count the cases which give degenerate charged lepton masses. We also list the number (without bracket) of models which contain up to four independent terms in the charged lepton superpotential $\mathcal{W}_e$. }
\end{table}

\subsection{\label{subsec:charge-lepton} Charged lepton sector }

We assume the three generations of left-handed lepton doublets $L\equiv(L_1, L_2, L_3)^{T}$   transform as a triplet $\mathbf{3}$ or $\mathbf{3'}$, and the right-handed charged leptons $E^{c}_{1,2,3}$ are assigned to the direct product $\mathbf{1}\oplus\mathbf{1}\oplus\mathbf{1}$, $\mathbf{2}\oplus\mathbf{1}$, $\mathbf{2'}\oplus\mathbf{1}$ or a triplet $\mathbf{3}$, $\mathbf{3'}$.
The modular weights of $L$ and $E^{c}_{1,2,3}$ are denoted as $k_L$ and $k_{E^c_{1,2,3}}$ respectively. For each representation assignment, there are in principle infinite possible weight assignments for the fields, and the number of the independent couplings in the superpotential $\mathcal{W}_e$ of the charged lepton mass terms generally increases with the weight of the involved modular forms. We consider modular forms up to weight 6 in the present work, the numbers of possible models in the charged lepton sector are shown in table~\ref{tab:lepton-rep}. Due to limitation of space we will list the models for which $\mathcal{W}_e$ contains at most four independent terms in the following.
\begin{description}[labelindent=-0.8em, leftmargin=0.3em]

\item[~~(\romannumeral1)]{$\rho_{L}=\mathbf{3}$, $\rho_{E^c_{1,2,3}}=\mathbf{1}$ }  \\
In this case, the three right-handed charged leptons are distinguished from each other by their different modular weights. For the weight assignments fulfilling $k_{E_{1,2,3}^c}+k_L=2,4,6$, the modular invariant superpotential is of the form
\begin{equation}
C^i_1: ~\mathcal{W}_{e}=\alpha \left( Y_{\mathbf{3}}^{(2)} E^{c}_{1}L H_{d}\right)_{\mathbf{1}}+\beta \left( Y_{\mathbf{3}}^{(4)} E^{c}_{2}L H_{d}\right)_{\mathbf{1}}+\gamma \left( Y_{\mathbf{3}I}^{(6)} E^{c}_{3}L H_{d}\right)_{\mathbf{1}}+\delta \left( Y_{\mathbf{3}II}^{(6)} E^{c}_{3}L H_{d}\right)_{\mathbf{1}}\,.
\end{equation}
The charged lepton mass matrix can be straightforwardly read out and its explicit form is listed in the supplementary file~\cite{Yao:2020sup}.

\item[~~(\romannumeral2)]{$\rho_{L}=\mathbf{3}$, $\rho_{E^c_{1,2,3}}=\mathbf{1}$ }

For the lowest weight assignment $k_{E_{1,2,3}^c}+k_L=2,4,6$, the charged lepton superpotential reads as
\begin{equation}
C^{ii}_1:~~\mathcal{W}_{e}=\alpha \left( Y_{\mathbf{3}'}^{(2)} E^{c}_{1}L H_{d}\right)_{\mathbf{1}}+\beta \left( Y_{\mathbf{3}'}^{(4)} E^{c}_{2}L H_{d}\right)_{\mathbf{1}}+\gamma \left( Y_{\mathbf{3}'I}^{(6)} E^{c}_{3}L H_{d}\right)_{\mathbf{1}}+\delta \left( Y_{\mathbf{3}'II}^{(6)} E^{c}_{3}L H_{d}\right)_{\mathbf{1}} \,.
\end{equation}
For both models $C^{ii}_1$ and $C^{ii}_2$, the phases of $\alpha$, $\beta$, $\gamma$ can be absorbed into the right-handed charged leptons and they can be taken to be real while the parameter $\delta$ is generally complex.

\item[~~(\romannumeral3)]{$\rho_{L}=\mathbf{3}$, $\rho_{E^c_{D}}=\mathbf{2}$,  $\rho_{E^c_{3}}=\mathbf{1}$ }

The first two generations of the right-handed charged leptons are assigned to a doublet of $A'_5$ and we denote $E^{c}_D=(E^{c}_1, E^{c}_2)$. There are six possible values of modular weights for which up to four terms are involved in the charged lepton Yukawa couplings, and accordingly the superpotential is of the following form
\begin{small}
  \begin{eqnarray}
\nonumber C^{iii}_1:~~ \mathcal{W}_{e}&=&\alpha \left(Y_{\mathbf{4}'}^{(3)} E^{c}_{D}L H_{d} \right)_{\mathbf{1}}+\beta \left(Y_{\mathbf{3}}^{(2)} E^{c}_{3}L H_{d} \right)_{\mathbf{1}}\,,~\text{for}~k_{E_{D,3}^c}+k_L=3, 2\,,\\
\nonumber C^{iii}_2:~~ \mathcal{W}_{e}&=&\alpha \left(Y_{\mathbf{4}'}^{(3)} E^{c}_{D}L H_{d} \right)_{\mathbf{1}}+\beta \left(Y_{\mathbf{3}}^{(4)} E^{c}_{3}L H_{d} \right)_{\mathbf{1}}\,,~\text{for}~k_{E_{D,3}^c}+k_L=3, 4\,,\\
\nonumber C^{iii}_3:~~ \mathcal{W}_{e}&=&\alpha \left( Y_{\mathbf{2}}^{(5)} E^{c}_{D}L H_{d}\right)_{\mathbf{1}}+\beta \left( Y_{\mathbf{4}'}^{(5)} E^{c}_{D}L H_{d}\right)_{\mathbf{1}}+\gamma \left( Y_{\mathbf{3}}^{(2)} E^{c}_{3}L H_{d}\right)_{\mathbf{1}}\,,~\text{for}~k_{E_{D,3}^c}+k_L=5, 2\,,    \\
\nonumber C^{iii}_4:~~ \mathcal{W}_{e}&=&\alpha \left( Y_{\mathbf{2}}^{(5)} E^{c}_{D}L H_{d}\right)_{\mathbf{1}}+\beta \left( Y_{\mathbf{4}'}^{(5)} E^{c}_{D}L H_{d}\right)_{\mathbf{1}}+\gamma \left( Y_{\mathbf{3}}^{(4)} E^{c}_{3}L H_{d}\right)_{\mathbf{1}}\,,~\text{for}~k_{E_{D,3}^c}+k_L=5, 4\,, \\
\nonumber C^{iii}_5:~~ \mathcal{W}_{e}&=&\alpha \left( Y_{\mathbf{4}'}^{(3)} E^{c}_{D}L H_{d}\right)_{\mathbf{1}}+\beta \left( Y_{\mathbf{3}I}^{(6)} E^{c}_{3}L H_{d}\right)_{\mathbf{1}}+\gamma \left( Y_{\mathbf{3}II}^{(6)} E^{c}_{3}L H_{d}\right)_{\mathbf{1}}\,,~~\text{for}~~k_{E_{D,3}^c}+k_L=3, 6\,, \\
\nonumber C^{iii}_6:~~ \mathcal{W}_{e}&=&\alpha \left(Y_{\mathbf{2}}^{(5)} E^{c}_{D}L H_{d} \right)_{\mathbf{1}}+\beta \left(Y_{\mathbf{4}'}^{(5)} E^{c}_{D}L H_{d} \right)_{\mathbf{1}}+\gamma \left(Y_{\mathbf{3}I}^{(6)} E^{c}_{3}L H_{d} \right)_{\mathbf{1}}+\delta \left(Y_{\mathbf{3}II}^{(6)} E^{c}_{3}L H_{d} \right)_{\mathbf{1}}\,,\\
    &&\qquad\qquad\qquad\qquad\qquad\qquad\qquad\qquad\qquad\qquad\qquad\text{for}~~k_{E_{D,3}^c}+k_L=5, 6\,.
\end{eqnarray}
\end{small}
\item[~~(\romannumeral4)]{$\rho_{L}=\mathbf{3}$, $\rho_{E^c_{D}}=\mathbf{2'}$,  $\rho_{E^c_{3}}=\mathbf{1}$ }

Similar to the previous case, the charged lepton Yukawa coupling can also take eight possible forms,
\begin{small}
\begin{eqnarray}
\nonumber C^{iv}_1:~~ \mathcal{W}_{e}&=&\alpha \left(Y_{\mathbf{6}}^{(1)} E^{c}_{D}L H_{d} \right)_{\mathbf{1}}+\beta \left(Y_{\mathbf{3}}^{(2)} E^{c}_{3}L H_{d} \right)_{\mathbf{1}}\,,~\text{for}~k_{E_{D,3}^c}+k_L=1, 2\,,\\
\nonumber C^{iv}_2:~~ \mathcal{W}_{e}&=&\alpha \left(Y_{\mathbf{6}}^{(1)} E^{c}_{D}L H_{d} \right)_{\mathbf{1}}+\beta \left(Y_{\mathbf{3}}^{(4)} E^{c}_{3}L H_{d} \right)_{\mathbf{1}}\,,~\text{for}~k_{E_{D,3}^c}+k_L=1, 4\,,\\
\nonumber C^{iv}_3:~~ \mathcal{W}_{e}&=&\alpha \left( Y_{\mathbf{6}I}^{(3)} E^{c}_{D}L H_{d}\right)_{\mathbf{1}}+\beta \left( Y_{\mathbf{6}II}^{(3)} E^{c}_{D}L H_{d}\right)_{\mathbf{1}}+\gamma \left( Y_{\mathbf{3}}^{(2)} E^{c}_{3}L H_{d}\right)_{\mathbf{1}}\,,~\text{for}~k_{E_{D,3}^c}+k_L=3, 2\,,    \\
\nonumber C^{iv}_4:~~ \mathcal{W}_{e}&=& \alpha \left( Y_{\mathbf{6}I}^{(3)} E^{c}_{D}L H_{d}\right)_{\mathbf{1}}+\beta \left( Y_{\mathbf{6}II}^{(3)} E^{c}_{D}L H_{d}\right)_{\mathbf{1}}+\gamma \left( Y_{\mathbf{3}}^{(4)} E^{c}_{3}L H_{d}\right)_{\mathbf{1}}\,,~\text{for}~k_{E_{D,3}^c}+k_L=3, 4\,, \\
\nonumber C^{iv}_5:~~ \mathcal{W}_{e}&=&\alpha \left( Y_{\mathbf{6}}^{(1)} E^{c}_{D}L H_{d}\right)_{\mathbf{1}}+\beta \left( Y_{\mathbf{3}I}^{(6)} E^{c}_{3}L H_{d}\right)_{\mathbf{1}}+\gamma \left( Y_{\mathbf{3}II}^{(6)} E^{c}_{3}L H_{d}\right)_{\mathbf{1}}\,,~\text{for}~k_{E_{D,3}^c}+k_L=1, 6\,,\\
  \nonumber C^{iv}_6:~~ \mathcal{W}_{e}&=&\alpha \left(Y_{\mathbf{6}I}^{(5)} E^{c}_{D}L H_{d} \right)_{\mathbf{1}}+\beta \left(Y_{\mathbf{6}II}^{(5)} E^{c}_{D}L H_{d} \right)_{\mathbf{1}}+\gamma \left(Y_{\mathbf{6}III}^{(5)} E^{c}_{D}L H_{d} \right)_{\mathbf{1}}+\delta \left(Y_{\mathbf{3}}^{(2)} E^{c}_{3}L H_{d} \right)_{\mathbf{1}}\,,\\
 \nonumber &&\qquad\qquad\qquad\qquad\qquad\qquad\qquad\qquad\qquad\qquad\qquad~\text{for}~k_{E_{D,3}^c}+k_L=5, 2\,,\\
  \nonumber C^{iv}_7:~~ \mathcal{W}_{e}&=&\alpha \left(Y_{\mathbf{6}I}^{(3)} E^{c}_{D}L H_{d} \right)_{\mathbf{1}}+\beta \left(Y_{\mathbf{6}II}^{(3)} E^{c}_{D}L H_{d} \right)_{\mathbf{1}}+\gamma \left(Y_{\mathbf{3}II}^{(6)} E^{c}_{3}L H_{d} \right)_{\mathbf{1}}+\delta \left(Y_{\mathbf{3}I}^{(6)} E^{c}_{3}L H_{d} \right)_{\mathbf{1}}\,,\\
  \nonumber &&\qquad\qquad\qquad\qquad\qquad\qquad\qquad\qquad\qquad\qquad\qquad~\text{for}~k_{E_{D,3}^c}+k_L=3, 6\,,\\
  \nonumber C^{iv}_8:~~ \mathcal{W}_{e}&=&\alpha \left(Y_{\mathbf{6}I}^{(5)} E^{c}_{D}L H_{d} \right)_{\mathbf{1}}+\beta \left(Y_{\mathbf{6}II}^{(5)} E^{c}_{D}L H_{d} \right)_{\mathbf{1}}+\gamma \left(Y_{\mathbf{6}III}^{(5)} E^{c}_{D}L H_{d} \right)_{\mathbf{1}}+\delta \left(Y_{\mathbf{3}}^{(4)} E^{c}_{3}L H_{d} \right)_{\mathbf{1}}\,,\\
  &&\qquad\qquad\qquad\qquad\qquad\qquad\qquad\qquad\qquad\qquad\qquad~\text{for}~k_{E_{D,3}^c}+k_L=5, 4\,.
\end{eqnarray}
\end{small}
\item[~~(\romannumeral5)]{$\rho_{L}=\mathbf{3'}$, $\rho_{E^c_{D}}=\mathbf{2}$,  $\rho_{E^c_{3}}=\mathbf{1}$ }

Depending on the weights of lepton fields, we find that the charged lepton superpotential can be
\begin{small}
\begin{eqnarray}
\nonumber C^{v}_1:~~ \mathcal{W}_{e}&=&\alpha \left(Y_{\mathbf{6}}^{(1)} E^{c}_{D}L H_{d} \right)_{\mathbf{1}}+\beta \left(Y_{\mathbf{3}'}^{(2)} E^{c}_{3}L H_{d} \right)_{\mathbf{1}}\,,~\text{for}~k_{E_{D,3}^c}+k_L=1, 2\,,\\
\nonumber C^{v}_2:~~ \mathcal{W}_{e}&=&\alpha \left(Y_{\mathbf{6}}^{(1)} E^{c}_{D}L H_{d} \right)_{\mathbf{1}}+\beta \left(Y_{\mathbf{3}'}^{(4)} E^{c}_{3}L H_{d} \right)_{\mathbf{1}}\,,~\text{for}~k_{E_{D,3}^c}+k_L=1, 4\,,\\
\nonumber C^{v}_3:~~ \mathcal{W}_{e}&=&\alpha \left( Y_{\mathbf{6}I}^{(3)} E^{c}_{D}L H_{d}\right)_{\mathbf{1}}+\beta \left( Y_{\mathbf{6}II}^{(3)} E^{c}_{D}L H_{d}\right)_{\mathbf{1}}+\gamma \left( Y_{\mathbf{3}'}^{(2)} E^{c}_{3}L H_{d}\right)_{\mathbf{1}}\,,~\text{for}~k_{E_{D,3}^c}+k_L=3, 2\,, \\
\nonumber C^{v}_4:~~ \mathcal{W}_{e}&=& \alpha \left( Y_{\mathbf{6}I}^{(3)} E^{c}_{D}L H_{d}\right)_{\mathbf{1}}+\beta \left( Y_{\mathbf{6}II}^{(3)} E^{c}_{D}L H_{d}\right)_{\mathbf{1}}+\gamma \left( Y_{\mathbf{3}'}^{(4)} E^{c}_{3}L H_{d}\right)_{\mathbf{1}}\,,~\text{for}~k_{E_{D,3}^c}+k_L=3, 4\,, \\
\nonumber C^{v}_5:~~ \mathcal{W}_{e}&=&\alpha \left( Y_{\mathbf{6}}^{(1)} E^{c}_{D}L H_{d}\right)_{\mathbf{1}}+\beta \left( Y_{\mathbf{3}'I}^{(6)} E^{c}_{3}L H_{d}\right)_{\mathbf{1}}+\gamma \left( Y_{\mathbf{3}'II}^{(6)} E^{c}_{3}L H_{d}\right)_{\mathbf{1}}\,,~\text{for}~k_{E_{D,3}^c}+k_L=1, 6\,,\\
  \nonumber C^{v}_6:~~ \mathcal{W}_{e}&=&\alpha \left(Y_{\mathbf{6}I}^{(5)} E^{c}_{D}L H_{d} \right)_{\mathbf{1}}+\beta \left(Y_{\mathbf{6}II}^{(5)} E^{c}_{D}L H_{d} \right)_{\mathbf{1}}+\gamma \left(Y_{\mathbf{6}III}^{(5)} E^{c}_{D}L H_{d} \right)_{\mathbf{1}}+\delta \left(Y_{\mathbf{3}'}^{(2)} E^{c}_{3}L H_{d} \right)_{\mathbf{1}}\,,\\
  \nonumber&&\qquad\qquad\qquad\qquad\qquad\qquad\qquad\qquad\qquad\qquad\qquad~\text{for}~k_{E_{D,3}^c}+k_L=5, 2\,,\\
  \nonumber C^{v}_7:~~ \mathcal{W}_{e}&=&\alpha \left(Y_{\mathbf{6}I}^{(3)} E^{c}_{D}L H_{d} \right)_{\mathbf{1}}+\beta \left(Y_{\mathbf{6}II}^{(3)} E^{c}_{D}L H_{d} \right)_{\mathbf{1}}+\gamma \left(Y_{\mathbf{3}'II}^{(6)} E^{c}_{3}L H_{d} \right)_{\mathbf{1}}+\delta \left(Y_{\mathbf{3}'I}^{(6)} E^{c}_{3}L H_{d} \right)_{\mathbf{1}}\,,\\
\nonumber&&\qquad\qquad\qquad\qquad\qquad\qquad\qquad\qquad\qquad\qquad\qquad ~\text{for}~k_{E_{D,3}^c}+k_L=3, 6\,,\\
  \nonumber C^{v}_8:~~ \mathcal{W}_{e}&=&\alpha \left(Y_{\mathbf{6}I}^{(5)} E^{c}_{D}L H_{d} \right)_{\mathbf{1}}+\beta \left(Y_{\mathbf{6}II}^{(5)} E^{c}_{D}L H_{d} \right)_{\mathbf{1}}+\gamma \left(Y_{\mathbf{6}III}^{(5)} E^{c}_{D}L H_{d} \right)_{\mathbf{1}}+\delta \left(Y_{\mathbf{3}'}^{(4)} E^{c}_{3}L H_{d} \right)_{\mathbf{1}}\,,\\
\nonumber&&\qquad\qquad\qquad\qquad\qquad\qquad\qquad\qquad\qquad\qquad\qquad  ~\text{for}~k_{E_{D,3}^c}+k_L=5, 4\,.\\
\end{eqnarray}
\end{small}
\item[~~(\romannumeral6)]{$\rho_{L}=\mathbf{3'}$, $\rho_{E^c_{D}}=\mathbf{2'}$,  $\rho_{E^c_{3}}=\mathbf{1}$ }

In the same fashion as previous cases, we can read out the superpotential for charged lepton masses as follows,
\begin{small}
\begin{eqnarray}
\nonumber C^{vi}_1:~~ \mathcal{W}_{e}&=&\alpha \left(Y_{\mathbf{4}'}^{(3)} E^{c}_{D}L H_{d} \right)_{\mathbf{1}}+\beta \left(Y_{\mathbf{3}'}^{(2)} E^{c}_{3}L H_{d} \right)_{\mathbf{1}}\,,~\text{for}~k_{E_{D,3}^c}+k_L=3, 2\,,\\
\nonumber C^{vi}_2:~~ \mathcal{W}_{e}&=&\alpha \left(Y_{\mathbf{4}'}^{(3)} E^{c}_{D}L H_{d} \right)_{\mathbf{1}}+\beta \left(Y_{\mathbf{3}'}^{(4)} E^{c}_{3}L H_{d} \right)_{\mathbf{1}}\,,~\text{for}~k_{E_{D,3}^c}+k_L=3, 4\,,\\
\nonumber C^{vi}_3:~~ \mathcal{W}_{e}&=&\alpha \left( Y_{\mathbf{2}'}^{(5)} E^{c}_{D}L H_{d}\right)_{\mathbf{1}}+\beta \left( Y_{\mathbf{4}'}^{(5)} E^{c}_{D}L H_{d}\right)_{\mathbf{1}}+\gamma \left( Y_{\mathbf{3}'}^{(2)} E^{c}_{3}L H_{d}\right)_{\mathbf{1}}\,,~\text{for}~k_{E_{D,3}^c}+k_L=5, 2\,, \\
\nonumber C^{vi}_4:~~ \mathcal{W}_{e}&=&\alpha \left( Y_{\mathbf{2}'}^{(5)} E^{c}_{D}L H_{d}\right)_{\mathbf{1}}+\beta \left( Y_{\mathbf{4}'}^{(5)} E^{c}_{D}L H_{d}\right)_{\mathbf{1}}+\gamma \left( Y_{\mathbf{3}'}^{(4)} E^{c}_{3}L H_{d}\right)_{\mathbf{1}}\,,~\text{for}~k_{E_{D,3}^c}+k_L=5, 4\,, \\
\nonumber C^{vi}_5:~~ \mathcal{W}_{e}&=&\alpha \left( Y_{\mathbf{4}'}^{(3)} E^{c}_{D}L H_{d}\right)_{\mathbf{1}}+\beta \left( Y_{\mathbf{3}'I}^{(6)} E^{c}_{3}L H_{d}\right)_{\mathbf{1}}+\gamma \left( Y_{\mathbf{3}'II}^{(6)} E^{c}_{3}L H_{d}\right)_{\mathbf{1}}\,,~\text{for}~k_{E_{D,3}^c}+k_L=3, 6\,,\\
  \nonumber C^{vi}_6:~~ \mathcal{W}_{e}&=&\alpha \left(Y_{\mathbf{2}'}^{(5)} E^{c}_{D}L H_{d} \right)_{\mathbf{1}}+\beta \left(Y_{\mathbf{4}'}^{(5)} E^{c}_{D}L H_{d} \right)_{\mathbf{1}}+\gamma \left(Y_{\mathbf{3}'I}^{(6)} E^{c}_{3}L H_{d} \right)_{\mathbf{1}}+\delta \left(Y_{\mathbf{3}'II}^{(6)} E^{c}_{3}L H_{d} \right)_{\mathbf{1}}\,,\\
&&\qquad\qquad\qquad\qquad\qquad\qquad\qquad\qquad\qquad\qquad\qquad ~\text{for}~k_{E_{D,3}^c}+k_L=5, 6\,.
\end{eqnarray}
\end{small}
For the cases $C^{iii, iv, v, vi}_{1,2}$, both $\alpha$ and $\beta$ can be taken as real parameters, while for the cases $C^{iii, iv, v, vi}_{3,4,5}$, the phases of $\alpha$ and $\gamma$ are unphysical, but the parameter $\beta$ is complex. For the cases $C^{iii,vi}_{6}, C^{iv,v}_{7}$, $\alpha$ and $\gamma$ are real parameters, $\beta$ and $\delta$ are complex parameters. For the cases $C^{iv,v}_{6,8}$, $\alpha$ and $\delta$ are real parameters, $\beta$ and $\gamma$ are complex parameters.

\item[~~(\romannumeral7)]{$\rho_{L}=\mathbf{3}$, $\rho_{E^c}=\mathbf{3}$}

The superpotential for the charged lepton masses are given by
  \begin{eqnarray}
 \nonumber  C^{vii}_{1}:~~W_{e}&=&\alpha \left(Y_{\mathbf{3}}^{(2)}E^{c}L H_{d}\right)_{\mathbf{1}}+\beta \left(Y_{\mathbf{5}}^{(2)}E^{c}L H_{d}\right)_{\mathbf{1}}\,,~\text{for}~k_{E^{c}}+k_{L}=2\\
 \nonumber C^{vii}_{2}:~~W_{e}&=&\alpha \left(Y_{\mathbf{1}}^{(4)}E^{c}L H_{d}\right)_{\mathbf{1}}+\beta \left(Y_{\mathbf{3}}^{(4)}E^{c}L H_{d}\right)_{\mathbf{1}}+\gamma \left(Y_{\mathbf{5}I}^{(4)}E^{c}L H_{d}\right)_{\mathbf{1}}+\delta \left(Y_{\mathbf{5}II}^{(4)}E^{c}L H_{d}\right)_{\mathbf{1}}\,,\\
  &&\qquad\qquad\qquad\qquad\qquad\qquad\qquad\qquad\qquad\qquad\qquad~\text{for}~k_{E^{c}}+k_{L}=4\,.
\end{eqnarray}
\item[~~(\romannumeral8)]{$\rho_{L}=\mathbf{3}$, $\rho_{E^c}=\mathbf{3'}$}

There are three possible forms of the superpotential in this case,
\begin{eqnarray}
 \nonumber  C^{viii}_{1}:~~W_{e}&=&\alpha \left(Y_{\mathbf{5}}^{(2)}E^{c}L H_{d}\right)_{\mathbf{1}}\,,~\text{for}~k_{E^{c}}+k_{L}=2\,.\\
 \nonumber  C^{viii}_{2}:~~W_{e}&=&\alpha \left(Y_{\mathbf{4}}^{(4)}E^{c}L H_{d}\right)_{\mathbf{1}}+\beta \left(Y_{\mathbf{5}I}^{(4)}E^{c}L H_{d}\right)_{\mathbf{1}}+\gamma \left(Y_{\mathbf{5}II}^{(4)}E^{c}L H_{d}\right)_{\mathbf{1}}\,,~\text{for}~k_{E^{c}}+k_{L}=4\,.\\
  \nonumber  C^{viii}_{3}:~~W_{e}&=&\alpha \left(Y_{\mathbf{4}I}^{(6)}E^{c}L H_{d}\right)_{\mathbf{1}}+\beta \left(Y_{\mathbf{4}II}^{(6)}E^{c}L H_{d}\right)_{\mathbf{1}}+\gamma \left(Y_{\mathbf{5}I}^{(6)}E^{c}L H_{d}\right)_{\mathbf{1}}+\delta \left(Y_{\mathbf{5}II}^{(6)}E^{c}L H_{d}\right)_{\mathbf{1}}\,,\\
 &&\qquad\qquad\qquad\qquad\qquad\qquad\qquad\qquad\qquad\qquad\qquad ~\text{for}~k_{E^{c}}+k_{L}=6\,.
\end{eqnarray}
\item[~~(\romannumeral9)]{$\rho_{L}=\mathbf{3'}$, $\rho_{E^c}=\mathbf{3}$}

In this case, the charged lepton superpotential can take three possible forms,
\begin{eqnarray}
 \nonumber  C^{ix}_{1}:~~W_{e}&=&\alpha \left(Y_{\mathbf{5}}^{(2)}E^{c}L H_{d}\right)_{\mathbf{1}}\,,~\text{for}~k_{E^{c}}+k_{L}=2\,.\\
 \nonumber  C^{ix}_{2}:~~W_{e}&=&\alpha \left(Y_{\mathbf{4}}^{(4)}E^{c}L H_{d}\right)_{\mathbf{1}}+\beta \left(Y_{\mathbf{5}I}^{(4)}E^{c}L H_{d}\right)_{\mathbf{1}}+\gamma \left(Y_{\mathbf{5}II}^{(4)}E^{c}L H_{d}\right)_{\mathbf{1}}\,,~\text{for}~k_{E^{c}}+k_{L}=4\,.\\
  \nonumber  C^{ix}_{3}:~~W_{e}&=&\alpha \left(Y_{\mathbf{4}I}^{(6)}E^{c}L H_{d}\right)_{\mathbf{1}}+\beta \left(Y_{\mathbf{4}II}^{(6)}E^{c}L H_{d}\right)_{\mathbf{1}}+\gamma \left(Y_{\mathbf{5}I}^{(6)}E^{c}L H_{d}\right)_{\mathbf{1}}+\delta \left(Y_{\mathbf{5}II}^{(6)}E^{c}L H_{d}\right)_{\mathbf{1}}\,,\\
&& \qquad\qquad\qquad\qquad\qquad\qquad\qquad\qquad\qquad\qquad\qquad ~\text{for}~k_{E^{c}}+k_{L}=6\,.
\end{eqnarray}
\item[~~(\romannumeral10)]{$\rho_{L}=\mathbf{3'}$, $\rho_{E^c}=\mathbf{3'}$}

Similar to previous case, we find that the charged lepton superpotential can be
\begin{eqnarray}
\nonumber  C^{x}_{1}:~~W_{e}&=&\alpha \left(Y_{\mathbf{3}'}^{(2)}E^{c}L H_{d}\right)_{\mathbf{1}}+\beta \left(Y_{\mathbf{5}}^{(2)}E^{c}L H_{d}\right)_{\mathbf{1}}\,,~\text{for}~k_{E^{c}}+k_{L}=2\\
\nonumber  C^{x}_{2}:~~W_{e}&=&\alpha \left(Y_{\mathbf{1}}^{(4)}E^{c}L H_{d}\right)_{\mathbf{1}}+\beta \left(Y_{\mathbf{3}'}^{(4)}E^{c}L H_{d}\right)_{\mathbf{1}}+\gamma \left(Y_{\mathbf{5}I}^{(4)}E^{c}L H_{d}\right)_{\mathbf{1}}+\delta \left(Y_{\mathbf{5}II}^{(4)}E^{c}L H_{d}\right)_{\mathbf{1}}\,,\\
 &&\qquad\qquad\qquad\qquad\qquad\qquad\qquad\qquad\qquad\qquad\qquad ~\text{for}~k_{E^{c}}+k_{L}=4\,.
\end{eqnarray}
In the above four cases, the left-handed lepton $L$ as well as the right-handed charged lepton fileds $E^{c}$ are assigned to a triplet of $A'_5$, there is only one freedom to absorb the complex phase. As a result, only the parameter $\alpha$ can be taken as real and all other parameters should be complex.
\end{description}

\subsection{\label{subsec:neutrino}Neutrino sector }

\begin{table}[!ht]
  \centering
  \begin{tabular}{|c|c|c|c|c|c|c|c|c|c|c|c|c|}
    \hline\hline
        & $S^{i}$ & $S^{ii}$ & $S^{iii}$ & $S^{iv}$ & $T^{i}$ & $T^{ii}$ & $T^{iii}$ & $T^{iv}$ & $W^{i}$ & $W^{ii}$ \\ \hline
    $\rho_L$ & $\mathbf{3}$ & $\mathbf{3}$ & $\mathbf{3'}$ & $\mathbf{3'}$ & $\mathbf{3}$ & $\mathbf{3}$ & $\mathbf{3'}$ & $\mathbf{3'}$ & $\mathbf{3}$ & $\mathbf{3'}$ \\ \hline
    $\rho_{N^c}$ & $\mathbf{3}$ & $\mathbf{3'}$ & $\mathbf{3}$ & $\mathbf{3'}$ & $\mathbf{2}$ & $\mathbf{2'}$ & $\mathbf{2}$ & $\mathbf{2'}$ & $-$ & $-$ \\ \hline
    \# of models & $3$ & $2$ & $2$ & $3$ & $5$ & $5$ & $5$ & $5$ & $3$ & $3$ \\
   (total number) & $(15)$ & $(12)$ & $(12)$ & $(15)$ & $(6)$ & $(9)$ & $(9)$ & $(6)$ & $(3)$ & $(3)$    \\ \hline\hline
  \end{tabular}
\caption{ \label{tab:neutrino-models} The number (in bracket) of possible neutrino models for different representation assignments of $L$ and $N^{c}$ under the finite modular group $A'_5$ up to weight 6 modular forms, where the case without modular forms is not counted. We also list the number (without bracket) of models which contain up to three independent terms in the neutrino superpotential $\mathcal{W}_\nu$. }
\end{table}

In this work, neutrinos are assumed to be Majorana particles. Their masses can
arise from the type I seesaw mechanism. We consider seesaw models with two and three right-handed neutrinos. The Weinberg operator can also induce the effective neutrino masses, where the right-handed neutrinos are absent. The numbers of possible neutrino model are listed in table~\ref{tab:neutrino-models} for different assignments of lepton doublets $L$ and right-handed neutrinos $N^c$. We are concerned with the phenomenological viable models with the smallest number of free parameters, and consequently in the following we give the explicit form of the neutrino superpotential for the cases that only two or three independent terms are present. For the models without right-handed neutrinos, similarly we consider the cases with at most three independent terms in the superpotential.

\subsubsection{Seesaw models with three right-handed neutrinos}

If the three right-handed neutrinos are assigned to $A_5$ singlets or the direct sum $\mathbf{2^{(')}}\oplus\mathbf{1}$, at least four terms would be involved so that the resulting models are not so predictive. Hence we assign the three right-handed neutrinos as well as lepton doublets $L$ to a triplet $\mathbf{3}$ or $\mathbf{3}'$ of $A_5$.

\begin{description}[labelindent=-0.8em, leftmargin=0.3em]
\item[~~(\romannumeral1)]{$\rho_{L}=\mathbf{3}$, $\rho_{N^c}=\mathbf{3}$ }

We find three possible values of the modular weights $k_{N^c}$ and $k_{L}$: $(k_{N^c}, k_L)=(1, -1), (0, 2), (1, 1)$\,, for which the superpotential of the seesaw Lagrangian has two or three modular invariant terms.
\begin{small}
\begin{eqnarray}
\nonumber S^{i}_1:~~ \mathcal{W}_{\nu}&=& g\left(N^{c}L H_{u}\right)_{\mathbf{1}}+\Lambda\left(Y_{\mathbf{5}}^{(2)}N^{c}N^{c}\right)_{\mathbf{1}}\,,~~\text{for}~~k_{N^c}=1,    k_{L}=-1\,, \\
\nonumber S^{i}_2:~~ \mathcal{W}_{\nu} &=& g_1\left(Y_{\mathbf{3}}^{(2)}N^{c}L H_{u}\right)_{\mathbf{1}}+g_2\left(Y_{\mathbf{5}}^{(2)}N^{c}L H_{u}\right)_{\mathbf{1}}+\Lambda \left(N^{c}N^{c}\right)_{\mathbf{1}} \,,~~\text{for}~~k_{N^c}=0,      k_{L}=2\,,\\
S^{i}_3:~~ \mathcal{W}_{\nu}&=&g_1\left(Y_{\mathbf{3}}^{(2)}N^{c}L H_{u}\right)_{\mathbf{1}}+g_2\left(Y_{\mathbf{5}}^{(2)}N^{c}L H_{u}\right)_{\mathbf{1}}+\Lambda\left(Y_{\mathbf{5}}^{(2)}N^{c}N^{c}\right)_{\mathbf{1}} \,,~~\text{for}~~k_{N^c}=1,       k_{L}=1\,.
\end{eqnarray}
\end{small}
For the model $S^{i}_1$, the light neutrino mass matrix only depends on the complex modulus and an overall scale factor $g^2v^2_u/\Lambda$, while it also depends on another complex parameter $g_2/g_1$ for $S^{i}_2$ and $S^{i}_3$. The predicted Dirac and Majorana neutrino mass matrices are given in the auxiliary file~\cite{Yao:2020sup}.

\item[~~(\romannumeral2)]{$\rho_{L}=\mathbf{3}$, $\rho_{N^c}=\mathbf{3'}$ }

For this kind of representation assignment, the neutrino superpotential can take the following two simple forms
\begin{eqnarray}
\nonumber S^{ii}_1:~~  \mathcal{W}_{\nu}&=&g\left(Y_{\mathbf{5}}^{(2)}N^{c}L H_{u}\right)_{\mathbf{1}}+\Lambda \left(N^{c}N^{c}\right)_{\mathbf{1}}\,,~~\text{for}~~k_{N^c}=0,    k_{L}=2\,,\\
\label{eq:Sii-2}S^{ii}_2:~~ \mathcal{W}_{\nu}&=&g\left(Y_{\mathbf{5}}^{(2)}N^{c}L H_{u}\right)_{\mathbf{1}}+\Lambda\left(Y_{\mathbf{5}}^{(2)}N^{c}N^{c}\right)_{\mathbf{1}}\,,~~\text{for}~~k_{N^c}=1,     k_{L}=1\,.
\end{eqnarray}
\item[~~(\romannumeral3)]{$\rho_{L}=\mathbf{3'}$, $\rho_{N^c}=\mathbf{3}$ }

Similar to the previous case, the superpotential for neutrino masses can also take two possible forms
\begin{eqnarray}
\nonumber S^{iii}_1:~~  \mathcal{W}_{\nu}&=&g\left(Y_{\mathbf{5}}^{(2)}N^{c}L H_{u}\right)_{\mathbf{1}}+\Lambda \left(N^{c}N^{c}\right)_{\mathbf{1}} \,,~~\text{for}~~k_{N^c}=0,    k_{L}=2\,,\\
\label{eq:Siii-2}S^{iii}_2:~~  \mathcal{W}_{\nu}&=&g\left(Y_{\mathbf{5}}^{(2)}N^{c}L H_{u}\right)_{\mathbf{1}}+\Lambda\left(Y_{\mathbf{5}}^{(2)}N^{c}N^{c}\right)_{\mathbf{1}} \,,~~\text{for}~~k_{N^c}=1,     k_{L}=1\,.
\end{eqnarray}
Although the superpotentials in Eq.~\eqref{eq:Sii-2} and Eq.~\eqref{eq:Siii-2} appear the same, using the CG coefficient to expand the contractions,
we find that they give rise to different terms and neutrino mass matrices.
For the neutrino models $S^{ii, iii}_{1,2}$, the effective light neutrino mass matrices are completely determined by the modulus $\tau$ up to the overall scale factor $g^2v^2_u/\Lambda $.
\item[~~(\romannumeral4)]{$\rho_{L}=\mathbf{3'}$, $\rho_{N^c}=\mathbf{3'}$}

The neutrino superpotential in this case can be
\begin{small}
\begin{eqnarray}
\nonumber S^{iv}_1:~~  \mathcal{W}_{\nu}&=&g\left(N^{c}L H_{u}\right)_{\mathbf{1}}+\Lambda\left(Y_{\mathbf{5}}^{(2)}N^{c}N^{c}\right)_{\mathbf{1}}\,,~~\text{for}~~k_{N^c}=1,    k_{L}=-1\,,\\
\nonumber S^{iv}_2:~~  \mathcal{W}_{\nu}&=&g_1\left(Y_{\mathbf{3}'}^{(2)}N^{c}L H_{u}\right)_{\mathbf{1}}+g_2\left(Y_{\mathbf{5}}^{(2)}N^{c}L H_{u}\right)_{\mathbf{1}}+\Lambda \left(N^{c}N^{c}\right)_{\mathbf{1}}\,,~~\text{for}~~k_{N^c}=0,      k_{L}=2\,,\\
S^{iv}_3:~~  \mathcal{W}_{\nu}&=&g_1\left(Y_{\mathbf{3}'}^{(2)}N^{c}L H_{u}\right)_{\mathbf{1}}+g_2\left(Y_{\mathbf{5}}^{(2)}N^{c}L H_{u}\right)_{\mathbf{1}}+\Lambda\left(Y_{\mathbf{5}}^{(2)}N^{c}N^{c}\right)_{\mathbf{1}}\,,~~\text{for}~~k_{N^c}=1,       k_{L}=1\,.
\end{eqnarray}
\end{small}

\end{description}

\subsubsection{Seesaw models with two right-handed neutrinos}

If the two right-handed neutrinos are assigned to $A'_5$ singlets distinguished by modular weights, generally more modular invariant terms are involved. As a consequence, we assume that the two right-handed neutrinos transform as a doublet $\mathbf{2}$ or $\mathbf{2'}$ under $A'_5$.

\begin{description}[labelindent=-0.8em, leftmargin=0.3em]

\item[~~(\romannumeral1)]{$\rho_{L}=\mathbf{3}$, $\rho_{N^c}=\mathbf{2}$ }

We find that only five sets of values of the modular weights $k_L$ and $k_{N^c}$ such that there are two or three independent terms in the neutrino superpotential.
\begin{eqnarray}
\nonumber  T^{i}_1:~~ \mathcal{W}_{\nu}&=& g\left(Y_{\mathbf{4}'}^{(3)}N^{c}L H_{u}\right)_{\mathbf{1}}+\Lambda\left(Y_{\mathbf{3}}^{(2)}N^{c}N^{c}\right)_{\mathbf{1}} \,,~\text{for}~k_{N}=1,     k_{L}=2\,,\\
\nonumber  T^{i}_2:~~ \mathcal{W}_{\nu}&=&g\left(Y_{\mathbf{4}'}^{(3)}N^{c}L H_{u}\right)_{\mathbf{1}}+\Lambda\left(Y_{\mathbf{3}}^{(4)}N^{c}N^{c}\right)_{\mathbf{1}} \,,~\text{for}~k_{N}=2,      k_{L}=1\,,\\
\nonumber  T^{i}_3:~~ \mathcal{W}_{\nu}&=&g_1\left(Y_{\mathbf{2}}^{(5)}N^{c}L H_{u}\right)_{\mathbf{1}}+g_2\left(Y_{\mathbf{4}'}^{(5)}N^{c}L H_{u}\right)_{\mathbf{1}}+\Lambda\left(Y_{\mathbf{3}}^{(2)}N^{c}N^{c}\right)_{\mathbf{1}} \,,~\text{for}~k_{N}=1,       k_{L}=4\,,\\
\nonumber  T^{i}_4:~~ \mathcal{W}_{\nu}&=&g_1\left(Y_{\mathbf{2}}^{(5)}N^{c}L H_{u}\right)_{\mathbf{1}}+g_2\left(Y_{\mathbf{4}'}^{(5)}N^{c}L H_{u}\right)_{\mathbf{1}}+\Lambda\left(Y_{\mathbf{3}}^{(4)}N^{c}N^{c}\right)_{\mathbf{1}} \,,~\text{for}~k_{N}=2,       k_{L}=3\,,\\
T^{i}_5:~~ \mathcal{W}_{\nu}&=& g\left(Y_{\mathbf{4}'}^{(3)}N^{c}L H_{u}\right)_{\mathbf{1}}+\Lambda _1\left(Y_{\mathbf{3}I}^{(6)}N^{c}N^{c}\right)_{\mathbf{1}}+\Lambda _2\left(Y_{\mathbf{3}II}^{(6)}N^{c}N^{c}\right)_{\mathbf{1}}\,,~\text{for}~k_{N}=3,      k_{L}=0\,.
\end{eqnarray}
The light neutrino mass matrix is fixed by modulus $\tau$ and an overall mass scale for $T^{i}_{1,2}$, and it depends on a third complex parameter $g_2/g_1$ for $T^{i}_{3,4}$ and $\Lambda_2/\Lambda_1$ for $T^{i}_{5}$ respectively.

\item[~~(\romannumeral2)]{$\rho_{L}=\mathbf{3}$, $\rho_{N^c}=\mathbf{2'}$ }

Analogously we also find five possible cases as follow,
\begin{small}
\begin{eqnarray}
\nonumber  T^{ii}_1:~~ \mathcal{W}_{\nu}&=& g\left(Y_{\mathbf{6}}^{(1)}N^{c}L H_{u}\right)_{\mathbf{1}}+\Lambda\left(Y_{\mathbf{3}'}^{(2)}N^{c}N^{c}\right)_{\mathbf{1}} \,,~\text{for}~k_{N^c}=1,     k_{L}=0\,, \\
\nonumber  T^{ii}_2:~~ \mathcal{W}_{\nu}&=&g\left(Y_{\mathbf{6}}^{(1)}N^{c}L H_{u}\right)_{\mathbf{1}}+\Lambda\left(Y_{\mathbf{3}'}^{(4)}N^{c}N^{c}\right)_{\mathbf{1}} \,,~\text{for}~k_{N^c}=2,     k_{L}=-1\,, \\
\nonumber  T^{ii}_3:~~ \mathcal{W}_{\nu}&=&g_1\left(Y_{\mathbf{6}I}^{(3)}N^{c}L H_{u}\right)_{\mathbf{1}}+g_2\left(Y_{\mathbf{6}II}^{(3)}N^{c}L H_{u}\right)_{\mathbf{1}}+\Lambda\left(Y_{\mathbf{3}'}^{(2)}N^{c}N^{c}\right)_{\mathbf{1}} \,,~\text{for}~k_{N^c}=1,       k_{L}=2\,, \\
\nonumber  T^{ii}_4:~~ \mathcal{W}_{\nu}&=&g_1\left(Y_{\mathbf{6}I}^{(3)}N^{c}L H_{u}\right)_{\mathbf{1}}+g_2\left(Y_{\mathbf{6}II}^{(3)}N^{c}L H_{u}\right)_{\mathbf{1}}+\Lambda\left(Y_{\mathbf{3}'}^{(4)}N^{c}N^{c}\right)_{\mathbf{1}} \,,~\text{for}~k_{N^c}=2,       k_{L}=1\,, \\
 T^{ii}_5:~~ \mathcal{W}_{\nu}&=& g\left(Y_{\mathbf{6}}^{(1)}N^{c}L H_{u}\right)_{\mathbf{1}}+\Lambda _1\left(Y_{\mathbf{3}'I}^{(6)}N^{c}N^{c}\right)_{\mathbf{1}}+\Lambda _2\left(Y_{\mathbf{3}'II}^{(6)}N^{c}N^{c}\right)_{\mathbf{1}}\,,~\text{for}~k_{N^c}=3,      k_{L}=-2\,.
\end{eqnarray}
\end{small}
\item[~~(\romannumeral3)]{$\rho_{L}=\mathbf{3'}$, $\rho_{N^c}=\mathbf{2}$}

The neutrino superpotential in this case can be
\begin{small}
\begin{eqnarray}
\nonumber  T^{iii}_1:~~ \mathcal{W}_{\nu}&=&g\left(Y_{\mathbf{6}}^{(1)}N^{c}L H_{u}\right)_{\mathbf{1}}+\Lambda\left(Y_{\mathbf{3}}^{(2)}N^{c}N^{c}\right)_{\mathbf{1}} \,,~\text{for}~k_{N^c}=1,     k_{L}=0\,, \\
\nonumber  T^{iii}_2:~~ \mathcal{W}_{\nu}&=&g\left(Y_{\mathbf{6}}^{(1)}N^{c}L H_{u}\right)_{\mathbf{1}}+\Lambda\left(Y_{\mathbf{3}}^{(4)}N^{c}N^{c}\right)_{\mathbf{1}} \,,~\text{for}~k_{N^c}=2,     k_{L}=-1\,, \\
\nonumber  T^{iii}_3:~~ \mathcal{W}_{\nu}&=&g_1\left(Y_{\mathbf{6}I}^{(3)}N^{c}L H_{u}\right)_{\mathbf{1}}+g_2\left(Y_{\mathbf{6}II}^{(3)}N^{c}L H_{u}\right)_{\mathbf{1}}+\Lambda\left(Y_{\mathbf{3}}^{(2)}N^{c}N^{c}\right)_{\mathbf{1}}\,,~\text{for}~k_{N^c}=1,       k_{L}=2\,, \\
\nonumber  T^{iii}_4:~~ \mathcal{W}_{\nu}&=&g_1\left(Y_{\mathbf{6}I}^{(3)}N^{c}L H_{u}\right)_{\mathbf{1}}+g_2\left(Y_{\mathbf{6}II}^{(3)}N^{c}L H_{u}\right)_{\mathbf{1}}+\Lambda\left(Y_{\mathbf{3}}^{(4)}N^{c}N^{c}\right)_{\mathbf{1}}\,,~\text{for}~k_{N^c}=2,       k_{L}=1\,, \\
  T^{iii}_5:~~ \mathcal{W}_{\nu}&=&g\left(Y_{\mathbf{6}}^{(1)}N^{c}L H_{u}\right)_{\mathbf{1}}+\Lambda _1\left(Y_{\mathbf{3}I}^{(6)}N^{c}N^{c}\right)_{\mathbf{1}}+\Lambda _2\left(Y_{\mathbf{3}II}^{(6)}N^{c}N^{c}\right)_{\mathbf{1}}\,,~\text{for}~k_{N^c}=3,      k_{L}=-2\,.
\end{eqnarray}
\end{small}
\item[~~(\romannumeral4)]{$\rho_{L}=\mathbf{3'}$, $\rho_{N^c}=\mathbf{2'}$}

The modular weight $k_L$ and $k_{N^c}$ can also take five values: $(k_{N^c}, k_L)=(1, 2), (2, 1), (1,4), (2, 3), (3, 0)$, and the superpotential for neutrino masses are given by
\begin{small}
\begin{eqnarray}
\nonumber  T^{iv}_1:~~ \mathcal{W}_{\nu}&=&g\left(Y_{\mathbf{4}'}^{(3)}N^{c}L H_{u}\right)_{\mathbf{1}}+\Lambda\left(Y_{\mathbf{3}'}^{(2)}N^{c}N^{c}\right)_{\mathbf{1}}\,,~\text{for}~k_{N^c}=1,     k_{L}=2\,, \\
\nonumber  T^{iv}_2:~~ \mathcal{W}_{\nu}&=&g\left(Y_{\mathbf{4}'}^{(3)}N^{c}L H_{u}\right)_{\mathbf{1}}+\Lambda\left(Y_{\mathbf{3}'}^{(4)}N^{c}N^{c}\right)_{\mathbf{1}}\,,~\text{for}~k_{N^c}=2,     k_{L}=1\,, \\
\nonumber  T^{iv}_3:~~ \mathcal{W}_{\nu}&=&g_1\left(Y_{\mathbf{2}'}^{(5)}N^{c}L H_{u}\right)_{\mathbf{1}}+g_2\left(Y_{\mathbf{4}'}^{(5)}N^{c}L H_{u}\right)_{\mathbf{1}}+\Lambda\left(Y_{\mathbf{3}'}^{(2)}N^{c}N^{c}\right)_{\mathbf{1}}\,,~\text{for}~k_{N^c}=1,       k_{L}=4\,, \\
\nonumber  T^{iv}_4:~~ \mathcal{W}_{\nu}&=&g_1\left(Y_{\mathbf{2}'}^{(5)}N^{c}L H_{u}\right)_{\mathbf{1}}+g_2\left(Y_{\mathbf{4}'}^{(5)}N^{c}L H_{u}\right)_{\mathbf{1}}+\Lambda\left(Y_{\mathbf{3}'}^{(4)}N^{c}N^{c}\right)_{\mathbf{1}}\,,~\text{for}~k_{N^c}=2,       k_{L}=3\,, \\
  T^{iv}_5:~~ \mathcal{W}_{\nu}&=&g\left(Y_{\mathbf{4}'}^{(3)}N^{c}L H_{u}\right)_{\mathbf{1}}+\Lambda _1\left(Y_{\mathbf{3}'I}^{(6)}N^{c}N^{c}\right)_{\mathbf{1}}+\Lambda _2\left(Y_{\mathbf{3}'II}^{(6)}N^{c}N^{c}\right)_{\mathbf{1}}\,,~\text{for}~k_{N^c}=3,      k_{L}=0\,.
\end{eqnarray}
\end{small}
The explicit forms of the Dirac neutrino mass matrix and the right-handed Majorana neutrino mass matrix are given in the supplementary file~\cite{Yao:2020sup}.

\end{description}

\subsubsection{Models without right-handed neutrino}

The light neutrino masses are described by the effective Weinberg operator in this scenario. For the triplet assignment of the left-handed leptons $\rho_L\sim\mathbf{3}\,,\mathbf{3'}$ under $A'_5$, there are two classes of neutrino models.

\begin{description}[labelindent=-0.8em, leftmargin=0.3em]

\item[~~(\romannumeral1)]{$\rho_{L}=\mathbf{3}$ }

We find that only three allowed values of the modular weights $k_L$ such that at most three independent terms are present in the superpotential of neutrino masses.
\begin{eqnarray}
\nonumber  W^{i}_1:~~ \mathcal{W}_{\nu}&=&\frac{1}{\Lambda_1}\left(Y_{\mathbf{5}}^{(2)}L^2 H_{u}^2\right)_{\mathbf{1}} \,,~\text{for}~k_{L}=1\,,\\
\nonumber  W^{i}_2:~~ \mathcal{W}_{\nu}&=&\frac{1}{\Lambda _1}\left(Y_{\mathbf{1}}^{(4)}L^2 H_{u}^2 \right)_{\mathbf{1}}+\frac{1}{\Lambda _2}\left( Y_{\mathbf{5}I}^{(4)}L^2 H_{u}^2\right)_{\mathbf{1}}+\frac{1}{\Lambda _3}\left(Y_{\mathbf{5}II}^{(4)}L^2 H_{u}^2\right)_{\mathbf{1}} \,,~\text{for}~k_{L}=2\,,\\
 W^{i}_3:~~ \mathcal{W}_{\nu}&=&\frac{1}{\Lambda _1}\left(Y_{\mathbf{1}}^{(6)}L^2 H_{u}^2\right)_{\mathbf{1}}+\frac{1}{\Lambda _2}\left(Y_{\mathbf{5}I}^{(6)}L^2 H_{u}^2\right)_{\mathbf{1}}+\frac{1}{\Lambda _3}\left(Y_{\mathbf{5}II}^{(6)}L^2 H_{u}^2\right)_{\mathbf{1}} \,,~\text{for}~k_{L}=3\,.
\end{eqnarray}
The light neutrino mass matrix is fixed by modulus $\tau$ and an overall
scale factor $v_u^2/\Lambda_1$ for $W^{i}_{1}$, and it depends on another two complex parameters $\Lambda_1/\Lambda_2$ and $\Lambda_1/\Lambda_3$ for $W^{i}_{2,3}$.

\item[~~(\romannumeral2)]{$\rho_{L}=\mathbf{3'}$}

Analogously we also find three possible cases as follow,
\begin{eqnarray}
\nonumber  W^{ii}_1:~~ \mathcal{W}_{\nu}&=& \frac{1}{\Lambda _1}\left(Y_{\mathbf{5}}^{(2)}L^2 H_{u}^2 \right)_{\mathbf{1}}\,,~\text{for}~k_{L}=1\,, \\
\nonumber  W^{ii}_2:~~ \mathcal{W}_{\nu}&=& \frac{1}{\Lambda _1}\left(Y_{\mathbf{1}}^{(4)}L^2 H_{u}^2 \right)_{\mathbf{1}}+\frac{1}{\Lambda _2}\left(Y_{\mathbf{5}I}^{(4)}L^2 H_{u}^2 \right)_{\mathbf{1}}+\frac{1}{\Lambda _3}\left(Y_{\mathbf{5}II}^{(4)}L^2 H_{u}^2 \right)_{\mathbf{1}} \,,~\text{for}~k_{L}=2\,, \\
 W^{ii}_3:~~ \mathcal{W}_{\nu}&=& \frac{1}{\Lambda _1}\left(Y_{\mathbf{1}}^{(6)}L^2 H_{u}^2 \right)_{\mathbf{1}}+\frac{1}{\Lambda _2}\left(Y_{\mathbf{5}I}^{(6)}L^2 H_{u}^2 \right)_{\mathbf{1}}+\frac{1}{\Lambda _3}\left(Y_{\mathbf{5}II}^{(6)}L^2 H_{u}^2 \right)_{\mathbf{1}} \,,~\text{for}~k_{L}=3\,.
\end{eqnarray}
The explicit form of the effective neutrino mass matrix can be straightforwardly read out and it is given in the supplementary file~\cite{Yao:2020sup}.
\end{description}

\subsection{Numerical results for lepton masses and mixing }

Combining the possible constructions of charged lepton sector listed in table~\ref{tab:lepton-rep} with those of neutrino sector in table~\ref{tab:neutrino-models}, we can obtain 720 possible lepton models with small number of free parameters based on the homogeneous finite modular group $A'_5$. These 720 lepton models are named as
$\mathcal{L}_{1}, \dots, \mathcal{L}_{720}$, which can be found in
table~3 of the supplementary material~\cite{Yao:2020sup}.
In order to determine which models are compatible with the current experiment, we perform a conventional $\chi^2$ analysis to search for the optimal values of the input parameters and quantitatively evaluate how well a model can accommodate the experimental data. In order to facilitate the numerical analysis, we divide the input parameters of each model into dimensionless parameters and overall mass scales. The dimensionless parameters include the ratios of the coupling constants and the VEV of the complex modulus $\tau$. As in Ref.~\cite{Feruglio:2017spp}, we shall not resort to certain modulus stabilization mechanism to dynamically select the value of the modulus $\tau$, and it is freely varied to match the experimental data.
The overall mass scales of the charged lepton and neutrino mass scales determine the magnitudes of the charged lepton masses and the absolute scale of neutrino masses, and their values can be fixed by the precisely measured electron mass and the neutrino mass squared difference $\Delta m_{21}^2$.  Therefore we construct the $\chi^2$ function based on the neutrino mixing angles $\theta_{12}$, $\theta_{13}$, $\theta_{23}$ and the mass ratios $m_e/m_{\mu}$, $m_{\mu}/m_{\tau}$,
$\Delta m_{21}^{2}/\Delta m_{31}^{2}$($\Delta m_{21}^{2}/\Delta m_{32}^{2}$) for NO (IO) neutrino mass ordering. Since the leptonic Dirac CP phase
$\delta^{l}_{CP}$ is not accurately measured and the indication of a preferred value of $\delta^{l}_{CP}$ from global data fit is quite weak, we did not include the contribution of $\delta^l_{CP}$ in the $\chi^2$ function.

The experimental data of the neutrino oscillation parameters are taken from NuFIT v5.0 with Super-Kamiokanda atmospheric data~\cite{Esteban:2020cvm}. For the normal ordering (NO) neutrino masses, the best fit values and the $1\sigma$ ranges of the three mixing angles, Dirac CP phase, and neutrino mass squared differences are as follows
\begin{equation}
\begin{array}{c}
\sin ^{2} \theta_{12}=0.304_{-0.012}^{+0.012}\,, \quad \sin ^{2} \theta_{13}=0.02219_{-0.00063}^{+0.00062}\,, \quad \sin ^{2} \theta_{23}=0.573_{-0.020}^{+0.016}\,, \\
\delta_{C P}^{l} / \pi=1.0944_{-0.1333}^{+0.1500}\,, \quad \frac{\Delta m_{21}^{2}}{10^{-5} \mathrm{eV}^{2}}=7.42_{-0.20}^{+0.21}\,, \quad \frac{\Delta m_{31}^{2}}{10^{-3} \mathrm{eV}^{2}}=2.517_{-0.028}^{+0.026}\,,
\end{array}
\end{equation}
For the inverted ordering (IO) mass spectrum, values of the oscillation
parameters are given by
\begin{equation}
\begin{array}{c}
\sin ^{2} \theta_{12}=0.304_{-0.012}^{+0.013}\,, \quad \sin ^{2} \theta_{13}=0.02238_{-0.00062}^{+0.00063}\,, \quad \sin ^{2} \theta_{23}=0.575_{-0.019}^{+0.016}\,, \\
\delta_{C P}^{l} / \pi=1.5667_{-0.1667}^{+0.1444}\,, \quad \frac{\Delta m_{21}^{2}}{10^{-5} \mathrm{eV}^{2}}=7.42_{-0.20}^{+0.21}\,, \quad \frac{\Delta m_{32}^{2}}{10^{-3} \mathrm{eV}^{2}}=-2.498_{-0.028}^{+0.028}\,.
\end{array}
\end{equation}
The ratios of the charged lepton masses are taken from
Ref.~\cite{Antusch:2013jca},
\begin{equation}
m_{e}/m_{\mu}=0.004737 \pm 0.000040 , \quad m_{\mu}/m_{\tau}=0.05857 \pm 0.00047\,.
\end{equation}
We have used the well-known package \texttt{TMinuit} developed by CERN to
search for the global minimum of the $\chi^2$ function.
The absolute values of all dimensionless parameters are assumed to be uniformly distributed in the range of $[0,10^7]$, their phases freely vary between $0$ and $2\pi$, and the VEV of modulus $\tau$ is restricted in the fundamental domain
$\mathcal{D}=\{ \tau ~|~ |\tau |\geq1, -0.5\leq\texttt{Re}(\tau) \leq0.5,~\texttt{Im}(\tau)>0 \}$.

We numerically diagonalize the charged lepton and neutrino mass matrices, and scan over the parameter space of each lepton model for both NO and IO. We find that 25 NO (49 IO) models with up to 10 parameters can accommodate the
experimental data at $3\sigma$ level, all the fitting
results of the 720 models are summarized in our supplementary material~\cite{Yao:2020sup}, and
we label the models which are compatible with the experimental data at the $3\sigma$ level or better with a star ``\ding{72}''. In the main text,  we have summarized the 25 NO models in
table~\ref{tab:lepton_models}. It can be seen that there are 15 phenomenologically viable models with 9 real free parameters including $\texttt{Re}(\tau)$ and $\texttt{Im}(\tau)$, while the other 10 models use 10 free parameters to describe the experimental data. We see that the right-handed charged leptons transform as $\mathbf{2}\oplus \mathbf{1}$ or $\mathbf{2}^{'}\oplus \mathbf{1}$ for the minimal models with 9 free parameters.
Notice that the homogeneous finite modular group was studied in the preprint~\cite{Wang:2020lxk} which appeared on the arXiv
during the final preparations of this article, and an example model corresponding to our model $\mathcal{L}_{210}$ with 10 parameters was constructed.

\tabcolsep=0.1cm
\begin{table}[t]
\centering
  \begin{tabular}{|c|c|c|c|c|c|c|c|}
\hline\hline
 Models & \#P & Combinations & $(\rho_{E^c},\rho_{L},\rho_{N^c})$ & $k_{E^c}$ & $k_L$ & $k_{N^c}$ &  with gCP\\ \hline\hline
$\mathcal{L}_{2}$ & $10$ & $C^{i}_{1},S^{i}_{2}$ & $(\mathbf{1}\oplus \mathbf{1}\oplus \mathbf{1},\mathbf{3},\mathbf{3})$ & $0, 2, 4$ & $2$ & $0$ & \ding{52} \\ \hline
$\mathcal{L}_{3}$ & $10$ & $C^{i}_{1},S^{i}_{3}$ & $(\mathbf{1}\oplus \mathbf{1}\oplus \mathbf{1},\mathbf{3},\mathbf{3})$ & $1, 3, 5$ & $1$ & $1$ & \ding{52} \\ \hline
$\mathcal{L}_{9}$ & $10$ & $C^{ii}_{1},S^{iv}_{2}$ & $(\mathbf{1}\oplus \mathbf{1}\oplus \mathbf{1},\mathbf{3}',\mathbf{3}')$ & $0, 2, 4$ & $2$ & $0$ & \ding{52} \\ \hline
$\mathcal{L}_{10}$ & $10$ & $C^{ii}_{1},S^{iv}_{3}$ & $(\mathbf{1}\oplus \mathbf{1}\oplus \mathbf{1},\mathbf{3}',\mathbf{3}')$ & $1, 3, 5$ & $1$ & $1$ & \ding{52} \\ \hline
$\mathcal{L}_{22}$ & $9$ & $C^{iii}_{3},S^{i}_{2}$ & $(\mathbf{2}\oplus \mathbf{1},\mathbf{3},\mathbf{3})$ & $3, 0$ & $2$ & $0$ & \ding{52} \\ \hline
$\mathcal{L}_{66}$ & $9$ & $C^{iv}_{6},S^{i}_{1}$ & $(\mathbf{2}'\oplus \mathbf{1},\mathbf{3},\mathbf{3})$ & $6, 3$ & $-1$ & $1$ & \ding{52} \\ \hline
$\mathcal{L}_{71}$ & $9$ & $C^{iv}_{7},S^{i}_{1}$ & $(\mathbf{2}'\oplus \mathbf{1},\mathbf{3},\mathbf{3})$ & $4, 7$ & $-1$ & $1$ & \ding{56} \\ \hline
$\mathcal{L}_{76}$ & $9$ & $C^{iv}_{8},S^{i}_{1}$ & $(\mathbf{2}'\oplus \mathbf{1},\mathbf{3},\mathbf{3})$ & $6, 5$ & $-1$ & $1$ & \ding{56} \\ \hline
$\mathcal{L}_{94}$ & $9$ & $C^{v}_{3},S^{iv}_{2}$ & $(\mathbf{2}\oplus \mathbf{1},\mathbf{3}',\mathbf{3}')$ & $1, 0$ & $2$ & $0$ & \ding{52} \\ \hline
$\mathcal{L}_{95}$ & $9$ & $C^{v}_{3},S^{iv}_{3}$ & $(\mathbf{2}\oplus \mathbf{1},\mathbf{3}',\mathbf{3}')$ & $2, 1$ & $1$ & $1$ & \ding{52} \\ \hline
$\mathcal{L}_{99}$ & $9$ & $C^{v}_{4},S^{iv}_{2}$ & $(\mathbf{2}\oplus \mathbf{1},\mathbf{3}',\mathbf{3}')$ & $1, 2$ & $2$ & $0$ & \ding{52} \\ \hline
$\mathcal{L}_{100}$ & $9$ & $C^{v}_{4},S^{iv}_{3}$ & $(\mathbf{2}\oplus \mathbf{1},\mathbf{3}',\mathbf{3}')$ & $2, 3$ & $1$ & $1$ & \ding{52} \\ \hline
$\mathcal{L}_{105}$ & $9$ & $C^{v}_{5},S^{iv}_{3}$ & $(\mathbf{2}\oplus \mathbf{1},\mathbf{3}',\mathbf{3}')$ & $0, 5$ & $1$ & $1$ & \ding{56} \\ \hline
$\mathcal{L}_{108}$ & $9$ & $C^{v}_{6},S^{iv}_{1}$ & $(\mathbf{2}\oplus \mathbf{1},\mathbf{3}',\mathbf{3}')$ & $6, 3$ & $-1$ & $1$ & \ding{52} \\ \hline
$\mathcal{L}_{113}$ & $9$ & $C^{v}_{7},S^{iv}_{1}$ & $(\mathbf{2}\oplus \mathbf{1},\mathbf{3}',\mathbf{3}')$ & $4, 7$ & $-1$ & $1$ & \ding{52} \\ \hline
$\mathcal{L}_{118}$ & $9$ & $C^{v}_{8},S^{iv}_{1}$ & $(\mathbf{2}\oplus \mathbf{1},\mathbf{3}',\mathbf{3}')$ & $6, 5$ & $-1$ & $1$ & \ding{52} \\ \hline
$\mathcal{L}_{134}$ & $9$ & $C^{vi}_{3},S^{iv}_{2}$ & $(\mathbf{2}'\oplus \mathbf{1},\mathbf{3}',\mathbf{3}')$ & $3, 0$ & $2$ & $0$ & \ding{56} \\ \hline
$\mathcal{L}_{203}$ & $10$ & $C^{i}_{1},T^{i}_{3}$ & $(\mathbf{1}\oplus \mathbf{1}\oplus \mathbf{1},\mathbf{3},\mathbf{2})$ & $-2, 0, 2$ & $4$ & $1$ & \ding{52} \\ \hline
$\mathcal{L}_{208}$ & $10$ & $C^{i}_{1},T^{ii}_{3}$ & $(\mathbf{1}\oplus \mathbf{1}\oplus \mathbf{1},\mathbf{3},\mathbf{2}')$ & $0, 2, 4$ & $2$ & $1$ & \ding{52} \\ \hline
$\mathcal{L}_{209}$ & $10$ & $C^{i}_{1},T^{ii}_{4}$ & $(\mathbf{1}\oplus \mathbf{1}\oplus \mathbf{1},\mathbf{3},\mathbf{2}')$ & $1, 3, 5$ & $1$ & $2$ & \ding{52} \\ \hline
$\mathcal{L}_{210}$ & $10$ & $C^{i}_{1},T^{ii}_{5}$ & $(\mathbf{1}\oplus \mathbf{1}\oplus \mathbf{1},\mathbf{3},\mathbf{2}')$ & $4, 6, 8$ & $-2$ & $3$ & \ding{52} \\ \hline
$\mathcal{L}_{213}$ & $10$ & $C^{ii}_{1},T^{iii}_{3}$ & $(\mathbf{1}\oplus \mathbf{1}\oplus \mathbf{1},\mathbf{3}',\mathbf{2})$ & $0, 2, 4$ & $2$ & $1$ & \ding{52} \\ \hline
$\mathcal{L}_{214}$ & $10$ & $C^{ii}_{1},T^{iii}_{4}$ & $(\mathbf{1}\oplus \mathbf{1}\oplus \mathbf{1},\mathbf{3}',\mathbf{2})$ & $1, 3, 5$ & $1$ & $2$ & \ding{52} \\ \hline
$\mathcal{L}_{310}$ & $9$ & $C^{iv}_{3},T^{ii}_{5}$ & $(\mathbf{2}'\oplus \mathbf{1},\mathbf{3},\mathbf{2}')$ & $5, 4$ & $-2$ & $3$ & \ding{56} \\ \hline
$\mathcal{L}_{654}$ & $9$ & $C^{v}_{2},W^{ii}_{3}$ & $(\mathbf{2}\oplus \mathbf{1},\mathbf{3}',-)$ & $-2, 1$ & $3$ & $-$ & \ding{56}\\ \hline\hline
  \end{tabular}
\caption{\label{tab:lepton_models}
The models that can accommodate the
experimental data at $3\sigma$ level for normal ordering neutrino masses.  The details of these models can be found in the supplemental material~\cite{Yao:2020sup}, which provides the complete results for all $A'_5$ modular lepton models with small number of free parameters.
We have listed the number of real free parameters involved in each
model in the second column. The constructions in the charged lepton
and neutrino sectors are given in the third column.
After generalized CP (gCP) is incorporated in these models, two more free parameters would be reduced and most of them can
still accommodate the $3\sigma$ experimental data, which will be marked
with ``\ding{52}'' in the last column, otherwise it will be marked
with ``\ding{56}''. }
\end{table}

We have displayed the fit results of the 25 NO models in
table~\ref{tab:model_Aa}, table~\ref{tab:model_Ab} and table~\ref{tab:model_Ac}, where we have
presented the best fit values of the input parameters and the corresponding
predictions for neutrino masses, mixing angles and CP violating phases.
Here we adopt the convention for lepton mixing angles and CP phases in the
standard parametrization~\cite{Zyla:2020zbs},
\begin{equation}
\label{eq:UPMNS-PDG}U=\left(\begin{array}{ccc}
c_{12} c_{13} ~& s_{12} c_{13} ~& s_{13} e^{-i \delta_{C P}} \\
-s_{12} c_{23}-c_{12} s_{13} s_{23} e^{i \delta_{C P}} ~& c_{12} c_{23}-s_{12} s_{13} s_{23} e^{i \delta_{C P}} ~& c_{13} s_{23} \\
s_{12} s_{23}-c_{12} s_{13} c_{23} e^{i \delta_{C P}} ~& -c_{12} s_{23}-s_{12} s_{13} c_{23} e^{i \delta_{C P}} ~& c_{13} c_{23}
\end{array}\right) \operatorname{diag}(1, e^{i \frac{\alpha_{21}}{2}}, e^{i \frac{\alpha_{31}}{2}})\,,
\end{equation}
where $c_{ij}\equiv\cos \theta_{ij}, s_{ij}\equiv\sin\theta_{ij}$, $\delta_{CP}$ is Dirac CP violation phase, and $\alpha_{21},\alpha_{31}$ are called Majorana CP phases. In the models with two right-handed neutrinos, the lightest neutrino would be massless such that there is only one Majorana phase\footnote{For example, the phase $\alpha_{31}$ is unphysical in the case of $m_3=0$.}. Then the phase matrix $\operatorname{diag}(1, e^{i \frac{\alpha_{21}}{2}}, e^{i \frac{\alpha_{31}}{2}})$ should be replaced by $\operatorname{diag}(1, e^{i \phi/2}, 1)$, where $\phi$ is the Majorana CP phase. Our numerical scan shows that almost all mixing angles for these 25
lepton models are predicted to fall within the $1\sigma$ ranges, except that
$\sin^2\theta_{23}$ for the models $\mathcal{L}_{22},\, \mathcal{L}_{95},\, \mathcal{L}_{100},\, \mathcal{L}_{105},\, \mathcal{L}_{108},\, \mathcal{L}_{113},\, \mathcal{L}_{118},\, \mathcal{L}_{208},\,\mathcal{L}_{654}$ and $\sin^2\theta_{12}$ for the models $\mathcal{L}_{113},\,\mathcal{L}_{208}$ are beyond the $1\sigma$ region but within $3\sigma$ region. It is notable that the Dirac CP phase $\delta_{CP}$ is close to
$1.5\pi$ in the models $\mathcal{L}_{9},\,\mathcal{L}_{22},\,\mathcal{L}_{71},\,\mathcal{L}_{94},\,\mathcal{L}_{99},\,\mathcal{L}_{105},\,\mathcal{L}_{134}$
and $\mathcal{L}_{213}$. Because the modular forms satisfy the identity $Y^{(k)}_{\mathbf{r}}(-\tau^{*})=[Y^{(k)}_{\mathbf{r}}(\tau)]^{*}$, therefore both neutrino and charged lepton mass matrices which are functions of modular forms and coupling constants, would become their complex conjugate under the    transformation $\tau\rightarrow-\tau^{*}, g_i\rightarrow g^{*}_i$. Hence the pair of input parameters $\{-\tau^{*}, g^{*}_i\}$ and $\{\tau, g_i\}$ lead to the same predictions for lepton mixing angles while the overall signs of CP violating phases are reversed. Therefore the numerical results in table~\ref{tab:model_Aa}, table~\ref{tab:model_Ab} and table~\ref{tab:model_Ac} should come in pair with opposite CP phases.

The right-handed charged leptons are usually assumed to transform as singlets under the finite modular group, such that at least one free parameter is introduced for each generation of charged leptons and hierarchical charged lepton masses can be accommodated. If both left-handed leptons $L$ and right-handed charged leptons $E^c$ are assigned to be irreducible triplets of the finite modular group, all the coupling constants in $\mathcal{W}_e$ are generally relevant to the three charged lepton masses and thus some fine-tuning is necessary to reproduce the observed hierarchical masses. Indeed we notice that the models for triplet assignments $L, E^{c}\sim \mathbf{3},\mathbf{3'}$ can not accommodate the experimental data of lepton masses and mixing angles with less than eleven free parameters if neutrino mass spectrum is normal ordering. Nevertheless we are lucky enough to find a viable model $\mathcal{L}_{546}$ with 10 parameters which can fit the experimental data well for inverted ordering neutrino masses. The representation assignment of the lepton fields are $L\sim\mathbf{3}, E^c\sim \mathbf{3'}, N^{c}\sim\mathbf{2'}$, and the structure of the charged lepton and neutrino sectors are given by $C^{viii}_{3}$ and $T^{ii}_{1}$ respectively. By scanning the parameter space of this model, we identify the best fit point of the input parameters as follow:
\begin{eqnarray}
  \nonumber&&\langle\tau\rangle=-0.499466+3.22253i\,,~~ \beta/\alpha=0.207601e^{-0.00419297i}\,,~~ \gamma/\alpha= 0.1878e^{0.970918i}\,,\\
  &&\delta/\alpha= 0.0310921 e^{4.79438i}\,,~~ \alpha v_d=0.130959~\text{MeV}\,,~~ gv_u^2/\Lambda=65.6432~\text{meV}\,.
\end{eqnarray}
The corresponding predictions for masses and mixing parameters are determined to be
\begin{eqnarray}
\nonumber&& \sin^2\theta_{13}=0.02238\,,\quad \sin^2\theta_{12}=0.3040\,,\quad \sin^2\theta_{23}=0.5750\,,\quad \delta_{CP}=1.4826\pi\,,\\
\nonumber&& \alpha_{21}=1.5621\pi\,,\quad \alpha_{31}=0.4328\pi\,,\quad m_e/m_{\mu}=0.004737\,,\quad m_{\mu}/m_{\tau}=0.05857\,,\\
\nonumber&&
m_1=49.2324~\text{meV}\,,\quad m_2=49.9803~\text{meV}\,,\quad m_3=0~\text{meV}\,,\\
&& \sum_i m_i = 99.2126~\text{meV}\,,\quad  m_{\beta}=48.9043~\text{meV}\,,\quad m_{\beta\beta}=39.1869~\text{meV}\,,
\end{eqnarray}
which are in the experimentally preferred $3\sigma$ ranges. It is remarkable that the light neutrino mass matrix only depends on the complex modulus $\tau$ besides the overall scale $gv_u^2/\Lambda$, and the parameters in the charged lepton superpotential are almost of the same order of magnitude to reproduce the hierarchical masses of charged leptons.

{\tabcolsep=0.05cm
\begin{table}[t]
  \centering
  \begin{adjustbox}{width=\textwidth}
  \begin{tabulary}{1.0\linewidth}{|c|c|c|c|c|c|c|c|c|c|}
\hline\hline
Model & $\mathcal{L}_{2}$ & $\mathcal{L}_{3}$ & $\mathcal{L}_{9}$ & $\mathcal{L}_{10}$ & $\mathcal{L}_{203}$ & $\mathcal{L}_{208}$ & $\mathcal{L}_{209}$ & $\mathcal{L}_{213}$ & $\mathcal{L}_{214}$ \\\hline
$\texttt{Re}(\tau)$ & $-0.4301$ & $0.01635$ & $-0.4744$ & $-0.2201$ & $0.1301$ & $-0.1617$ & $-0.4154$ & $-0.2126$ & $-0.4533$ \\
$\texttt{Im}(\tau)$ & $1.5662$ & $1.1309$ & $1.5138$ & $1.0857$ & $1.0105$ & $1.0261$ & $1.3825$ & $1.1659$ & $1.0272$ \\
$\beta/\alpha$ & $15.3310$ & $16.4102$ & $10.6903$ & $17.0830$ & $20.2623$ & $15.2426$ & $6.1506$ & $310.8679$ & $4.2487$ \\
$\gamma/\alpha$ & $7.7360$ & $30.0197$ & $31.3607$ & $174.0123$ & $148.5829$ & $34.0370$ & $24.7109$ & $16.1121$ & $51.1912$ \\
$|\delta/\alpha|$ & $36.8022$ & $167.6753$ & $6.9087$ & $49.3413$ & $250.3694$ & $141.1909$ & $28.6349$ & $5.3042$ & $3.5954$ \\
$\arg(\delta/\alpha)/\pi$ & $0.5159$ & $-0.6076$ & $1.9128$ & $-0.04841$ & $1.9887$ & $-0.7929$ & $-0.5655$ & $-0.05895$ & $0.06108$ \\
$|g_2/g_1|$ & $0.2085$ & $0.4039$ & $0.1511$ & $0.1871$ & $0.8829$ & $0.1324$ & $0.8124$ & $1.3425$ & $0.1275$ \\
$\arg(g_2/g_1)/\pi$ & $0.9580$ & $-0.8764$ & $1.8735$ & $0.6548$ & $0.4599$ & $0.01922$ & $0.5882$ & $-0.6071$ & $-0.9575$ \\
$\alpha v_d/\text{MeV}$ & $0.2999$ & $0.07157$ & $1.0088$ & $0.1832$ & $0.04549$ & $0.06238$ & $0.3007$ & $0.1732$ & $0.6087$ \\
$(g_1^2 v_u^2/\Lambda)/\text{meV}$ & $0.6347$ & $6.4174$ & $2.6552$ & $13.2093$ & $0.003466$ & $0.3803$ & $0.3326$ & $0.03614$ & $30.5547$ \\\hline
$\sin^2\theta_{13}$ & $0.02219$ & $0.02219$ & $0.02219$ & $0.02219$ & $0.02219$ & $0.02222$ & $0.02219$ & $0.02219$ & $0.02219$ \\
$\sin^2\theta_{12}$ & $0.3040$ & $0.3040$ & $0.3040$ & $0.3040$ & $0.3040$ & $0.3304$ & $0.3040$ & $0.3040$ & $0.3040$ \\
$\sin^2\theta_{23}$ & $0.5730$ & $0.5730$ & $0.5730$ & $0.5730$ & $0.5730$ & $0.5268$ & $0.5730$ & $0.5730$ & $0.5730$ \\
$\delta_{CP}/\pi$ & $1.9530$ & $1.1723$ & $1.5987$ & $1.2714$ & $1.0917$ & $1.8661$ & $1.0537$ & $1.5926$ & $1.8659$ \\
$\alpha_{21}/\pi\text{~or~}\phi/\pi$ & $0.9819$ & $1.9604$ & $0.5319$ & $1.8453$ & $0.8918$ & $1.8616$ & $1.6567$ & $0.1986$ & $0.8342$ \\
$\alpha_{31}/\pi$ & $0.2069$ & $1.5915$ & $0.3322$ & $0.7125$ & $-$ & $-$ & $-$ & $-$ & $-$ \\\hline
$m_1/\text{meV}$ & $1.5466$ & $2.6701$ & $29.4704$ & $8.5857$ & $0$ & $0$ & $0$ & $0$ & $0$ \\
$m_2/\text{meV}$ & $8.7517$ & $9.0183$ & $30.7035$ & $12.1620$ & $8.6139$ & $8.6139$ & $8.6139$ & $8.6139$ & $8.6139$ \\
$m_3/\text{meV}$ & $50.1937$ & $50.2408$ & $58.1852$ & $50.8989$ & $50.1699$ & $50.0286$ & $50.1698$ & $50.1698$ & $50.1704$ \\
$\sum_{i}m_{i}/\text{meV}$ & $60.4920$ & $61.9292$ & $118.3591$ & $71.6466$ & $58.7838$ & $58.6426$ & $58.7838$ & $58.7838$ & $58.7843$ \\
$m_{\beta}/\text{meV}$ & $8.9611$ & $9.2216$ & $30.7638$ & $12.3135$ & $8.8266$ & $8.9211$ & $8.8266$ & $8.8266$ & $8.8267$ \\
$m_{\beta\beta}/\text{meV}$ & $1.3783$ & $3.8458$ & $19.8973$ & $10.0743$ & $1.5008$ & $3.2825$ & $3.4644$ & $2.4001$ & $2.5732$\\ \hline\hline
  \end{tabulary}
\end{adjustbox}
\caption{\label{tab:model_Aa}
The best fit values of the free parameters and the corresponding predictions for lepton mixing parameters and neutrino masses for the phenomenologically viable lepton models $\mathcal{L}_2$, $\mathcal{L}_3$, $\mathcal{L}_9$, $\mathcal{L}_{10}$, $\mathcal{L}_{203}$, $\mathcal{L}_{208}$, $\mathcal{L}_{209}$, $\mathcal{L}_{213}$} and $\mathcal{L}_{214}$. Here we only show the results for NO neutrino masses, similar results can be obtained for IO.
\end{table}
}

{\begin{table}[tp!]
    \centering
{\tabcolsep=0.225cm
    \begin{tabulary}{1.0\linewidth}{|c|c|c|c|c|c|c|}
\hline\hline
Model & $\mathcal{L}_{22}$ & $\mathcal{L}_{94}$ & $\mathcal{L}_{95}$ & $\mathcal{L}_{99}$ & $\mathcal{L}_{100}$ & $\mathcal{L}_{134}$ \\\hline
$\texttt{Re}(\tau)$ & $-0.4910$ & $0.3839$ & $-0.4778$ & $0.4747$ & $-0.4668$ & $0.4990$ \\
$\texttt{Im}(\tau)$ & $1.1953$ & $2.4844$ & $1.1283$ & $2.5061$ & $0.8867$ & $0.8858$ \\
$|\beta/\alpha|$ & $1.4037$ & $0.2154$ & $0.2309$ & $0.2183$ & $0.01022$ & $0.3598$ \\
$\gamma/\alpha$ & $0.00786$ & $0.001115$ & $0.1315$ & $0.0002241$ & $0.00013$ & $0.01207$ \\
$\arg(\beta/\alpha)/\pi$ & $0.9985$ & $0.06929$ & $1.0002$ & $1.9289$ & $0.01887$ & $1.0296$ \\
$|g_2/g_1|$ & $0.1863$ & $0.1603$ & $0.2239$ & $0.1582$ & $0.1734$ & $0.09765$ \\
$\arg(g_2/g_1)/\pi$ & $2.0000$ & $0.6789$ & $1.5045$ & $1.3016$ & $0.0618$ & $0.6328$ \\
$\alpha v_d/\text{MeV}$ & $6.3649$ & $84.0053$ & $137.6646$ & $83.5208$ & $89.8671$ & $7.6514$ \\
$(g_1^2 v_u^2/\Lambda)/\text{meV}$ & $2.4388$ & $3.5208$ & $19.4457$ & $3.5300$ & $63.2376$ & $4.6820$ \\\hline
$\sin^2\theta_{13}$ & $0.02218$ & $0.02255$ & $0.02219$ & $0.02208$ & $0.02214$ & $0.02222$ \\
$\sin^2\theta_{12}$ & $0.3049$ & $0.2975$ & $0.3040$ & $0.3003$ & $0.3036$ & $0.3031$ \\
$\sin^2\theta_{23}$ & $0.5139$ & $0.5687$ & $0.5730$ & $0.5768$ & $0.5261$ & $0.5806$ \\
$\delta_{CP}/\pi$ & $1.3747$ & $1.3886$ & $1.8457$ & $1.4930$ & $1.9651$ & $1.5105$ \\
$\alpha_{21}/\pi$ & $1.4775$ & $0.2152$ & $0.6612$ & $0.2212$ & $1.3186$ & $0.6445$ \\
$\alpha_{31}/\pi$ & $0.6216$ & $0.9762$ & $0.5052$ & $1.3185$ & $0.3786$ & $1.4878$ \\\hline
$m_1/\text{meV}$ & $30.5138$ & $53.9796$ & $27.9151$ & $52.7775$ & $72.9690$ & $103.2841$ \\
$m_2/\text{meV}$ & $31.7063$ & $54.6626$ & $29.2139$ & $53.4758$ & $73.4756$ & $103.6427$ \\
$m_3/\text{meV}$ & $58.7318$ & $73.9091$ & $57.4135$ & $72.2129$ & $88.5617$ & $114.8190$ \\
$\sum_{i}m_{i}/\text{meV}$ & $120.9519$ & $182.5514$ & $114.5425$ & $178.4662$ & $235.0063$ & $321.7459$ \\
$m_ {\beta}/\text{meV}$ & $31 .7657$ & $54 .7069$ & $29 .2773$ & $53 .4873$ & $73 .5001$ & $103 .6605$ \\
$m_{\beta\beta}/\text{meV}$ & $23.4781$ & $51.9764$ & $16.0374$ & $50.2940$ & $41.6132$ & $64.6318$\\ \hline
    \end{tabulary}}\\
\begin{tabular}{ccc}
    \begin{tabulary}{1.0\linewidth}{|c|c|}
\hline
Model & $\mathcal{L}_{105}$ \\\hline
$\texttt{Re}(\tau)$ & $0.4962$ \\
$\texttt{Im}(\tau)$ & $0.9177$ \\
$\beta/\alpha$ & $0.2504$ \\
$|\gamma/\alpha|$ & $0.1404$ \\
$\arg(\gamma/\alpha)/\pi$ & $1.2437\times 10^{-6}$ \\
$|g_2/g_1|$ & $0.2464$ \\
$\arg(g_2/g_1)/\pi$ & $0.5020$ \\
$\alpha v_d/\text{MeV}$ & $81.9172$ \\
$(g_1^2 v_u^2/\Lambda)/\text{meV}$ & $31.1447$ \\\hline
$\sin^2\theta_{13}$ & $0.02266$ \\
$\sin^2\theta_{12}$ & $0.3036$ \\
$\sin^2\theta_{23}$ & $0.5898$ \\
$\delta_{CP}/\pi$ & $1.4790$ \\
$\alpha_{21}/\pi$ & $1.9410$ \\
$\alpha_{31}/\pi$ & $0.9605$ \\\hline
$m_1/\text{meV}$ & $71.9302$ \\
$m_2/\text{meV}$ & $72.4442$ \\
$m_3/\text{meV}$ & $87.5970$ \\
$\sum_{i}m_{i}/\text{meV}$ & $231.9714$ \\
$m_ {\beta}/\text{meV}$ & $72 .4748$ \\
$m_{\beta\beta}/\text{meV}$ & $72.1775$\\ \hline\hline
    \end{tabulary} &
\begin{tabulary}{1.0\linewidth}{|c|c|}
\hline
Model & $\mathcal{L}_{210}$ \\\hline
$\texttt{Re}(\tau)$ & $-0.1451$ \\
$\texttt{Im}(\tau)$ & $1.3560$ \\
$\beta/\alpha$ & $6.9431$ \\
$\gamma/\alpha$ & $43.2744$ \\
$|\delta/\alpha|$ & $48.3050$ \\
$\arg(\delta/\alpha)/\pi$ & $0.2958$ \\
$|\Lambda_2/\Lambda_1|$ & $1.5523$ \\
$\arg(\Lambda_2/\Lambda_1)/\pi$ & $0.09099$ \\
$\alpha v_d/\text{MeV}$ & $0.2300$ \\
$(g^2 v_u^2/\Lambda_1)/\text{meV}$ & $486.6774$ \\\hline
$\sin^2\theta_{13}$ & $0.02219$ \\
$\sin^2\theta_{12}$ & $0.3040$ \\
$\sin^2\theta_{23}$ & $0.5730$ \\
$\delta_{CP}/\pi$ & $1.0114$ \\
$\phi/\pi$ & $1.9742$ \\\hline

$m_1/\text{meV}$ & $0$ \\
$m_2/\text{meV}$ & $8.6139$ \\
$m_3/\text{meV}$ & $50.1696$ \\
$\sum_{i}m_{i}/\text{meV}$ & $58.7835$ \\
$m_{\beta}/\text{meV}$ & $8.8266$ \\
$m_{\beta\beta}/\text{meV}$ & $3.6738$\\ \hline\hline
\end{tabulary} &
\raisebox{0.255cm}{\begin{tabulary}{1.0\linewidth}{|c|c|}
\hline
Model & $\mathcal{L}_{310}$ \\\hline
$\texttt{Re}(\tau)$ & $-0.1214$ \\
$\texttt{Im}(\tau)$ & $1.3386$ \\
$|\beta/\alpha|$ & $0.6978$ \\
$\gamma/\alpha$ & $53.2957$ \\
$\arg(\beta/\alpha)/\pi$ & $1.0011$ \\
$|\Lambda_2/\Lambda_1|$ & $0.4808$ \\
$\arg(\Lambda_2/\Lambda_1)/\pi$ & $0.04589$ \\
$\alpha v_d/\text{MeV}$ & $3.6866$ \\
$(g^2 v_u^2/\Lambda_1)/\text{meV}$ & $53.3074$ \\\hline
$\sin^2\theta_{13}$ & $0.02219$ \\
$\sin^2\theta_{12}$ & $0.3040$ \\
$\sin^2\theta_{23}$ & $0.5729$ \\
$\delta_{CP}/\pi$ & $1.6779$ \\
$\phi/\pi$ & $0.02598$ \\\hline
$m_1/\text{meV}$ & $0$ \\
$m_2/\text{meV}$ & $8.6139$ \\
$m_3/\text{meV}$ & $50.1703$ \\
$\sum_{i}m_{i}/\text{meV}$ & $58.7842$ \\
$m_ {\beta}/\text{meV}$ & $8 .8266$ \\
$m_{\beta\beta}/\text{meV}$ & $2.3928$\\ \hline\hline
\end{tabulary}}
\end{tabular}
\caption{\label{tab:model_Ab}
The best fit values of the free parameters and the corresponding predictions for lepton mixing parameters and neutrino masses for the phenomenologically viable lepton models $\mathcal{L}_{22}$, $\mathcal{L}_{94}$, $\mathcal{L}_{95}$, $\mathcal{L}_{99}$, $\mathcal{L}_{100}$, $\mathcal{L}_{105}$, $\mathcal{L}_{134}$, $\mathcal{L}_{210}$ and $\mathcal{L}_{310}$, Here we only show the results for NO spectrum, similar results can be obtained for IO.}
\end{table}
}

\begin{table}[htp]
  \centering
  {\small
    \begin{tabular}{cc}
\tabcolsep=0.05cm
  \begin{tabulary}{0.6\linewidth}{|c|c|c|c|c|c|c|}
\hline
Model & $\mathcal{L}_{66}$ & $\mathcal{L}_{71}$ & $\mathcal{L}_{76}$ & $\mathcal{L}_{108}$ & $\mathcal{L}_{113}$ & $\mathcal{L}_{118}$ \\\hline
$\texttt{Re}(\tau)$ & $0.4294$ & $-0.261$ & $0.4294$ & $-0.1348$ & $-0.2062$ & $-0.1348$ \\
$\texttt{Im}(\tau)$ & $0.9908$ & $1.1280$ & $0.9908$ & $1.3023$ & $0.9786$ & $1.3023$ \\
$|\beta/\alpha|$ & $0.5076$ & $0.3257$ & $0.5075$ & $5.2661$ & $0.2303$ & $5.2670$ \\
$|\gamma/\alpha|$ & $0.7506$ & $1.8503$ & $0.7504$ & $4.4250$ & $1.8075$ & $4.4257$ \\
$\delta/\alpha$ & $0.009538$ & $1.8444$ & $0.0004353$ & $0.01518$ & $2.4523$ & $0.004519$ \\
$\arg(\beta/\alpha)/\pi$ & $-0.9523$ & $0.9492$ & $-0.9525$ & $1.8857$ & $1.0011$ & $1.8858$ \\
$\arg(\gamma/\alpha)/\pi$ & $-0.951$ & $-0.1346$ & $-0.9512$ & $1.8899$ & $0.00007229$ & $1.8899$ \\
$\alpha v_d/\text{MeV}$ & $13.6924$ & $5.3249$ & $13.6953$ & $5.8555$ & $6.9077$ & $5.8545$ \\
$(g^2 v_u^2/\Lambda)/\text{meV}$ & $452.8878$ & $318.7155$ & $452.9001$ & $310.6406$ & $299.6404$ & $310.6424$ \\\hline
$\sin^2\theta_{13}$ & $0.0222$ & $0.02213$ & $0.0222$ & $0.0222$ & $0.02265$ & $0.0222$ \\
$\sin^2\theta_{12}$ & $0.3057$ & $0.3086$ & $0.3057$ & $0.3072$ & $0.2906$ & $0.3072$ \\
$\sin^2\theta_{23}$ & $0.5731$ & $0.5738$ & $0.573$ & $0.5324$ & $0.5525$ & $0.5324$ \\
$\delta_{CP}/\pi$ & $1.0015$ & $1.5379$ & $1.0016$ & $1.0370$ & $1.0001$ & $1.0369$ \\
$\alpha_{21}/\pi$ & $1.1130$ & $1.4056$ & $1.1130$ & $1.1604$ & $0.9998$ & $1.1604$ \\
$\alpha_{31}/\pi$ & $0.4217$ & $0.212$ & $0.4219$ & $0.5879$ & $1.9994$ & $0.5879$ \\\hline
$m_1/\text{meV}$ & $14.8481$ & $10.9731$ & $14.8485$ & $11.7901$ & $8.7593$ & $11.7902$ \\
$m_2/\text{meV}$ & $17.1658$ & $13.9502$ & $17.1662$ & $14.6016$ & $12.2852$ & $14.6017$ \\
$m_3/\text{meV}$ & $52.2585$ & $51.3274$ & $52.2591$ & $51.8982$ & $51.3966$ & $51.8987$ \\
$\sum_{i}m_{i}/\text{meV}$ & $84.2725$ & $76.2507$ & $84.2737$ & $78.2899$ & $72.4411$ & $78.2906$ \\
$m_{\beta}/\text{meV}$ & $17.2737$ & $14.0865$ & $17.2741$ & $14.7647$ & $12.4857$ & $14.7648$ \\
$m_{\beta\beta}/\text{meV}$ & $5.6017$ & $6.8406$ & $5.6022$ & $4.2079$ & $3.7470$ & $4.2073$\\ \hline\hline
  \end{tabulary} &
  \begin{tabulary}{0.1\linewidth}{|c|c|}
\hline
Model & $\mathcal{L}_{654}$ \\\hline
$\texttt{Re}(\tau)$ & $0.0007672$ \\
$\texttt{Im}(\tau)$ & $1.0033$ \\
$\beta/\alpha$ & $1.5040$ \\
$|\Lambda_1/\Lambda_2|$ & $0.0005601$ \\
$|\Lambda_1/\Lambda_3|$ & $0.0007015$ \\
$\arg(\Lambda_1/\Lambda_2)/\pi$ & $0.9717$ \\
$\arg(\Lambda_1/\Lambda_3)/\pi$ & $-0.1754$ \\
$\alpha v_d/\text{MeV}$ & $29.9295$ \\
$(v_u^2/\Lambda_1)/\text{meV}$ & $89.3207$ \\\hline
$\sin^2\theta_{13}$ & $0.02211$ \\
$\sin^2\theta_{12}$ & $0.303$ \\
$\sin^2\theta_{23}$ & $0.5191$ \\
$\delta_{CP}/\pi$ & $1.7986$ \\
$\alpha_{21}/\pi$ & $0.1196$ \\
$\alpha_{31}/\pi$ & $0.221$ \\\hline
$m_1/\text{meV}$ & $159.9097$ \\
$m_2/\text{meV}$ & $160.1415$ \\
$m_3/\text{meV}$ & $167.5600$ \\
$\sum_{i}m_{i}/\text{meV}$ & $487.6113$ \\
$m_{\beta}/\text{meV}$ & $160.1515$ \\
$m_{\beta\beta}/\text{meV}$ & $153.1525$ \\ \hline\hline
  \end{tabulary}
 \end{tabular}
}
  \caption{\label{tab:model_Ac}The best fit values of the free parameters and the corresponding predictions for lepton mixing parameters and neutrino masses for the phenomenologically viable lepton models $\mathcal{L}_{66}$, $\mathcal{L}_{71}$, $\mathcal{L}_{76}$, $\mathcal{L}_{108}$, $\mathcal{L}_{113}$, $\mathcal{L}_{118}$ and $\mathcal{L}_{654}$. We only show the results for NO neutrino masses, similar results can be obtained for IO. }
\end{table}

\begin{table}[htp]
  \centering
  {\tabcolsep=0.007cm
\begin{tabulary}{1.0\linewidth}{|c|c|c|c|c|c|c|c|c|c|}
\hline\hline
Model & $\mathcal{L}_{2}$ & $\mathcal{L}_{3}$ & $\mathcal{L}_{9}$ & $\mathcal{L}_{10}$ & $\mathcal{L}_{203}$ & $\mathcal{L}_{208}$ & $\mathcal{L}_{209}$ & $\mathcal{L}_{213}$ & $\mathcal{L}_{214}$ \\\hline
$\texttt{Re}(\tau)$ & $-0.2138$ & $-0.3522$ & $0.1065$ & $-0.3594$ & $0.1243$ & $0.1652$ & $-0.4151$ & $-0.2433$ & $-0.4476$ \\
$\texttt{Im}(\tau)$ & $1.7278$ & $1.1894$ & $1.5264$ & $0.9641$ & $0.9922$ & $0.9863$ & $1.3481$ & $1.2336$ & $1.0189$ \\
$\beta/\alpha$ & $10.9859$ & $15.6084$ & $29.5919$ & $12.9879$ & $19.7339$ & $13.9606$ & $6.3027$ & $233.8248$ & $5.4382$ \\
$\gamma/\alpha$ & $17.4072$ & $41.1052$ & $418.1776$ & $183.8761$ & $144.6948$ & $33.5498$ & $63.8737$ & $13.5794$ & $69.6730$ \\
$\delta/\alpha$ & $31.2268$ & $-74.7463$ & $202.3009$ & $35.4668$ & $245.0666$ & $-118.8148$ & $55.8680$ & $3.9524$ & $5.9153$ \\
$g_2/g_1$ & $-0.5546$ & $0.2770$ & $-0.1202$ & $-0.1693$ & $3.7779$ & $0.1298$ & $2.7708$ & $-5.4091$ & $-0.1275$ \\
$\alpha v_d/\text{MeV}$ & $0.4969$ & $0.1067$ & $0.2396$ & $0.1714$ & $0.04456$ & $0.06268$ & $0.2220$ & $0.2418$ & $0.4664$ \\
$(g_1^2 v_u^2/\Lambda)/\text{meV}$ & $0.2144$ & $24.9728$ & $2.2785$ & $27.6420$ & $0.0003282$ & $0.3518$ & $0.02329$ & $0.002565$ & $30.2048$ \\\hline
$\sin^2\theta_{13}$ & $0.02219$ & $0.02219$ & $0.02219$ & $0.02219$ & $0.02228$ & $0.02222$ & $0.02219$ & $0.02219$ & $0.02219$ \\
$\sin^2\theta_{12}$ & $0.3040$ & $0.3040$ & $0.3040$ & $0.3040$ & $0.3004$ & $0.3311$ & $0.3040$ & $0.3040$ & $0.3040$ \\
$\sin^2\theta_{23}$ & $0.5730$ & $0.5730$ & $0.5730$ & $0.5730$ & $0.5783$ & $0.5277$ & $0.5730$ & $0.5730$ & $0.5730$ \\
$\delta_{CP}/\pi$ & $1.7651$ & $1.5038$ & $1.5499$ & $1.4679$ & $1.0000$ & $2.0000$ & $1.4054$ & $1.2309$ & $1.7173$ \\
$\alpha_{21}/\pi\text{~or~}\phi/\pi$ & $1.2072$ & $1.6230$ & $1.8248$ & $1.8967$ & $2.0000$ & $0.$ & $1.0998$ & $0.3718$ & $0.9520$ \\
$\alpha_{31}/\pi$ & $1.1691$ & $0.7682$ & $0.05413$ & $1.7955$ & $-$ & $-$ & $-$ & $-$ & $-$ \\\hline
$m_1/\text{meV}$ & $3.5140$ & $7.9171$ & $17.7241$ & $21.1711$ & $0$ & $0$ & $0$ & $0$ & $0$ \\
$m_2/\text{meV}$ & $9.3031$ & $11.6996$ & $19.7064$ & $22.8564$ & $8.6139$ & $8.6139$ & $8.6139$ & $8.6139$ & $8.6139$ \\
$m_3/\text{meV}$ & $50.2927$ & $50.7906$ & $53.2085$ & $54.4539$ & $50.1720$ & $50.0307$ & $50.1699$ & $50.1699$ & $50.1701$ \\
$\sum_{i}m_{i}/\text{meV}$ & $63.1098$ & $70.4073$ & $90.6391$ & $98.4814$ & $58.7859$ & $58.6446$ & $58.7838$ & $58.7839$ & $58.7840$ \\
$m_{\beta}/\text{meV}$ & $9.5004$ & $11.8570$ & $19.8003$ & $22.9374$ & $8.8248$ & $8.9242$ & $8.8266$ & $8.8266$ & $8.8266$ \\
$m_{\beta\beta}/\text{meV}$ & $2.7678$ & $8.5148$ & $16.1483$ & $19.8234$ & $3.6479$ & $3.9003$ & $3.6433$ & $1.6898$ & $3.1280$\\ \hline
  \end{tabulary}
  }\\
  {\begin{tabular}{cc}
      \tabcolsep=0.1cm
  \begin{tabulary}{1.0\linewidth}{|c|c|c|c|c|c|}
\hline
Model & $\mathcal{L}_{22}$ & $\mathcal{L}_{94}$ & $\mathcal{L}_{95}$ & $\mathcal{L}_{99}$ & $\mathcal{L}_{100}$ \\\hline
$\texttt{Re}(\tau)$ & $-0.491$ & $0.4671$ & $-0.4682$ & $0.4672$ & $0.4601$ \\
$\texttt{Im}(\tau)$ & $1.1952$ & $2.7528$ & $0.8847$ & $2.7493$ & $0.8900$ \\
$\beta/\alpha$ & $-1.4038$ & $-0.2449$ & $0.01109$ & $-0.2178$ & $0.005634$ \\
$\gamma/\alpha$ & $0.00786$ & $0.0005718$ & $0.008922$ & $0.0001079$ & $0.0001036$ \\
$g_2/g_1$ & $0.1863$ & $0.1428$ & $0.1713$ & $0.1428$ & $-0.1709$ \\
$\alpha v_d/\text{MeV}$ & $6.3631$ & $157.9035$ & $89.2429$ & $167.4996$ & $92.0670$ \\
$(g_1^2 v_u^2/\Lambda)/\text{meV}$ & $2.4392$ & $7.9148$ & $66.7014$ & $7.8966$ & $58.9994$ \\\hline
$\sin^2\theta_{13}$ & $0.0222$ & $0.02214$ & $0.02218$ & $0.02215$ & $0.02216$ \\
$\sin^2\theta_{12}$ & $0.3045$ & $0.3039$ & $0.3045$ & $0.304$ & $0.3043$ \\
$\sin^2\theta_{23}$ & $0.5139$ & $0.4716$ & $0.5336$ & $0.4751$ & $0.5239$ \\
$\delta_{CP}/\pi$ & $1.3742$ & $1.4922$ & $1.7430$ & $1.4925$ & $1.4035$ \\
$\alpha_{21}/\pi$ & $1.4774$ & $1.9757$ & $0.7991$ & $1.9765$ & $1.6386$ \\
$\alpha_{31}/\pi$ & $0.6209$ & $0.9833$ & $1.6663$ & $0.9839$ & $0.7474$ \\\hline
$m_1/\text{meV}$ & $30.5220$ & $96.4660$ & $75.2809$ & $96.2227$ & $64.9946$ \\
$m_2/\text{meV}$ & $31.7142$ & $96.8498$ & $75.7721$ & $96.6074$ & $65.5629$ \\
$m_3/\text{meV}$ & $58.7392$ & $108.7335$ & $90.4731$ & $108.5176$ & $82.1105$ \\
$\sum_{i}m_{i}/\text{meV}$ & $120.9755$ & $302.0493$ & $241.5261$ & $301.3477$ & $212.6681$ \\
$m_{\beta}/\text{meV}$ & $31.7741$ & $96.8683$ & $75.7967$ & $96.6261$ & $65.5910$ \\
$m_{\beta\beta}/\text{meV}$ & $23.4883$ & $96.7927$ & $37.5455$ & $96.5544$ & $57.1477$\\ \hline\hline
  \end{tabulary}&
  \begin{tabulary}{1.0\linewidth}{|c|c|}
\hline
Model & $\mathcal{L}_{210}$ \\\hline
$\texttt{Re}(\tau)$ & $0.1167$ \\
$\texttt{Im}(\tau)$ & $1.3753$ \\
$\beta/\alpha$ & $6.4493$ \\
$\gamma/\alpha$ & $62.2329$ \\
$\delta/\alpha$ & $65.1171$ \\
$\Lambda_2/\Lambda_1$ & $1.6840$ \\
$\alpha v_d/\text{MeV}$ & $0.2222$ \\
$(g^2 v_u^2/\Lambda_1)/\text{meV}$ & $531.2511$ \\\hline
$\sin^2\theta_{13}$ & $0.02219$ \\
$\sin^2\theta_{12}$ & $0.3040$ \\
$\sin^2\theta_{23}$ & $0.5730$ \\
$\delta_{CP}/\pi$ & $1.7314$ \\
$\phi/\pi$ & $0.9433$ \\\hline
$m_1/\text{meV}$ & $0$ \\
$m_2/\text{meV}$ & $8.6139$ \\
$m_3/\text{meV}$ & $50.1698$ \\
$\sum_{i}m_{i}/\text{meV}$ & $58.7837$ \\
$m_ {\beta}/\text{meV}$ & $8.8266$ \\
$m_{\beta\beta}/\text{meV}$ & $3.0747$\\ \hline\hline
  \end{tabulary}
  \end{tabular}
}
\caption{\label{tab:model_gCP_A}The best fit values of the free parameters and the corresponding predictions for lepton mixing parameters and neutrino masses for the phenomenologically viable lepton models $\mathcal{L}_{2}$, $\mathcal{L}_{3}$, $\mathcal{L}_{9}$, $\mathcal{L}_{10}$, $\mathcal{L}_{22}$, $\mathcal{L}_{94}$, $\mathcal{L}_{95}$, $\mathcal{L}_{99}$, $\mathcal{L}_{100}$, $\mathcal{L}_{203}$, $\mathcal{L}_{208}$, $\mathcal{L}_{209}$, $\mathcal{L}_{210}$, $\mathcal{L}_{213}$ and $\mathcal{L}_{214}$ after CP invariance is incorporated. We only show the results for NO neutrino masses, similar results can be obtained for IO. }
\end{table}

\begin{table}[htp]
  \centering
  \begin{tabulary}{0.6\linewidth}{|c|c|c|c|c|}
\hline
Model & $\mathcal{L}_{66}$ & $\mathcal{L}_{108}$ & $\mathcal{L}_{113}$ & $\mathcal{L}_{118}$ \\\hline
$\texttt{Re}(\tau)$ & $-0.4473$ & $-0.1018$ & $0.2062$ & $-0.1019$ \\
$\texttt{Im}(\tau)$ & $0.9923$ & $1.3295$ & $0.9785$ & $1.3295$ \\
$\beta/\alpha$ & $-0.4859$ & $6.7614$ & $-0.2299$ & $6.7613$ \\
$\gamma/\alpha$ & $-0.7149$ & $5.6934$ & $1.8047$ & $5.6933$ \\
$\delta/\alpha$ & $0.01136$ & $0.01965$ & $2.4496$ & $0.005307$ \\
$\alpha v_d/\text{MeV}$ & $14.4960$ & $4.5445$ & $6.9183$ & $4.5446$ \\
$(g^2 v_u^2/\Lambda)/\text{meV}$ & $448.6740$ & $315.4090$ & $299.5957$ & $315.4089$ \\\hline
$\sin^2\theta_{13}$ & $0.02238$ & $0.02218$ & $0.02261$ & $0.02218$ \\
$\sin^2\theta_{12}$ & $0.3238$ & $0.3051$ & $0.2903$ & $0.3051$ \\
$\sin^2\theta_{23}$ & $0.5918$ & $0.5264$ & $0.5521$ & $0.5264$ \\
$\delta_{CP}/\pi$ & $1.2289$ & $1.2707$ & $1.0000$ & $1.2707$ \\
$\alpha_{21}/\pi$ & $1.0148$ & $1.2751$ & $1.0000$ & $1.2751$ \\
$\alpha_{31}/\pi$ & $1.7057$ & $0.5202$ & $0.0000$ & $0.5202$ \\\hline
$m_1/\text{meV}$ & $14.7337$ & $12.1497$ & $8.7568$ & $12.1496$ \\
$m_2/\text{meV}$ & $17.0670$ & $14.8934$ & $12.2834$ & $14.8934$ \\
$m_3/\text{meV}$ & $51.3132$ & $52.0157$ & $51.3632$ & $52.0157$ \\
$\sum_{i}m_{i}/\text{meV}$ & $83.1140$ & $79.0588$ & $72.4033$ & $79.0588$ \\
$m_{\beta}/\text{meV}$ & $17.1650$ & $15.0495$ & $12.4761$ & $15.0495$ \\
$m_{\beta\beta}/\text{meV}$ & $3.6797$ & $7.3807$ & $3.7492$ & $7.3807$\\ \hline\hline
  \end{tabulary}
  \caption{\label{tab:model_gCP_B}The best fit values of the free parameters and the corresponding predictions for lepton mixing parameters and neutrino masses for the phenomenologically viable lepton models $\mathcal{L}_{66}$, $\mathcal{L}_{108}$, $\mathcal{L}_{113}$ and $\mathcal{L}_{118}$ after CP invariance is incorporated. We only show the results for NO neutrino masses, similar results can be obtained for IO.}
\end{table}

Furthermore, we notice that the predictions for the CP phases $\delta_{CP}$, $\alpha_{21}$ and $\alpha_{31}$ ($\phi$ for $m_1=0$) scatter in wide ranges. The future long baseline neutrino experiments DUNE and Hyper-Kamiokande, if running in both neutrino and anti-neutrino modes, will allow for a measurement of the Dirac CP phase with a certain precision and have thus the potential to rule out some of our model. The Majorana phase could be probed or at least constrained by neutrinoless double beta decay ($0\nu\beta\beta$) experiments. The dependence of the $0\nu\beta\beta$ decay amplitude on the mixing parameters enter through the effective Majorana neutrino mass $m_{\beta\beta}$ with
\begin{equation}
m_{\beta\beta}=\left|m_{1} \cos ^{2} \theta_{12} \cos ^{2} \theta_{13}+m_{2} \sin ^{2} \theta_{12} \cos ^{2} \theta_{13} e^{i \alpha_{21}}+m_{3} \sin ^{2} \theta_{13} e^{i\left(\alpha_{31}-2 \delta_{C P}\right)}\right|\,,
\end{equation}
which involves all mixing parameters except $\theta_{23}$. If the lightest neutrino is massless, $m_{\beta\beta}$ takes a simpler form,
\begin{equation}
m_{\beta\beta}=\left\{\begin{array}{lc}\left|m_{2} \sin ^{2} \theta_{12} \cos ^{2} \theta_{13} e^{i \phi}+m_{3} \sin ^{2} \theta_{13} e^{-i2 \delta_{C P}}\right|,  ~&~  m_1=0\,,\\[0.1in]
\left|m_{1} \cos ^{2} \theta_{12} \cos ^{2} \theta_{13}+m_{2} \sin ^{2} \theta_{12} \cos ^{2} \theta_{13} e^{i\phi}\right|,  ~&~  m_3=0 \,.
\end{array}
\right.
\end{equation}
The most stringent bound on the effective Majorana neutrino mass is $m_{\beta\beta}<(61\sim 165)$ meV from KamLAND-Zen~\cite{KamLAND-Zen:2016pfg}. We see that these viable models predict $m_{\beta\beta}<80$ meV except the model $\mathcal{L}_{654}$ which gives $m_{\beta\beta}=153.1525$ meV. All these values of the effective Majorana neutrino mass $m_{\beta\beta}$ are below the upper bound of KamLAND-Zen. In the two right-handed neutrino models, $m_{\beta\beta}$ is determined to be few meV and it is far below the sensitivity of future $0\nu\beta\beta$ experiments. It is well-known that neutrino masses can be directly probed by kinematic studies of weak-interaction processes such as $\beta$ decay of tritium, and the effective neutrino mass measured in beta decay is
\begin{align}
&m_\beta=\sqrt{m^2_1 \cos^2\theta_{12}\cos^2\theta_{13}+m^2_2\sin^2\theta_{12}\cos^2\theta_{13}+m^2_3\sin^2\theta_{13} }\,,
\end{align}
which is independent of CP phases. The latest bound on $m_{\beta}$ is $m_{\beta}<1.1$ eV at $90\%$ confidence level from KATRIN~\cite{Aker:2019uuj}, and the sensitivity on $m_{\beta}$ is expected to be improved by one order of magnitude down to $0.2$ eV. We see that the KATRIN bound $m_{\beta}<1.1$ eV is safely fulfilled in all these 25 models.

\subsection{Lepton models with gCP}

As shown in section~\ref{sec:gCP-A5DC},  the generalized CP symmetry enforces all coupling constants to be real in our working basis. In other words, CP invariance requires the phases of coupling constants are equal to $0$ or $\pi$. Thus the VEV of complex modulus $\tau$ would be the unique
source of all CP violating phases. As a consequence, the free parameters of  the $25$ lepton models in table~\ref{tab:lepton_models} are reduced by two after incorporating CP invariance. We find that only 6 models are excluded by the experimental data after gCP symmetry is imposed, and the remaining 19 models were still in good agreement with the experiments. The minimal models with gCP and $A'_5$ modular symmetry are $\mathcal{L}_{22},\,\mathcal{L}_{66},\, \mathcal{L}_{94},\,\mathcal{L}_{95},\, \mathcal{L}_{99},\, \mathcal{L}_{100},\, \mathcal{L}_{108},\, \mathcal{L}_{113}$ and $\mathcal{L}_{118}$ which use 7 real free parameters to describe the 12 observables including 3 charged lepton masses, 3 light neutrino masses, 3 lepton mixing angles and 3 CP violation phases. The right-handed charged leptons transform as the direct sum of doublet and singlet of $A'_5$ in all these minimal models. The numerical results are summarized in table~\ref{tab:model_gCP_A} and table~\ref{tab:model_gCP_B}, we see that the current bounds on both $m_{\beta}$ and $m_{\beta\beta}$ are satisfied as well.

In the models $\mathcal{L}_{66}$, $\mathcal{L}_{108}$, $\mathcal{L}_{113}$ and $\mathcal{L}_{118}$, the charged lepton mass matrix depends on four coupling constants $\alpha$, $\beta$, $\gamma$ and $\delta$ while the light neutrino mass matrix only depends on the modulus $\tau$ up to the overall scale $g^2v^2_u/\Lambda$ such that the neutrino mass ratios are completely determined by the value of $\tau$. For the models $\mathcal{L}_{22}$, $\mathcal{L}_{94}$, $\mathcal{L}_{95}$, $\mathcal{L}_{99}$ and $\mathcal{L}_{100}$, the charged lepton mass hierarchies rely on cancellation of the comparable $\alpha$ and $\beta$ terms, this is a new feature in comparison with other modular models in the literature. It is notable that the VEVs of $\tau$ are very close to the CP conserved points $\texttt{Re}(\tau)=\pm 1/2$ in these models, and departure from $\texttt{Re}(\tau)=\pm 1/2$ leads to nontrivial CP violating phases. As regards the description of charged lepton masses, the model $\mathcal{L}_{22}$ is superior to the others, since the best fit value of the input parameter $\beta/\alpha=-1.4038$ is of order one. Moreover, the model $\mathcal{L}_{22}$ predicts the neutrino mass sum $\sum^3_{i=1} m_i=120.9755~\text{meV}$ and the Dirac CP phase $\delta_{CP}$ around $1.5\pi$. These predictions can be tested in future neutrino experiments.

\begin{figure}[ht!]
\centering
\includegraphics[width=6.5in]{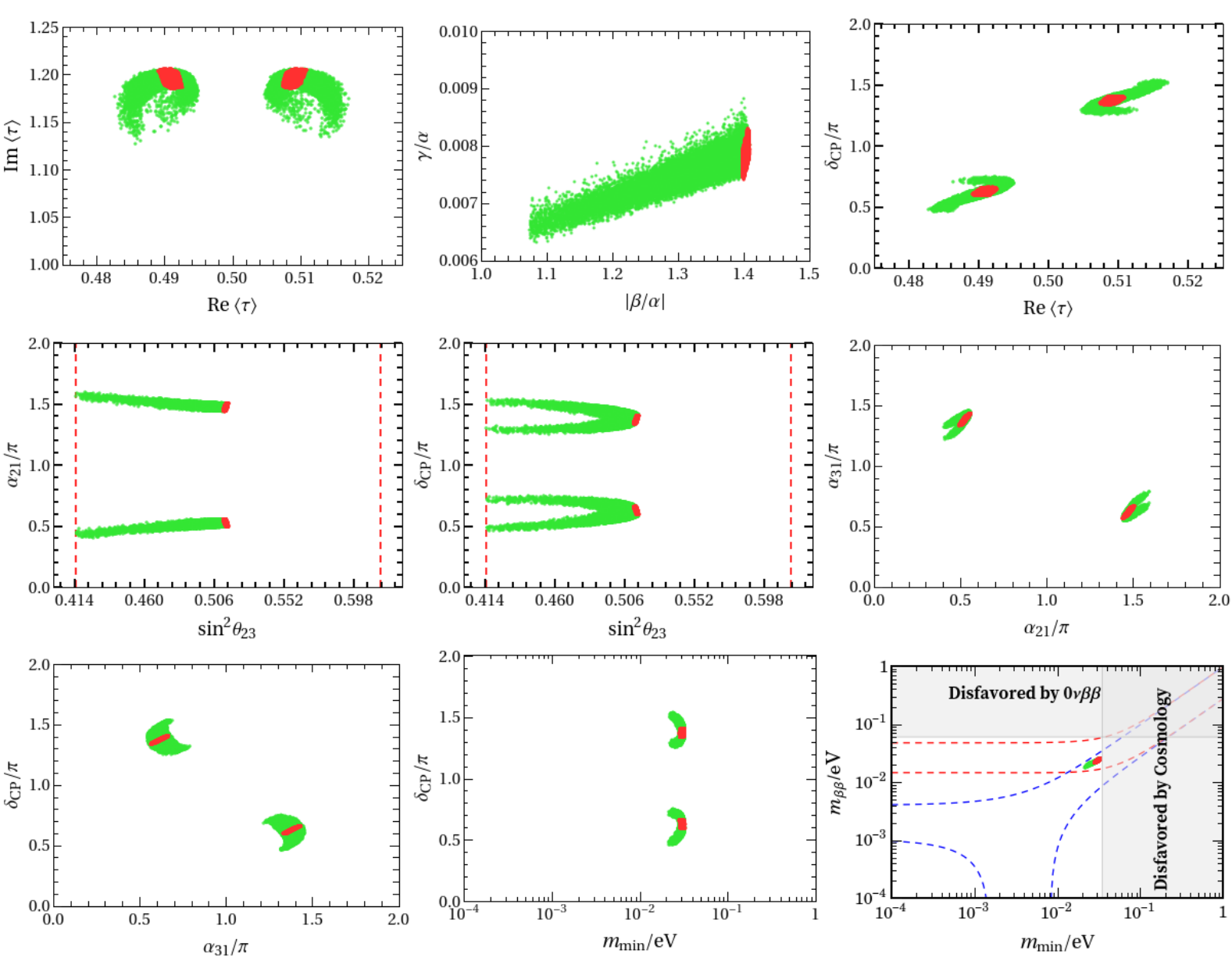}
\caption{The predicted correlations among the input free parameters, neutrino mixing angles and CP violating phases in the lepton model $\mathcal{L}_{22}$ with (red) and without (green) gCP symmetry.}
\label{fig:modelL22}
\end{figure}

In the following, we take the model $\mathcal{L}_{22}$ as an example for illustration. The charged lepton mass term depends on three couplings $\alpha$, $\beta$ and $\gamma$ while the neutrino superpotential depends on two coupling constants $g_1$ and $g_2$ besides the flavor scale $\Lambda$. Moreover, $\alpha$, $\gamma$ and $g_1$ can be taken real by field redefinition while the parameters $\beta$ and $g_2$ are generically complex parameters without gCP and they become real once gCP is imposed. We use the parameter scan tool \texttt{MultiNest}~\cite{Feroz:2007kg,Feroz:2008xx} to efficiently explore the parameter space, and we use a $\chi^2$ function defined as usual to serve as a test-statistic for the goodness-of-fit. The charged lepton masses, neutrino mass squared differences and the neutrino mixing angles are required to lie in $3\sigma$ regions. The correlations among input parameters and observables are plotted in figure~\ref{fig:modelL22}, where the green points and red points are for the scenarios without gCP and with gCP respectively. Obviously the allowed regions of the input parameters are reduced considerably after considering gCP, and accordingly the predictions for observables shrink to quite small regions.

\section{\label{sec:quark-lepton}Unified description of all fermions}

As shown in previous section, one can construct predictive lepton models based on the homogeneous finite modular group $A'_5$. In this section, we shall apply $A'_5$ to explain the quark masses and CKM mixing matrix. Analogous to what we have done for charged lepton sector in section~\ref{subsec:charge-lepton}, we can systematically analyze the possible quark models with $A'_5$ modular symmetry. We find that the measured quark masses and CKM mixing matrix can be explained in terms of a few free parameters if quark fields are embedded in doublet and singlet representations of $A'_5$. We have made a systematic classification of quark models for this kind of quark arrangement and generalize the strategy
of numerical analysis for leptons to the quark sector. The values of quark masses and CKM parameters are taken from~\cite{Antusch:2013jca},
\begin{eqnarray}
\nonumber&&m_{u} / m_{c}=(1.9286 \pm 0.6017) \times 10^{-3}\,, \quad m_{c} / m_{t}=(2.8213 \pm 0.1195) \times 10^{-3}\,,\\
\nonumber&&m_{d} / m_{s}=(5.0523 \pm 0.6191) \times 10^{-2}\,, \quad m_{s} / m_{b}=(1.8241 \pm 0.1005) \times 10^{-2}\,, \\
\nonumber&& m_{t}=87.4555 \mathrm{GeV}\,, \quad m_{b}=0.9682 \mathrm{GeV}\,, \quad \delta_{C P}^{q}/\pi=0.3845 \pm 0.0173\,, \\
&&\theta_{12}^{q}=0.22736 \pm 0.00073\,, \quad \theta_{13}^{q}=0.00349 \pm 0.00013\,,\quad \theta_{23}^{q}=0.04015 \pm 0.00064\,,
\end{eqnarray}
which are calculated at the GUT scale with $\tan\beta=10$ and the SUSY breaking scale $M_{\text{SUSY}}=10$~TeV. In the end, we find four models out of 892 possibilities can describe the experimental data of quarks with 11 free parameters including $\texttt{Re}(\tau)$ and $\texttt{Im}(\tau)$.
A systematic classification of the quark models as well as the numerical results are included in the supplementary file~\cite{Yao:2020sup}.

Furthermore, we combine the $25$ viable lepton models collected in table~\ref{tab:lepton_models} with the two candidate quark models to give a unified description of quark and lepton in a single model. The crucial point is whether there exist common values of $\tau$ for which the experimental data of both quarks and leptons can be reproduced. It is remarkable that we really find a quark-lepton unification model: the lepton sector is the model $\mathcal{L}_2$ and the quark sector is $\mathcal{Q}_3$ which can be found in the supplementary material~\cite{Yao:2020sup}. In the model $\mathcal{Q}_3$, the quark fields and the Higgs fields transform under modular symmetry as follows,
\begin{eqnarray}
\nonumber &&\rho_Q=\mathbf{2'}\oplus \mathbf{1}\,,\quad \rho_{U^c}=\rho_{D^c}=\mathbf{2'}\oplus \mathbf{1} \,,\quad \rho_{H_u}=\rho_{H_d}=\mathbf{1}\,, \\
&& k_{Q_3}=k_{Q_D}-5=-k_{U^c_D}-3=-k_{U^c_3}=-k_{D^c_D}-1=-k_{D^c_3}\,,\quad k_{H_u}=k_{H_d}=0\,,
\end{eqnarray}
where we denote $Q_D\equiv (Q_1,~Q_2)^T,\,U^c_D\equiv (U_1^c,~U_2^c)^T,\,D^c_D\equiv (D^c_1,~D^c_2)^T$, and the modular weight $k_Q$ is a general integer. Thus the modular invariant superpotentials of quark sector are given by
\begin{align}
\nonumber&\mathcal{W}_{u}=\alpha_u \left(Y_{\mathbf{3}'}^{(2)} U^{c}_{D} Q_{D} H_{u}  \right)_{\mathbf{1}}+\beta_u \left(Y_{\mathbf{2}'}^{(5)} U^{c}_{3} Q_{D} H_{u}  \right)_{\mathbf{1}}+\gamma_u \left(U^{c}_{3} Q_{3} H_{u} \right)_{\mathbf{1}} \,,\\
&\mathcal{W}_{d}=\alpha_d \left(Y_{\mathbf{1}}^{(4)} D^{c}_{D} Q_{D} H_{d} \right)_{\mathbf{1}}+\beta_d \left(Y_{\mathbf{3}'}^{(4)} D^{c}_{D} Q_{D} H_{d} \right)_{\mathbf{1}}+\gamma_d \left(Y_{\mathbf{2}'}^{(5)} D^{c}_{3} Q_{D} H_{d} \right)_{\mathbf{1}}+\delta_d \left( D^{c}_{3} Q_{3} H_{d}\right)_{\mathbf{1}}\,.
\end{align}
The phases of $\alpha_u$, $\beta_u$, $\gamma_u$, $\alpha_d$ and $\gamma_d$ are unphysical and they can be absorbed by quark fields, nevertheless the phases of $\beta_d$ and $\delta_d$ can not be removed by field redefinition.  The corresponding up and down quark masses matrices can be written as
\begin{align}
&M_{u}=\begin{pmatrix}
 -\sqrt{2} \alpha_u  Y_{\mathbf{3}',3}^{(2)} & \alpha_u  Y_{\mathbf{3}',1}^{(2)} & 0 \\
 \alpha_u  Y_{\mathbf{3}',1}^{(2)} & \sqrt{2} \alpha_u  Y_{\mathbf{3}',2}^{(2)} & 0 \\
 -\beta_u  Y_{\mathbf{2}',2}^{(5)} & \beta_u  Y_{\mathbf{2}',1}^{(5)} & \gamma_u  \\
\end{pmatrix}v_{u}\,,~ M_{d}=\begin{pmatrix}
 -\sqrt{2} \beta_d  Y_{\mathbf{3}',3}^{(4)} & \beta_d  Y_{\mathbf{3}',1}^{(4)}-\alpha_d  Y_{\mathbf{1},1}^{(4)} & 0 \\
 \alpha_d  Y_{\mathbf{1},1}^{(4)}+\beta_d  Y_{\mathbf{3}',1}^{(4)} & \sqrt{2} \beta_d  Y_{\mathbf{3}',2}^{(4)} & 0 \\
 -\gamma_d  Y_{\mathbf{2}',2}^{(5)} & \gamma_d  Y_{\mathbf{2}',1}^{(5)} & \delta_d  \\
\end{pmatrix}v_{d}\,.
\end{align}
Notice that both (13) and (23) entries of $M_u$ and $M_d$ vanish exactly. This model uses 15 real dimensionless parameters to describe the mixing angles, CP violation phases and the masses ratios of both quarks and leptons. The 4 overall scale factors $\alpha v_d$, $g_1v_u^2/\Lambda$, $\alpha_u v_u$ and $\alpha_d v_d$ give the absolute masses scale of the charged leptons, neutrinos, up quarks and down quarks. By scanning the parameter space of the complete model, we identify the best fit point of the input parameters,
\begin{eqnarray}
\nonumber&&\langle\tau\rangle=-0.499996+0.894402i\,,~~ \beta/\alpha=24.5400\,,~~ \gamma/\alpha= 654.9320\,,~~ \delta/\alpha= 100.8429 e^{4.3652i}\,,\\
\nonumber&&  g_2/g_1=0.3818 e^{5.5110i}\,,~~ \beta_u/\alpha_u=32.6296\,,~~ \gamma_u/\alpha_u= 11.1981\,,~~\beta_d/\alpha_d=6.3229e^{0.00068i}\,,\\
\nonumber&&\gamma_d/\alpha_d=1.7462\,,~~ \delta_d/\alpha_d=1.9895e^{5.4463i}\,,~~ \alpha v_d=0.04361~\text{MeV}\,,~~  g_1v_u^2/\Lambda=1.0895~\text{meV}\,,\\
&&\alpha_u v_u = 0.03432~\text{GeV}\,,\quad \alpha_d v_d = 0.00243~\text{GeV}\,.
\end{eqnarray}
The corresponding predictions for masses and mixing parameters of quarks and leptons are determined to be
\begin{eqnarray}
\nonumber&& \sin^2\theta^l_{13}=0.02231\,,\quad \sin^2\theta^l_{12}=0.3019\,,\quad \sin^2\theta^l_{23}=0.4570\,,\quad \delta^l_{CP}=1.1671\pi\,,\\
\nonumber&& \alpha_{21}=1.3593\pi\,,\quad \alpha_{31}=0.4762\pi\,,\quad m_e/m_{\mu}=0.00474\,,\quad m_{\mu}/m_{\tau}=0.05857\,,\\
\nonumber&&
m_1=77.2709~\text{meV}\,,\quad m_2=77.7495~\text{meV}\,,\quad m_3=92.1717~\text{meV}\,,\\
\nonumber&& \sum_i m_i = 247.1921~\text{meV}\,,\quad   m_{\beta}=77.7755~\text{meV}\,,\quad m_{\beta\beta}=48.9783~\text{meV}\,,\\
\nonumber&& \theta^q_{13}=0.003498\,,\quad \theta^q_{12}=0.22764\,,\quad \theta^q_{23}=0.04023\,,\quad \delta^q_{CP}=69.1740^\circ \,,\\
&& m_u / m_c=0.00021\,,\quad m_c / m_t=0.00282\,,\quad m_d/m_s=0.03457\,,\quad m_s / m_b=0.01769\,,
\end{eqnarray}
which are in the experimentally preferred $3\sigma$ ranges. It is interesting to note that the common value of $\tau$ is very close to the fixed point $\tau_{ST}=e^{2\pi i/3}$ which preserves a $Z^{ST}_3$ residual symmetry.

\section{\label{sec:rational-weight-MF}Extension to rational weight modular forms at level 5}

The modular weights could be rational numbers, as shown in the string construction~\cite{Ibanez:1992hc,Olguin-Trejo:2017zav,Nilles:2020nnc}. It is interesting to study the fractional weight modular forms from the bottom-up modular invariance approach. To discuss rational weight modular forms, it is convenient to consider the metaplectic cover of the modular group $SL(2,\mathbb{Z})$~\cite{Liu:2020msy}. For the concerned case, we should consider the
5-fold covering of $SL(2,\mathbb{Z})$ and it is denoted as $\widetilde{\Gamma}$ with
\begin{equation}
\widetilde{\Gamma}=\Big\{\widetilde{\gamma}=(\gamma, \phi(\gamma,\tau)) ~|~ \gamma \in \begin{pmatrix}
a & b \\ c & d  \end{pmatrix} \in SL(2,\mathbb{Z}),\quad \phi(\gamma,\tau)^5=(c\tau+d) \Big\}\,.
\end{equation}
The action of $\widetilde{\gamma}$ on $\tau$ is the same as that of $\gamma$, i.e. $\widetilde{\gamma}\tau=\gamma\tau$. The group multiplication of $\widetilde{\Gamma}$ is defined as
\begin{equation}
(\gamma_1,\phi_1(\gamma_1,\tau))(\gamma_2,\phi_2(\gamma_2,\tau))=(\gamma_1\gamma_2,\phi_1(\gamma_1,\gamma_2\tau)\phi_2(\gamma_2,\tau))\,,
\end{equation}
where $\phi_1=\epsilon_1(c\tau+d)^{1/5}$,  $\phi_2=\epsilon_2(c\tau+d)^{1/5}$ and $\epsilon_{1,2}\in \{1, \omega_5, \omega_5^2, \omega_5^3, \omega_5^4\}$. The principal branch of the quintic root is chosen here. Obviously each element $\gamma\in SL(2,\mathbb{Z})$ corresponds to five element $\tilde{\gamma}=(\gamma, \omega^j_5(c\tau+d)^{1/5})$ of the metaplectic group $\widetilde{\Gamma}$ with $\omega_5=e^{2\pi i /5}$ and $j=0, 1, 2, 3, 4$.
The group $\widetilde{\Gamma}$ can be generated by
\begin{equation}
\label{eq:gens_cover}
\widetilde{S}=\left(\begin{pmatrix}
0 & 1 \\ -1 & 0
\end{pmatrix}\,,~ \omega_5^2(-\tau)^{1/5} \right)\,,\quad \widetilde{T}=\left(\begin{pmatrix}
1 & 1 \\ 0 & 1
\end{pmatrix}\,,1 \right)\,,
\end{equation}
where
$(-\tau)^{1/5}$ denotes the principal branch of quintic root, possessing positive real part. It follows that
\begin{equation}
\widetilde{S}^2=\widetilde{R}\,,\quad \widetilde{R}^{10}=1\,,\quad (\widetilde{S}\widetilde{T})^3=1\,,
\end{equation}
with
\begin{equation}
\widetilde{R}=\left(\begin{pmatrix}
-1 & 0 \\ 0 & -1
\end{pmatrix}\,,-\omega_5 \right)\,.
\end{equation}
Therefore the elements $\widetilde{S}$ and $\widetilde{S}\widetilde{T}$ are of orders 20 and 3 respectively while $\widetilde{T}$ is of infinite order. Obviously the element $\widetilde{R}$ generates the center of $\widetilde{\Gamma}_5$.
The metaplectic principal congruence subgroup at level 5 is defined as
\begin{equation}
\widetilde{\Gamma}(5)=\Big\{\widetilde{h}=(h, v_5(h)J_{1/5}(h,\tau)) \Big| h \in \Gamma(5)  \Big\}\,,
\end{equation}
where $J_{1/5}(h,\tau)=(c\tau+d)^{1/5}$, $v_5(h)$ is the multiplier of the weight $1/5$ modular forms\footnote{The weight $1/5$ modular forms for $\Gamma(5)$ are holomorphic functions satisfying $f(\gamma\tau)=v_{5}(\gamma)(c\tau+d)^{1/5} f(\tau) $ for $\gamma \in \Gamma(5)$.} for $\Gamma(5)$, and the identity $v^5_5(h)=1$ is fulfilled for all $h\in\Gamma(5)$. The explicit expression of $v_5(h)$ has been given in~\cite{ibukiyama2000modular,Liu:2020msy}, it is too lengthy to present here. It is obvious that $\widetilde{\Gamma}(5)$ is isomorphic to $\Gamma(5)$, and it is a normal subgroup of $\widetilde{\Gamma}$. In particular, the element $\widetilde{T}^5$ belongs to $\widetilde{\Gamma}(5)$. Taking the group quotient, we can obtain the finite metaplectic group $\widetilde{\Gamma}_5=$
$\widetilde{\Gamma}/\Gamma(5) \cong A'_5 \times Z_5$ which can be obtained by imposing another condition $\widetilde{T}^5=1$. Thus the generator relations of $\widetilde{\Gamma}_5$ are
\begin{equation}
\widetilde{S}^2=\widetilde{R}\,,\quad \widetilde{R}^{10}=1\,,\quad (\widetilde{S}\widetilde{T})^3=\widetilde{T}^5=1\,, \quad \widetilde{R}\widetilde{T}=\widetilde{T}\widetilde{R}\,.
\end{equation}
The group ID of $\widetilde{\Gamma}_5$ in \texttt{GAP} system is $[600,54]$, and it is isomorphic to the direct product of $A'_5$ and $Z_5$, i.e., $\widetilde{\Gamma}_5\cong A'_5 \times Z_5$. Hence it is more convenient to choose another set of generators $S$, $T$ and $U$ which obey the relations
\begin{equation}
S^2=R, \quad (ST)^3=T^5=R^2=1,\quad U^5=1, \quad US=SU, \quad UT=TU\,.
\end{equation}
Notice that the generators $S$ and $T$ generate a $A'_5$ subgroup, and $U$ generates a $Z_5$ subgroup.
The generators $S$, $T$ and $U$ are related to $\widetilde{S}$, $\widetilde{T}$ as follows
\begin{eqnarray}
\nonumber&& S =  \widetilde{R}^{2}\widetilde{S}=\left(\begin{pmatrix}
0 & 1 \\ -1 & 0
\end{pmatrix}\,,~ \omega_5^4(-\tau)^{1/5} \right)\,,\\
\nonumber&&T=\widetilde{R}^{-2}\widetilde{T}=\left(\begin{pmatrix}
1 & 1 \\ 0 & 1
\end{pmatrix}\,,\omega^3_5 \right)\,,\\
&&U=\widetilde{S}^{12}=\left(\begin{pmatrix}
1 & 0 \\ 0 & 1
\end{pmatrix}\,,\omega_5 \right) \,,
\end{eqnarray}
and vice versa
\begin{equation}
\widetilde{S}= U^{-2} S, \quad \widetilde{T}= U^2 T, \quad \widetilde{R}=UR \,.
\end{equation}
The irreducible representations of $\widetilde{\Gamma}_5$ can be obtained from the tensor products of the irreducible representations of $A'_5$ and $Z^U_5$ where the superscript $U$ denotes the generator of the $Z_5$ subgroup.
The irreducible representations of $A'_5$ are summarized in table~\ref{tab:rep-matrices}, and abelian subgroup $Z^U_5$ has five one-dimensional irreducible representations and the generator $U$ is represented by $\omega^j_5$ with $j=0, 1, 2, 3, 4$. We shall denote the irreducible representation of $\widetilde{\Gamma}_5$ as $\mathbf{r}^j$ which is the tensor product of the $A'_5$ irreducible representation $\mathbf{r}$ with the $Z^U_5$ representation $\omega^j_5$. The generators $S$, $T$ and $U$ are represented by
\begin{align}
\mathbf{r}^j~: ~\rho_{\mathbf{r}^j}(S)=\rho_{\mathbf{r}}(S)\,,\quad \rho_{\mathbf{r}^j}(T)=\rho_{\mathbf{r}}(T)\,, \quad \rho_{\mathbf{r}^j}(U)=\omega_5^j\,,
\end{align}
where $\rho_{\mathbf{r}}$ is the representation matrix of the $A'_5$ irreducible representation $\mathbf{r}$ as shown in Appendix~\ref{sec:A5_group_DC}. The Kronecker products of $\widetilde{\Gamma}_5$ can be easily obtained from those of $A'_5$ given in Eq.~\eqref{eq:kronecker-Gamma'5} by including an extra index $j$ for each $A'_5$ representation. For instance, we have $\mathbf{2}^3 \otimes \mathbf{3}^4 = \mathbf{2}^2 \oplus \mathbf{4}'^2$, and the corresponding  Clebsch-Gordan coefficients coefficients remains the same as those of $A'_5$, which are listed in the Appendix~\ref{sec:A5_group_DC}.

It has been shown that the modular forms of weight $k/5$ ($k:$ non-negative integers) can be constructed for the principal congruence subgroup $\Gamma(5)$~\cite{ibukiyama2000modular}, and a multiplier $v_5(\gamma)$ is necessary so that $v(\gamma)(c\tau+d)^{k/5}$ is the correct automorphy factor satisfying the cocycle relation.
The graded ring of modular forms of weight $k/5$ at level 5 can be generated by two algebraically independent weight $1/5$ modular forms ~\cite{ibukiyama2000modular}:
\begin{equation}
\mathcal{M}(\Gamma(5))=\mathbb{C}[f_1(\tau),~f_2(\tau)]\,,
\end{equation}
with
\begin{equation}
\label{eq:basis_1/5MF}
f_1(\tau)=\dfrac{\theta_{(\frac{1}{10},\frac{1}{2})}(5\tau)}{\eta(\tau)^{3/5}}\,,~~~~
f_2(\tau)=\dfrac{\theta_{(\frac{3}{10},\frac{1}{2})}(5\tau)}{\eta(\tau)^{3/5}}\,.
\end{equation}
Notice that the theta constant is defined as~\cite{ibukiyama2000modular}
\begin{equation}
\theta_{(m',m'')}(\tau)=\sum_{m\in \mathbb{Z}} e^{2\pi i [\frac{1}{2}(m+m')^2\tau + (m+m')m'']}\,,
\end{equation}
and $\eta(\tau)$ is the well-known Dedekind eta function given in Eq.~\eqref{eq:eta_function}.
Hence the linear space of the modular forms of weight $k/5$ and level 5 has dimension $k+1$, and the linear independent basis vectors are $k+1$ polynomials of $f_1$ and $f_2$ with degree $k$. Under the action of the generators $S$ and $T$ of the modular group, $f_1$ and $f_2$ transform as~\cite{ibukiyama2000modular}
\begin{eqnarray}
\nonumber&&f_1(\tau) \xrightarrow{S}f_1(-\frac{1}{\tau})=\omega_5^2 (-\tau)^{1/5} \frac{1}{\sqrt{5}}\left[\left(\omega_5^3 - \omega_5\right)f_1(\tau)+\left(\omega_5^2 - \omega_5\right)f_2(\tau)\right],\,,\\
\nonumber&&f_2(\tau) \xrightarrow{S}f_2(-\frac{1}{\tau})=\omega_5^2 (-\tau)^{1/5} \frac{1}{\sqrt{5}}\left[\left(\omega_5^3 - \omega_5^2\right)f_1(\tau)+\left( \omega_5 - \omega_5^3\right)f_2(\tau)\right]\,,\\
\label{eq:transm-f12}&&f_1(\tau) \xrightarrow{T}f_1(\tau+1)=f_1(\tau),~~~~f_2(\tau) \xrightarrow{T}f_2(\tau+1)= \omega_5 f_2(\tau)\,.
\end{eqnarray}
Analogous to the half-integral weight modular forms~\cite{Liu:2020msy}, the modular forms of weight $r=k/5$ and level 5 can be arranged into different irreducible representations $Y^{(r)}_{\mathbf{r}^j}(\tau)$ of the finite metaplectic group $\widetilde{\Gamma}_5$. The modular multiplet $Y^{(r)}_{\mathbf{r}^j}(\tau)$ transform under $\widetilde{\Gamma}_5=\widetilde{\Gamma}/\widetilde{\Gamma}(5)$ as
\begin{equation}
\label{eq:decomp-MF}Y^{(r)}_{\mathbf{r}^j}(\widetilde{\gamma}\tau)=\phi^{5r}(\gamma,\tau)\rho_{\mathbf{r}^j}(\widetilde{\gamma})Y^{(r)}_{\mathbf{r}^j}(\tau)\,,
\end{equation}
where $\widetilde{\gamma}$ stands for a representative element of $\widetilde{\Gamma}_5$.
Applying Eq.~\eqref{eq:decomp-MF} to the element $\widetilde{\gamma}=U$, we find that the constraint $5r+j=0\,(\texttt{mod}~5)$ should be fulfilled.
In the group representation basis chosen above, the two weight $1/5$ modular forms can be arranged into a $\widetilde{\Gamma}_5$ doublet $\mathbf{2}^4$ as follows,
\begin{equation}
Y^{(\frac{1}{5})}_{\mathbf{2}^4}(\tau) \equiv \begin{pmatrix}
F_1(\tau) \\ F_2(\tau)
\end{pmatrix},~~~F_1(\tau)= e^{-\frac{\pi i}{10}}f_1(\tau),~~~F_2(\tau)= e^{-\frac{3\pi i}{10}}f_2(\tau)\,.
\end{equation}
The $q$-series expressions of $F_1(\tau)$ and $F_2(\tau)$ are given by
\begin{equation}
\label{eq:q-series_F1F2} F_1(\tau)\,\eta(\tau)^{3/5}=q^{1/40}\sum_{m\in \mathbb{Z}}(-1)^m q^{(5m^2+m)/2},~~~~F_2(\tau)\,\eta(\tau)^{3/5}=q^{9/40}\sum_{m\in \mathbb{Z}}(-1)^m q^{(5m^2+3m)/2}\,,
\end{equation}
with $q=e^{2\pi i \tau}$. Using the transformation rules in Eq.~\eqref{eq:transm-f12}, we find that
$Y^{(\frac{1}{5})}_{\mathbf{2}^4}(\tau)$ really transforms in the two-dimensional representation $\mathbf{2}^4$ of $\widetilde{\Gamma}_5$,
\begin{eqnarray}
\nonumber&&Y^{(\frac{1}{5})}_{\mathbf{2}^4}(\tau)\xrightarrow{S}Y^{(\frac{1}{5})}_{\mathbf{2}^4}(-\frac{1}{\tau})=\omega_5^4(-\tau)^{1/5}\,
\rho_{\mathbf{2}^4}(S)Y^{(\frac{1}{5})}_{\mathbf{2}^4}(\tau)\,,\\
&&Y^{(\frac{1}{5})}_{\mathbf{2}^4}(\tau)\xrightarrow{T}Y^{(\frac{1}{5})}_{\mathbf{2}^4}(\tau+1)=\omega^3_5 \, \rho_{\mathbf{2}^4}(T)Y^{(\frac{1}{5})}_{\mathbf{2}^4}(\tau)\,,
\end{eqnarray}
and it is invariant under the generator $U$.
Using the Clebsch-Gordan coefficients of $\widetilde{\Gamma}_5$, we can construct the higher rational weight modular forms through the tensor products of
$Y^{(\frac{1}{5})}_{\mathbf{2}^4}(\tau)$. At weight $r=2/5$, we have
\begin{eqnarray}
&& Y^{(\frac{2}{5})}_{\mathbf{3}^3}= -\frac{1}{\sqrt{2}} \left(Y^{(\frac{1}{5})}_{\mathbf{2}^4}Y^{(\frac{1}{5})}_{\mathbf{2}^4}\right)_{\mathbf{3}^3}=
\begin{pmatrix} \sqrt{2}F_1 F_2 \\   -F_2^2 \\ F_1^2 \end{pmatrix}\,,
\end{eqnarray}
where an overall constant $-\frac{1}{\sqrt{2}}$ is multiplied to make the resulting expression relatively simple. Note that another contraction $\left(Y^{(\frac{1}{5})}_{\mathbf{2}^4}Y^{(\frac{1}{5})}_{\mathbf{2}^4}\right)_{\mathbf{1}^3} =0$ because of the antisymmetric Clebsch-Gordan coefficients.
In exactly the same fashion, other rational weight modular forms can be obtained, and we present linearly independent rational weight modular forms up to weight $1$:
\begin{eqnarray}
&& r=3/5: ~~~~~~~~~
Y^{(\frac{3}{5})}_{\mathbf{4}'^2}= -\frac{1}{\sqrt{3}}\left(Y^{(\frac{1}{5})}_{\mathbf{2}^4}Y^{(\frac{2}{5})}_{\mathbf{3}^3}\right)_{\mathbf{4}'^2}=
\begin{pmatrix} F_1^3 \\
 -\sqrt{3} F_1^2 F_2 \\
 \sqrt{3} F_1 F_2^2 \\
 F_2^3 \\ \end{pmatrix}, \\[0.1in]
&& r=4/5: ~~~~~~~~~
Y^{(\frac{4}{5})}_{\mathbf{5}^1}= \frac{1}{2}\left(Y^{(\frac{1}{5})}_{\mathbf{2}^4}Y^{(\frac{3}{5})}_{\mathbf{4}'^2}\right)_{\mathbf{5}^1}=
\begin{pmatrix} \sqrt{6} F_1^2 F_2^2  \\  2 F_1 F_2^3 \\ F_2^4 \\ F_1^4 \\-2 F_1^3 F_2 \end{pmatrix}, \\[0.1in]
&& r=1: ~~~~~~~~~
Y^{(1)}_{\mathbf{6}^0}= -\left(Y^{(\frac{1}{5})}_{\mathbf{2}^4}Y^{(\frac{4}{5})}_{\mathbf{5}^1}\right)_{\mathbf{6}^0}=
\begin{pmatrix} F_1^5 + 2 F_2^5  \\  2 F_1^5 - F_2^5 \\ 5 F_1^4 F_2 \\ 5\sqrt{2}F_1^3 F_2^2 \\-5\sqrt{2} F_1^2 F_2^3 \\ 5F_1 F_2^4 \end{pmatrix}\,. \label{eq:wt1_MF}
\end{eqnarray}
From Eqs.~(\ref{eq:q-series_F1F2}, \ref{eq:eta_function}), we can read out the $q$-expansion of the $Y^{(1)}_{\mathbf{6}^0}$ as follow,
\begin{equation}
\label{eq:wt1-MF-1/5}Y^{(1)}_{\mathbf{6}^0}(\tau)=\begin{pmatrix}
1+5q+10q^3-5q^4+5q^5+ \dots \\
2+5 q+10 q^2+5 q^4+5 q^5+\dots \\
5q^{1/5}\left(1+2q+2q^2+q^3+2q^4+2q^5+\dots \right)\\
5\sqrt{2}q^{2/5}\left(1+q+q^2+q^3+2q^4+q^6+\dots \right)\\
-5\sqrt{2}q^{3/5} \left(1+q^2+q^3+q^4-q^5+\dots \right)\\
5q^{4/5} \left(1-q+2q^2+2q^6-2q^7+\dots \right)
\end{pmatrix}\,,
\end{equation}
which is exactly identical with the $q-$expansion expressions of weight 1 modular form obtained from the Dedekind eta function and Klein forms in Eq.~\eqref{eq:q_series_wt1}. Furthermore, comparing Eq.~\eqref{eq:wt1_MF} with Eq.~\eqref{eq:wt1_MF_Kelin}, we find that theta constants and the Klein forms are closely related as follows
\begin{equation}
\label{eq:identity}\theta_{(\frac{1}{10},\frac{1}{2})}(\tau)=e^{\frac{\pi i}{10}}\eta^3(\tau) \mathfrak{k}_{\frac{2}{5}, 0}(\tau) \,,~~~\theta_{(\frac{3}{10},\frac{1}{2})}(\tau)=e^{\frac{3\pi i}{10}}\eta^3(\tau) \mathfrak{k}_{\frac{1}{5}, 0}(\tau)\,.
\end{equation}
These two identities can be proved by using the famous Jacobi triple product identity as shown
in Appendix~\ref{sec:proof}.
As a consequence, the basis vectors $e_{1,2,3,4,5,6}(\tau)$ of weight 1 modular form at level 5 are
polynomials of degree 5 in $F_1(\tau)$ and $F_2(\tau)$,
\begin{eqnarray}
\nonumber&&e_1(\tau)=F_1^5(\tau)\,,~~~~ e_2(\tau)=F_1^4(\tau)F_2(\tau)\,,~~~~ e_3(\tau)=F_1^3(\tau)F_2^2(\tau)\,,~~\\
&& e_4(\tau)=F_1^2(\tau)F_2^3(\tau)\,,~~~~ e_5(\tau)=F_1(\tau)F_2^4(\tau)\,,~~~~ e_6(\tau)=F_2^5(\tau)\,.
\end{eqnarray}
Because the two weight $1/5$ modular forms $F_1(\tau)$ and $F_2(\tau)$ are algebraically independent, the use of $F_1(\tau)$ and $F_2(\tau)$ considerably simplifies the process of finding linearly independent modular multiplets, and it is not necessary to examine the constraints relating redundant higher weight multiplets when constructing modular forms in terms of $F_1(\tau)$ and $F_2(\tau)$. For instance, it is easy to check that the constraints Eqs.~(\ref{eq:constraint1_wt1MF}, \ref{eq:constraint2_wt1MF}) for the weight 1 modular forms are now trivially fulfilled.

\subsection{\label{sec:rational_model} A lepton model with rational weight modular forms at level 5 }

All $A'_5$ modular models, in particular the phenomenologically viable models discussed in sections~\ref{sec:lepton-models} and~\ref{sec:quark-lepton}, can be reproduced from the $\widetilde{\Gamma}_5$ modular symmetry if the modular weights of both matter fields and modular forms are integral. Obviously the metaplectic covering group $\widetilde{\Gamma}_5$ has more richer structure of modular forms which can be utilized to explain the flavor structure of  lepton and quark in the bottom-up modular invariance approach. In a similar fashion as we have done for the $A'_5$ modular symmetry, one can systematically classify the modular invariant lepton and quark models based on $\widetilde{\Gamma}_5$. As an example, we present a benchmark lepton model which involves rational weight modular forms at level 5.

\begin{figure}[ht!]
\centering
\includegraphics[width=6.5in]{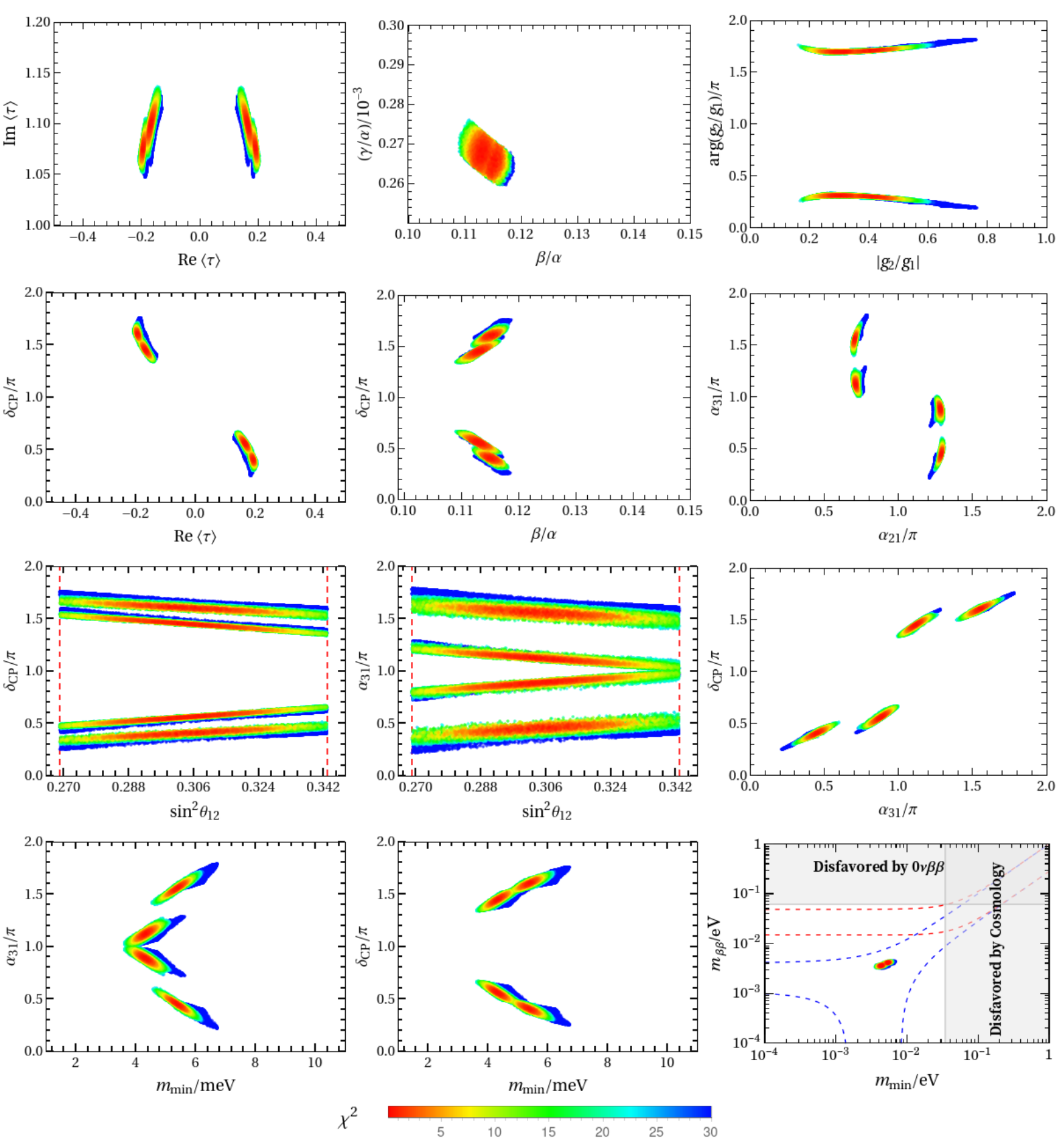}
\caption{The predicted correlations among the input parameters, neutrino mixing angles and CP violating phases in the lepton model with rational weight modular forms at level 5.
}
\label{fig:model1}
\end{figure}

In this model, the neutrino masses arises from the type-I seesaw mechanism. The three generations of left-handed lepton $L$ are assumed to transform as a triplet $\mathbf{3}^3$ under $\widetilde{\Gamma}_5$, the right-handed charged leptons $E^c_{1,2,3}$ are assigned to be $\widetilde{\Gamma}_5$ singlets $\mathbf{1}^2$, $\mathbf{1}^4$ and $\mathbf{1}^4$ respectively. Three right-handed neutrinos are introduced, and they transform as a triplet $\mathbf{3'}^0$. We summarize the modular transformation properties and weights assignment for the lepton and Higgs superfields as follow\footnote{Note that in order to ensure the consistency of the theory~\cite{Liu:2020msy}, the original modular transformation Eq.~\eqref{eq:modularTrs_Phi} should be adjusted to metaplectic modular transformation: $\tau\to \widetilde{\gamma}\tau,\quad \Phi_I \to \phi^{-5k_I}(\gamma,\tau)\rho_I(\widetilde{\gamma})\Phi_I$, where $\phi(\gamma,\tau)^5=c\tau+d$ , $\widetilde{\gamma}=(\gamma,\phi(\gamma,\tau))\in \widetilde{\Gamma}$ and $-k_I$ is the modular weight of superfield $\Phi_I$.},
\begin{eqnarray}
\nonumber &&\rho_L=\mathbf{3}^3\,,\quad \rho_{E^c_{1,2,3}}=\mathbf{1}^2 \oplus \mathbf{1}^4 \oplus \mathbf{1}^4\,, \quad \rho_{N^c}=\mathbf{3'}^0\,,\quad \rho_{H_u}=\rho_{H_d}=\mathbf{1}^0\,, \\
&& k_{L}=\frac{3}{5}\,,\quad k_{E^c_{1}}=\frac{7}{5}\,,\quad k_{E^c_{2}}=-\frac{1}{5}\,,\quad k_{E^c_{3}}=\frac{9}{5}\,, \quad k_{N^c}=1\,,\quad k_{H_u}=k_{H_d}=0\,.
\end{eqnarray}
Thus we can read out the following modular invariant superpotential for charged leptons and neutrinos,
\begin{align}
\nonumber
&\mathcal{W}_e = \alpha\left(Y^{(2)}_{\mathbf{3}^0} E_1^c L H_d \right)_{\mathbf{1}^0}+\beta\left(Y^{(\frac{2}{5})}_{\mathbf{3}^3} E_2^c L H_d \right)_{\mathbf{1}^0} + \gamma \left(Y^{(\frac{12}{5})}_{\mathbf{3}^3} E_3^c L H_d \right)_{\mathbf{1}^0}\,,\\
&\mathcal{W}_{\nu} = g_1\left(Y^{(\frac{8}{5})}_{\mathbf{4}^2} N^c L H_u \right)_{\mathbf{1}^0}+g_2\left(Y^{(\frac{8}{5})}_{\mathbf{5}^2} N^c L H_u \right)_{\mathbf{1}^0} + \Lambda \left(Y^{(2)}_{\mathbf{5}^0} N^c N^c\right)_{\mathbf{1}^0}\,.
\end{align}
Hence the charged lepton and neutrino mass matrices are given by
\begin{align}
\nonumber
&M_e=\begin{pmatrix}
\alpha Y^{(2)}_{\mathbf{3}^0, 1} ~&~ \alpha Y^{(2)}_{\mathbf{3}^0, 3} ~&~ \alpha Y^{(2)}_{\mathbf{3}^0, 2} \\[0.1in]
\beta Y^{(\frac{2}{5})}_{\mathbf{3}^3, 1} ~&~ \beta Y^{(\frac{2}{5})}_{\mathbf{3}^3, 3} ~&~ \beta Y^{(\frac{2}{5})}_{\mathbf{3}^3, 2} \\[0.1in]
\gamma Y^{(\frac{12}{5})}_{\mathbf{3}^3,1} ~&~ \gamma Y^{(\frac{12}{5})}_{\mathbf{3}^3,3} ~&~ \gamma Y^{(\frac{12}{5})}_{\mathbf{3}^3,2}
\end{pmatrix} v_d,\quad M_N= \Lambda \begin{pmatrix}
2 Y^{(2)}_{\mathbf{5}^0, 1} ~& -\sqrt{3}Y^{(2)}_{\mathbf{5}^0, 4} ~& -\sqrt{3}Y^{(2)}_{\mathbf{5}^0, 3} \\[0.1in]
-\sqrt{3}Y^{(2)}_{\mathbf{5}^0, 4} ~& \sqrt{6}Y^{(2)}_{\mathbf{5}^0, 2} ~& -Y^{(2)}_{\mathbf{5}^0, 1} \\[0.1in]
-\sqrt{3}Y^{(2)}_{\mathbf{5}^0, 3} ~& -Y^{(2)}_{\mathbf{5}^0, 1} ~& \sqrt{6}Y^{(2)}_{\mathbf{5}^0, 5}
\end{pmatrix} \,,\\
&M_D=\begin{pmatrix}
\sqrt{3}g_2 Y^{(\frac{8}{5})}_{\mathbf{5}^2, 1} ~&~ \sqrt{2}g_1 Y^{(\frac{8}{5})}_{\mathbf{4}^2, 4} + g_2  Y^{(\frac{8}{5})}_{\mathbf{5}^2, 5} ~&~ \sqrt{2}g_1 Y^{(\frac{8}{5})}_{\mathbf{4}^2, 1} + g_2  Y^{(\frac{8}{5})}_{\mathbf{5}^2, 2}  \\[0.1in]
-\sqrt{2}g_1 Y^{(\frac{8}{5})}_{\mathbf{4}^2, 3} + g_2  Y^{(\frac{8}{5})}_{\mathbf{5}^2, 4} ~&~ -g_1 Y^{(\frac{8}{5})}_{\mathbf{4}^2, 2} -\sqrt{2} g_2  Y^{(\frac{8}{5})}_{\mathbf{5}^2, 3} ~&~ g_1 Y^{(\frac{8}{5})}_{\mathbf{4}^2, 4} -\sqrt{2} g_2  Y^{(\frac{8}{5})}_{\mathbf{5}^2, 5} \,,\\[0.1in]
-\sqrt{2}g_1 Y^{(\frac{8}{5})}_{\mathbf{4}^2, 2} + g_2  Y^{(\frac{8}{5})}_{\mathbf{5}^2, 3} ~&~ g_1 Y^{(\frac{8}{5})}_{\mathbf{4}^2, 1} -\sqrt{2} g_2  Y^{(\frac{8}{5})}_{\mathbf{5}^2, 2} ~&~ -g_1 Y^{(\frac{8}{5})}_{\mathbf{4}^2, 3} -\sqrt{2} g_2  Y^{(\frac{8}{5})}_{\mathbf{5}^2, 4}
\end{pmatrix}v_u \,.
\end{align}
The parameters $\alpha$, $\beta$, $\gamma$ and $g_1$ can be taken real by redefining the unphysical phases of fields. Thus we have only two complex parameters $\tau$ and $g_2$ left. The charged lepton masses can be reproduced by adjusting the parameters $\alpha$, $\beta$ and $\gamma$. The light neutrino mass matrix $m_{\nu}=-M^{T}_DM^{-1}_NM_D$ depends on a single complex parameter $g_2/g_1$ and an overall scale $g^2_1v^2_u/\Lambda$ besides the complex modulus $\tau$. Hence this model effectively depends on 8 free real parameters at low energy. Numerically scanning over the parameter space, we find a good agreement between the model predictions and experimental data
can be achieved for the following values of the input parameters
\begin{equation}
\begin{gathered}
\langle\tau\rangle=-0.16639+1.09859i\,,\qquad \beta/\alpha=0.11310\,,\qquad \gamma/\alpha= 0.000268\,,\\
g_2/g_1=0.17295+0.24924i\,,~~\alpha v_d=639.62166~\text{MeV},~~\frac{g_1v_u^2}{\Lambda}=35.71373~\text{meV}\,.
\end{gathered}
\end{equation}
Accordingly the predictions for the lepton mixing parameters and neutrino masses are given by
\begin{eqnarray}
\nonumber&&\sin^2 \theta_{12}=0.30398\,,~~\sin^2 \theta_{13}=0.02219\,,~~ \sin^2 \theta_{23}=0.57297\,,~~ \delta_{CP}=1.4454\pi\,,\\
\nonumber&&\alpha_{21}=0.7135\pi\,,~~ \alpha_{31}=1.1178\pi\,,~~~m_e/m_\mu =0.00473 \,, ~~~m_\mu/m_\tau=0.05857\,,\\
\nonumber&& m_{1}=4.3331~\text{meV}\,, ~~ m_{2}=9.6424~\text{meV}\,,~~m_{3}=50.3562~\text{meV}\,,\\
&&m_\beta=9.8328~\text{meV} \,,~~~~ m_{\beta\beta}=3.5926~\text{meV}\,,
\end{eqnarray}
which are compatible with the experimental data at $1\sigma$ level~\cite{Esteban:2020cvm}.
The mass ordering is normal, and the
Planck bound $\sum_im_{i}<120$ meV~\cite{Aghanim:2018eyx} is fulfilled. The effective masses $m_{\beta}$ in beta decay and $m_{\beta\beta}$ in neutrinoless double beta decay are very tiny and consequently they are outside the reach of the next generation experiments. We comprehensively scan over the parameter space of this model, and all  lepton masses and mixing angles are required to lie in $3\sigma$ regions~\cite{Esteban:2020cvm}. Some interesting correlations between the input parameters and observables are shown in figure~\ref{fig:model1}. It can be seen that the three CP violating phases $\delta_{CP}$, $\alpha_{21}$ and $\alpha_{31}$ are strongly correlated with each other. Moreover, the allowed range of the lightest neutrino mass $m_1$ is [3.61meV, 6.75meV] in this model.

\section{\label{sec:conclusion} Conclusion}

The inhomogeneous finite modular groups $\Gamma_N$ together with even weight modular forms have been extensively studied
in bottom-up flavor model building, while the homogeneous finite modular groups $\Gamma'_N$ are less discussed. Generally $\Gamma_N$ is isomorphic to the quotient of $\Gamma'_N$ modulo $Z_2$ subgroup generated by $R=S^2$, or we say $\Gamma'_N$ is the double covering of $\Gamma_N$. The transformation of the general integral weight modular forms can be described by unitary representations of $\Gamma'_N$. In the present work, we perform a comprehensive study of the homogeneous finite modular group $\Gamma'_5\cong A'_5$ at level 5. The vacuum alignment problem of traditional $A'_5$ flavor symmetry models~\cite{Everett:2010rd,Hashimoto:2011tn,Chen:2011dn}
would be simplified considerably, and the flavor symmetry is spontaneously broken by the vacuum expectation value of $\tau$.

The integral weight modular forms of level 5 can be written as a polynomial $\sum_{i=0}^{5k} c_i F^i_{1}(\tau)F^{5k-i}_2(\tau)$, where $F_1(\tau)$ and $F_2(\tau)$ are two algebraically independent weight $1/5$ modular forms and they are expressed in terms of theta constants and Dedekind eta function. Hence the linear space of modular forms of weight $k$ and level 5 has dimension $5k+1$. One can also construct level 5 modular forms with Klein form and Dedekind eta function. We find that both construction methods give the same results by using the identities Eq.~\eqref{eq:identity} between theta constants and Klein form. There are six linearly independent lowest weight 1 and level 5 modular forms, and they can be arranged into a six dimensional representation of $A'_5$. Moreover, we explicitly give the expressions of $A'_5$ multiplets of modular forms up to weight 6.

It is known that the generalized CP symmetry can be consistently combined with modular symmetry, and the complex modulus $\tau$ is required to transform as $\tau\rightarrow -\tau^{*}$ up to a modular transformation under the action of generalized CP. Accordingly the modular generators $S$ and $T$ are mapped to their inverse $S^{-1}$ and $T^{-1}$ respectively under CP, and we find that it is a class-inverting automorphism of $A'_5$ so that the corresponding CP symmetry is indeed physically well defined in the context of $A'_5$ modular symmetry. In our representation basis, both $S$ and $T$ are represented by symmetric matrices in all irreducible representations of $A'_5$, therefore the CP transformation for $\tau\rightarrow-\tau^{*}$ is the canonical one $X_{\mathbf{r}}=1$. Moreover all CG coefficients are real in our basis, consequently invariance under generalized CP symmetry requires all coupling constants real.
As a consequence, the vacuum expectation value of $\tau$ would be the unique source of both modular and CP symmetry breaking if generalized CP symmetry is imposed on the $A'_5$ modular models.

According the representation assignments for the lepton fields, we systematically classify the lepton models based on $A'_5$ modular symmetry. The left-handed lepton doublets are assigned to a triplet of $A'_5$, the right-handed charged leptons transform as singlets or the direct sum of a doublet and a singlet under $A'_5$. The neutrino masses are generated by the type I seesaw mechanism, and both scenarios with two and three right-handed neutrinos are considered. We aim to find out all $A'_5$ modular models with minimal number of free parameters in this work. Numerically minimizing the $\chi^2$ function for each model, we find out 15 phenomenologically viable models with 9 real free parameters and 10 viable models with 10 real free parameters including the real and imaginary parts of $\tau$, and the results of fit are listed in tables~\ref{tab:model_Aa}, \ref{tab:model_Ab}, \ref{tab:model_Ac}. For all these 25 models, two more free parameters would be reduced by imposing CP invariance. We find 9 model with 7 free parameters and 10 models with 8 free parameters can accommodate the experimental data in lepton sector, as shown in table~\ref{tab:model_gCP_A} and table~\ref{tab:model_gCP_B}. Moreover, we use the $A'_5$ modular symmetry to understand the quark masses and CKM mixing matrix, and a quark-lepton unification model is presented.

Furthermore, we extend the framework to include rational weight modular forms at level 5, and the modular group $SL(2,\mathbb{Z})$ should be extended to its 5-fold metaplectic covering group. Rational weight modular forms of level 5 transform in representations of the finite metaplectic group $\widetilde{\Gamma}_5\cong A'_5\times Z_5$.
We find that the two weight $1/5$ modular forms $F_1(\tau)$ and $F_2(\tau)$ furnish a doublet $\mathbf{2}^4$ of $\widetilde{\Gamma}_5$ with $Y^{(\frac{1}{5})}_{\mathbf{2}^4}(\tau) =(F_1(\tau), F_2(\tau))^{T}$. The ring of modular forms of level 5 is constructed through the tensor products of $Y^{(\frac{1}{5})}_{\mathbf{2}^4}(\tau)$,
and remarkably it is unnecessary to consider the constraints among high weight modular forms in this way because $F_1(\tau)$ and $F_2(\tau)$ and algebraically independent and they are the minimal set of functions to construct the modular space of level 5. Finally we provide a concrete lepton model based on $\widetilde{\Gamma}_5$, the predictions for lepton masses, mixing angles and CP violating phases are studied numerically, and strong correlations between mixing parameters are found.

At lower levels $N=2, 3, 4$, the homogeneous (inhomogeneous) finite modular groups $\Gamma'_N$ ($\Gamma_N$) only have one-dimensional, two-dimensional and three-dimensional irreducible representations.
Nevertheless, $\Gamma'_5\cong A'_5$ at level 5 have four-dimensional, five-dimensional and six-dimensional irreducible representations besides the singlet, doublet and triplet representations, and the contractions among doublets and triplets can give rise to quartet, quintet and sextet such as $\mathbf{2}\otimes \mathbf{3}=\mathbf{2}\oplus \mathbf{4'}$, $\mathbf{2'}\otimes \mathbf{3'}=\mathbf{2'}\oplus \mathbf{4'}$, $\mathbf{2}\otimes \mathbf{3'}=\mathbf{2'}\otimes \mathbf{3}=\mathbf{6}$, $\mathbf{3}\otimes \mathbf{3}=\mathbf{1_s}\oplus \mathbf{3_a}\oplus \mathbf{5_s}$, $\mathbf{3'}\otimes \mathbf{3'}=\mathbf{1_s}\oplus \mathbf{3'_a}\oplus \mathbf{5_s}$ and $\mathbf{3}\otimes \mathbf{3'}=\mathbf{4}\oplus \mathbf{5}$. Hence level 5 opens up new model building possibilities not available for the levels 2, 3, 4, as shown in the paper. The three right-handed charged leptons are usually assigned to be singlets of the finite modular group such that one can tune the couplings of each charged lepton to easily reproduce the hierarchical charged lepton masses. As shown in table~\ref{tab:lepton_models}, we can see that the right-handed charged leptons prefer to transform as $\mathbf{2}\oplus\mathbf{1}$ or $\mathbf{2'}\oplus\mathbf{1}$ under $A'_5$ in the minimal lepton models which can explain data, this is a notable feature of $A'_5$ modular symmetry with respect to other modular invariant models of levels $N=2, 3, 4$.

In summary, we have performed a comprehensive and systematical analysis of the homogeneous finite modular groups $\Gamma'_5\cong A'_5$ and its metaplectic cover $\widetilde{\Gamma}_5\cong A'_5\times Z_5$ in the present work. The integral modular forms at level $N=5$ are expressed as polynomials of the weight $1/5$ modular forms $F_1(\tau)$ and $F_2(\tau)$, this construction naturally bypasses the need to look for constraints
relating redundant higher weight multiplets. The possible lepton and quark models are classified according to the representation assignments of the matter fields under the $A'_5$ modular group. For the first time, we consider the generalized CP symmetry in the context of $A'_5$. We find out all the minimal $A'_5$ modular models which can explain the experimental data of lepton masses and mixing angles with/without CP symmetries. Furthermore, we construct a complete model for quarks and leptons based on $A'_5$ modular symmetry in which the observed masses and mixing patterns of both quarks and leptons can be reproduced for a common value of modulus $\tau$. Motivated by the fact that some fields with fractional weights are required to be present in construction of string theory~\cite{Ibanez:1992hc,Olguin-Trejo:2017zav,Nilles:2020nnc}, we consider the modular forms of weight $k/5$ in the modular invariance approach, and a benchmark lepton model is constructed. The finite metaplectic group $\widetilde{\Gamma}_5$ has five singlet representations denoted as $\mathbf{1}^{j}$ with $j=0, 1, 2, 3, 4$, thus it is more convenient to realize the quark and charged lepton mass hierarchies by assigning the right-handed matter fields to singlets of $\widetilde{\Gamma}_5$, see the model in section~\ref{sec:rational_model}.

In the present work, we have taken the approach of bottom-up flavor model building, it will be interesting to see how the modular symmetries $A'_5$ and $\widetilde{\Gamma}_5$ together with the rational weight modular forms can be naturally derived in the top-down approach from string theory. Because both representations and modular weights of matter fields and modular forms are not subject to any constraint in the current bottom-up approach,
there are many possibilities for model constructions.
We expect this drawback could be overcome from top-down perspective such as the eclectic flavor scheme~\cite{Nilles:2020nnc,Nilles:2020kgo,Nilles:2020tdp,Baur:2020jwc,Nilles:2020gvu} and thus the framework of modular invariance becomes more constrained and predictive.

\vskip0.2in

\section*{Acknowledgements}

XGL and GJD are supported by the National Natural Science Foundation of China under Grant Nos 11975224,  11835013 and 11947301. CYY is supported in part by the Grants No.~NSFC-11975130, No.~NSFC-12035008, No.~NSFC-12047533, by the National Key Research and Development Program of China under Grant No. 2017YFA0402200 and the China Post-doctoral Science Foundation under Grant No. 2018M641621.

\newpage

\begin{appendix}

\setcounter{equation}{0}
\renewcommand{\theequation}{\thesection.\arabic{equation}}

\section{\label{sec:A5_group_DC}Group theory of $A'_5$}

As shown in~\cite{Liu:2019khw}, the double covering of the icosahedral group $A_5$ can be generated by three generators $S$, $T$ and $R$ which obey the rules,
\begin{equation}
S^2=R,\quad T^5=(ST)^3=R^2=1,\quad RT=TR\,,
\end{equation}
or equivalently
\begin{equation}
S^4=T^5=(ST)^3=1,\quad S^2T=TS^2\,.
\end{equation}
There are several different presentations for the group $A'_5$ in the literature. In the Shirai basis~\cite{Shirai:1992math,Everett:2010rd}, all group elements can be expressed in terms of the order four generator $U$ and the order ten generator $V$ which satisfy the following multiplication rules,
\begin{equation}
  U^2=V^5=R,\quad (V^2UV^3UV^{-1}UVUV^{-1})^3=1~~ [\text{or~}((UV^{-1})^2U)^3=1],\quad R^2=1\,.
\end{equation}
The generators $S$ and $T$ can be related to the Shirai generators $U$ and $V$ by
\begin{equation}
  S=U,\quad T=V^2,\quad \text{and vice versa }\quad U=S,\quad V=T^{-2}R\,.
\end{equation}
In the presentation of Threlfall~\cite{Coxeter1972}, $A'_5$ is also generated by two generators $a$ and $b$ with
\begin{equation}
  a^3=b^5=(ab)^2=R,\quad R^2=1\,,
\end{equation}
which are related to the modular generators $S$ and $T$ by
\begin{equation}
S=ab,\quad T=b^{-1}R\quad \text{and vice versa}\quad a=ST^{-1}, \quad b=TR\,.
\end{equation}
We would like to introduce another presentation as simple as Threlfall's, the two generators $P$ and $Q$ obey the relations
\begin{equation}
  P^2=Q^5=(PQ)^3=R,\quad R^2=1\,,
\end{equation}
which are related to $S$ and $T$ by
\begin{equation}
S=P,\quad T=QR,\quad \text{and vice versa}\quad P=S,\quad  Q=TR\,.
\end{equation}
The presentation described in the work of Cummins and Patera~\cite{Cummins:1988math}, involves three generators $A_{1,2,3}$ satisfying
\begin{equation}
A_1^3=A_2^2=A_3^2=R,\quad (A_1A_2)^3=(A_1A_3)^2=(A_2A_3)^3=R,\quad R^2=1\,,
\end{equation}
which are related to the modular generators as follows,
\begin{equation}
S=A_2,~ T=A_1A_2A_3^{-1},~~ \text{and vice versa} ~~A_1=TST^{-2},~ A_2=S,~ A_3=ST^{-2}ST^2S\,.
\end{equation}
The group $A'_5$ has 120 elements which is twice as many elements as $A_5$,
and all the elements can be divided into the 9 conjugacy classes as follows:
\begin{align}
\nonumber 1C_1&= \{1\}\,, \\
\nonumber 1C_2&= \{R\}=(1C_1)\cdot R\,, \\
\nonumber 20C_3&= \{ST,TS,S^3T^4,T^4S^3,T^2ST^4,T^2S^3T^2,T^3ST^3,T^3S^3T,T^4ST^2,TS^3T^3,\\
\nonumber &~~~~~ T^2ST^3S,T^3ST^2S,ST^2ST^3,ST^3ST^2,S^3T^3ST,TST^3S^3,S^3T^2ST^4,\\
\nonumber &~~~~~ T^4ST^2S^3,ST^2ST^2S,TST^3ST\}\,, \\
\nonumber 30C_4&= \{S,SR,T^2ST^3,T^2ST^3R,T^3ST^2,T^3ST^2R, T^4ST,T^4STR,TST^4,TST^4R,\\
\nonumber &~~~~~ST^2ST,ST^2STR, TST^2S,TST^2SR,ST^3ST^2S,ST^3ST^2SR,
ST^2ST^3S,ST^2ST^3SR,  \\
\nonumber &~~~~~ T^2ST^3ST^2, T^2ST^3ST^2R, T^3ST^3ST, T^3ST^3STR, TST^3ST^3, TST^3ST^3R, \\
\nonumber &~~~~~ ST^2ST^3ST,
ST^2ST^3STR, TST^3ST^2S, TST^3ST^2SR, T^2ST^3ST^2S, T^2ST^3ST^2SR \}\,, \\
\nonumber 12C_5&= \{T,T^4,T^3S,ST^3,S^3T^2,T^2S^3,T^2ST,TST^2,S^3TS,TS^3T,T^3S^3T^4,T^4S^3T^3\}\,,\\
\nonumber 12C_5'&= \{T^2,T^3,S^3T^2S,S^3T^3S,T^2ST^2S,ST^2ST^2,T^3ST^3S,ST^3ST^3,T^2ST^3ST,\\
\nonumber &~~~~~ TST^3ST^2,T^3ST^2ST^4,T^4ST^2ST^3\}\,, \\
\nonumber 20C_6&= \{STR,TSR,S^3T^4R,T^4S^3R,T^2ST^4R,T^2S^3T^2R,T^3ST^3R,T^3S^3TR,T^4ST^2R,\\
\nonumber &~~~~~ TS^3T^3R, T^2ST^3SR,T^3ST^2SR,ST^2ST^3R,ST^3ST^2R,S^3T^3STR,TST^3S^3R,\\
\nonumber &~~~~~ S^3T^2ST^4R, T^4ST^2S^3R,ST^2ST^2SR,TST^3STR \} = (20C_3)\cdot R \,,\\
\nonumber 12C_{10}&= \{TR,T^4R,T^3SR,ST^3R,S^3T^2R,T^2S^3R,T^2STR,TST^2R,S^3TSR,TS^3TR,\\
\nonumber&~~~~~ T^3S^3T^4R,T^4S^3T^3R\}=(12C_5)\cdot R\,, \\
\nonumber 12C_{10}'&=\{T^2R,T^3R,S^3T^2SR,S^3T^3SR,T^2ST^2SR,ST^2ST^2R,T^3ST^3SR,ST^3ST^3R,\\
&~~~~~ T^2ST^3STR, TST^3ST^2R,T^3ST^2ST^4R,T^4ST^2ST^3R\}=(12C_5')\cdot R\,,
\end{align}
where the number in front denotes the number of group elements contained in this class, and the subscript ``$n$'' of the notation $C_n$ is the order of the class. The number of conjugacy classes is the same as the inequivalent irreducible representations. In addition to the five inequivalent irreducible representations $\mathbf{1}$, $\mathbf{3}$, $\mathbf{3'}$, $\mathbf{4}$, $\mathbf{5}$ of the $A_5$ group, $A'_5$ has four inequivalent representations $\mathbf{2}$, $\mathbf{2'}$, $\mathbf{4'}$ and $\mathbf{6}$.
The explicit forms of the generators $S$, $T$ and $R$ in each of the irreducible representations are summarized in table~\ref{tab:rep-matrices}.  In the single-valued representations $\mathbf{1}$, $\mathbf{3}$, $\mathbf{3'}$, $\mathbf{4}$ and $\mathbf{5}$, the generator $R$ is represented by the identity matrix $I$, and the elements of $A'_5$ are described by the same matrices which represent the elements in $A_5$,  consequently the group $A'_5$ can not be distinguished from the group $A_5$ with these representations. The generator $R$ is $-I$ in the double-valued representations $\mathbf{2}$, $\mathbf{2'}$, $\mathbf{4'}$ and $\mathbf{6}$. Then one can easily read out the character table of the group $A'_5$ shown in table~\ref{tab:character-table}.

\begin{table}[!ht]
  \centering
  \begin{tabular}{|c|c|c|c|c|c|c|c|c|c|c|}
    \hline\hline
    Classes & $1C_1$ & $1C_2$ & $20C_3$ & $30C_4$ & $12C_5$ & $12C_5'$ & $12C_5''$ & $20C_6$ & $12C_{10}$ \\ \hline
    $G$ & $1$ & $R$ & $ST$ & $S$ & $T$ & $T^2$ & $S^3T$ & $RT$ & $T^2R$ \\ \hline
$\mathbf{1}$ & $1$ & $1$ & $1$ & $1$ & $1$ & $1$ & $1$ & $1$ & $1$\\ \hline
$\mathbf{2}$ & $2$ & $-2$ & $-1$ & $0$ & $-\phi$ & $\frac{1}{\phi }$ & $1$ & $\phi$ & $-\frac{1}{\phi }$\\ \hline
$\mathbf{2'}$ & $2$ & $-2$ & $-1$ & $0$ & $\frac{1}{\phi }$ & $-\phi$ & $1$ & $-\frac{1}{\phi }$ & $\phi$\\ \hline
$\mathbf{3}$ & $3$ & $3$ & $0$ & $-1$ & $\phi$ & $-\frac{1}{\phi }$ & $0$ & $\phi$ & $-\frac{1}{\phi }$\\ \hline
$\mathbf{3'}$ & $3$ & $3$ & $0$ & $-1$ & $-\frac{1}{\phi }$ & $\phi$ & $0$ & $-\frac{1}{\phi }$ & $\phi$\\ \hline
$\mathbf{4}$ & $4$ & $4$ & $1$ & $0$ & $-1$ & $-1$ & $1$ & $-1$ & $-1$\\ \hline
$\mathbf{4'}$ & $4$ & $-4$ & $1$ & $0$ & $-1$ & $-1$ & $-1$ & $1$ & $1$\\ \hline
$\mathbf{5}$ & $5$ & $5$ & $-1$ & $1$ & $0$ & $0$ & $-1$ & $0$ & $0$\\ \hline
$\mathbf{6}$ & $6$ & $-6$ & $0$ & $0$ & $1$ & $1$ & $0$ & $-1$ & $-1$ \\ \hline\hline
\end{tabular}
\caption{\label{tab:character-table}Character table of $A'_5$, where $\phi=(1+\sqrt{5})/2$ is the golden ratio and $G$ stands for the representative element of each conjugacy class.}
\end{table}

\begin{landscape}
{\footnotesize
\setlength{\LTcapwidth}{1.1\textwidth}{
\begin{longtable}{ |>{\centering\arraybackslash}p{0.6cm} |>{\centering\arraybackslash}p{10cm}|>{\centering\arraybackslash}p{5cm}|>{\centering\arraybackslash}p{6cm}| }
\hline\endhead
    Rep. & $\rho_{\mathbf{r}}(S)$ & $\rho_{\mathbf{r}}(T)$ & $\rho_{\mathbf{r}}(R)$\\ \hline
    $\mathbf{1}$ & 1 & 1 & 1 \\ \hline
    $\mathbf{2}$ & $i\sqrt{\frac{1}{\sqrt{5}\phi}}\begin{pmatrix}
 \phi  & 1 \\
 1 & -\phi  \\
\end{pmatrix}$ & $\begin{pmatrix}
 \omega _5^2 & 0 \\
 0 & \omega _5^3 \\
\end{pmatrix}$ & $\begin{pmatrix}
 -1 & 0 \\
 0 & -1 \\
\end{pmatrix}$ \\ \hline
    $\mathbf{2'}$ & $i\sqrt{\frac{1}{\sqrt{5}\phi}}\begin{pmatrix}
 1 & \phi  \\
 \phi  & -1 \\
\end{pmatrix}$ & $\begin{pmatrix}
 \omega _5 & 0 \\
 0 & \omega _5^4 \\
\end{pmatrix}$ & $\begin{pmatrix}
 -1 & 0 \\
 0 & -1 \\
\end{pmatrix}$ \\ \hline
    $\mathbf{3}$ & $\frac{1}{\sqrt{5}}\begin{pmatrix}
 1 & -\sqrt{2} & -\sqrt{2} \\
 -\sqrt{2} & -\phi  & \frac{1}{\phi } \\
 -\sqrt{2} & \frac{1}{\phi } & -\phi  \\
\end{pmatrix}$ & $\begin{pmatrix}
 1 & 0 & 0 \\
 0 & \omega _5 & 0 \\
 0 & 0 & \omega _5^4 \\
\end{pmatrix}$ & $\begin{pmatrix}
 1 & 0 & 0 \\
 0 & 1 & 0 \\
 0 & 0 & 1 \\
\end{pmatrix}$\\ \hline
  $\mathbf{3'}$ & $\frac{1}{\sqrt{5}}\begin{pmatrix}
 -1 & \sqrt{2} & \sqrt{2} \\
 \sqrt{2} & -\frac{1}{\phi } & \phi  \\
 \sqrt{2} & \phi  & -\frac{1}{\phi } \\
\end{pmatrix}$ & $\begin{pmatrix}
 1 & 0 & 0 \\
 0 & \omega _5^2 & 0 \\
 0 & 0 & \omega _5^3 \\
\end{pmatrix}$ & $\begin{pmatrix}
 1 & 0 & 0 \\
 0 & 1 & 0 \\
 0 & 0 & 1 \\
\end{pmatrix}$ \\ \hline
    $\mathbf{4}$ & $\frac{1}{\sqrt{5}}\begin{pmatrix}
 1 & \frac{1}{\phi } & \phi  & -1 \\
 \frac{1}{\phi } & -1 & 1 & \phi  \\
 \phi  & 1 & -1 & \frac{1}{\phi } \\
 -1 & \phi  & \frac{1}{\phi } & 1 \\
\end{pmatrix}$ & $\begin{pmatrix}
 \omega _5 & 0 & 0 & 0 \\
 0 & \omega _5^2 & 0 & 0 \\
 0 & 0 & \omega _5^3 & 0 \\
 0 & 0 & 0 & \omega _5^4 \\
\end{pmatrix}$ & $\begin{pmatrix}
 1 & 0 & 0 & 0 \\
 0 & 1 & 0 & 0 \\
 0 & 0 & 1 & 0 \\
 0 & 0 & 0 & 1 \\
\end{pmatrix}$ \\ \hline
    $\mathbf{4'}$ & $i\sqrt{\frac{1}{5\sqrt{5}\phi}}\begin{pmatrix}
 -\phi ^2 & \sqrt{3} \phi  & -\sqrt{3} & -\frac{1}{\phi } \\
 \sqrt{3} \phi  & \frac{1}{\phi } & -\phi ^2 & -\sqrt{3} \\
 -\sqrt{3} & -\phi ^2 & -\frac{1}{\phi } & -\sqrt{3} \phi  \\
 -\frac{1}{\phi } & -\sqrt{3} & -\sqrt{3} \phi  & \phi ^2 \\
\end{pmatrix}$ & $\begin{pmatrix}
 \omega _5 & 0 & 0 & 0 \\
 0 & \omega _5^2 & 0 & 0 \\
 0 & 0 & \omega _5^3 & 0 \\
 0 & 0 & 0 & \omega _5^4 \\
\end{pmatrix}$ & $\begin{pmatrix}
 -1 & 0 & 0 & 0 \\
 0 & -1 & 0 & 0 \\
 0 & 0 & -1 & 0 \\
 0 & 0 & 0 & -1 \\
\end{pmatrix}$ \\\hline
    $\mathbf{5}$& $\frac{1}{5}\begin{pmatrix}
 -1 & \sqrt{6} & \sqrt{6} & \sqrt{6} & \sqrt{6} \\
 \sqrt{6} & \frac{1}{\phi ^2} & -2 \phi  & \frac{2}{\phi } & \phi ^2 \\
 \sqrt{6} & -2 \phi  & \phi ^2 & \frac{1}{\phi ^2} & \frac{2}{\phi } \\
 \sqrt{6} & \frac{2}{\phi } & \frac{1}{\phi ^2} & \phi ^2 & -2 \phi  \\
 \sqrt{6} & \phi ^2 & \frac{2}{\phi } & -2 \phi  & \frac{1}{\phi ^2} \\
\end{pmatrix}$ & $\begin{pmatrix}
 1 & 0 & 0 & 0 & 0 \\
 0 & \omega _5 & 0 & 0 & 0 \\
 0 & 0 & \omega _5^2 & 0 & 0 \\
 0 & 0 & 0 & \omega _5^3 & 0 \\
 0 & 0 & 0 & 0 & \omega _5^4 \\
\end{pmatrix}$ & $\begin{pmatrix}
 1 & 0 & 0 & 0 & 0 \\
 0 & 1 & 0 & 0 & 0 \\
 0 & 0 & 1 & 0 & 0 \\
 0 & 0 & 0 & 1 & 0 \\
 0 & 0 & 0 & 0 & 1 \\
\end{pmatrix}$  \\ \hline
    $\mathbf{6}$ & $i\sqrt{\frac{1}{5\sqrt{5}\phi}}\begin{pmatrix}
 -1 & \phi  & \frac{1}{\phi } & \sqrt{2} \phi  & \sqrt{2} & \phi ^2 \\
 \phi  & 1 & \phi ^2 & \sqrt{2} & -\sqrt{2} \phi  & -\frac{1}{\phi } \\
 \frac{1}{\phi } & \phi ^2 & 1 & -\sqrt{2} & \sqrt{2} \phi  & -\phi  \\
 \sqrt{2} \phi  & \sqrt{2} & -\sqrt{2} & -\phi  & -1 & \sqrt{2} \phi  \\
 \sqrt{2} & -\sqrt{2} \phi  & \sqrt{2} \phi  & -1 & \phi  & \sqrt{2} \\
 \phi ^2 & -\frac{1}{\phi } & -\phi  & \sqrt{2} \phi  & \sqrt{2} & -1 \\
\end{pmatrix}$ & $\begin{pmatrix}
 1 & 0 & 0 & 0 & 0 & 0 \\
 0 & 1 & 0 & 0 & 0 & 0 \\
 0 & 0 & \omega _5 & 0 & 0 & 0 \\
 0 & 0 & 0 & \omega _5^2 & 0 & 0 \\
 0 & 0 & 0 & 0 & \omega _5^3 & 0 \\
 0 & 0 & 0 & 0 & 0 & \omega _5^4 \\
\end{pmatrix}$ & $\begin{pmatrix}
 -1 & 0 & 0 & 0 & 0 & 0 \\
 0 & -1 & 0 & 0 & 0 & 0 \\
 0 & 0 & -1 & 0 & 0 & 0 \\
 0 & 0 & 0 & -1 & 0 & 0 \\
 0 & 0 & 0 & 0 & -1 & 0 \\
 0 & 0 & 0 & 0 & 0 & -1 \\
\end{pmatrix}$ \\ \hline\hline
\caption{\label{tab:rep-matrices}The representation matrices of the generators $S$, $T$ and $R$ in different irreducible representations of $A'_5$, where $\omega_5$ is the quintic unit root $\omega_5=e^{2\pi i/5}$. }
\end{longtable}}}
\end{landscape}

From the character table, one can straightforwardly calculate the multiplication rules of the irreducible representations as follows,
\begin{eqnarray}
\nonumber && \mathbf{2}\otimes \mathbf{2}=\mathbf{1_a}\oplus \mathbf{3_s}\,,\quad \mathbf{2}\otimes \mathbf{2'}=\mathbf{4}\,,\quad \mathbf{2}\otimes \mathbf{3}=\mathbf{2}\oplus \mathbf{4'}\,,\quad \mathbf{2}\otimes \mathbf{3'}=\mathbf{2'}\otimes \mathbf{3}=\mathbf{6}\,,\\
\nonumber &&\mathbf{2}\otimes \mathbf{4}=\mathbf{2'}\oplus \mathbf{6}\,,\quad \mathbf{2}\otimes \mathbf{4'}=\mathbf{3}\oplus \mathbf{5}\,,\quad \mathbf{2}\otimes \mathbf{5}=\mathbf{2'}\otimes \mathbf{5}=\mathbf{4'}\oplus \mathbf{6}\,,\\
\nonumber && \mathbf{2}\otimes \mathbf{6}=\mathbf{3}\otimes \mathbf{4}=\mathbf{3'}\oplus \mathbf{4}\oplus \mathbf{5}\,,\quad \mathbf{2'}\otimes \mathbf{2'}=\mathbf{1_a}\oplus \mathbf{3'_s}\,,\quad \mathbf{2'}\otimes \mathbf{3'}=\mathbf{2'}\oplus \mathbf{4'}\,,\\
\nonumber && \mathbf{2'}\otimes \mathbf{4}=\mathbf{2}\oplus \mathbf{6}\,,\quad \mathbf{2'}\otimes \mathbf{4'}=\mathbf{3'}\oplus \mathbf{5}\,,\quad \mathbf{2'}\otimes \mathbf{6}=\mathbf{3'}\otimes \mathbf{4}=\mathbf{3}\oplus \mathbf{4}\oplus \mathbf{5}\,,\\
\nonumber &&\mathbf{3}\otimes \mathbf{3}=\mathbf{1_s}\oplus \mathbf{3_a}\oplus \mathbf{5_s}\,,\quad \mathbf{3}\otimes \mathbf{3'}=\mathbf{4}\oplus \mathbf{5}\,,\quad \mathbf{3}\otimes \mathbf{4'}=\mathbf{2}\oplus \mathbf{4'}\oplus \mathbf{6}\,,\\
\nonumber && \mathbf{3}\otimes \mathbf{5}=\mathbf{3'}\otimes \mathbf{5}=\mathbf{3}\oplus \mathbf{3'}\oplus \mathbf{4}\oplus \mathbf{5}\,,\quad \mathbf{3}\otimes \mathbf{6}=\mathbf{2'}\oplus \mathbf{4'}\oplus \mathbf{6_1}\oplus \mathbf{6_2}\,,\\
\nonumber &&\mathbf{3'}\otimes \mathbf{3'}=\mathbf{1_s}\oplus \mathbf{3'_a}\oplus \mathbf{5_s}\,,\quad \mathbf{3'}\otimes \mathbf{4'}=\mathbf{2'}\oplus \mathbf{4'}\oplus \mathbf{6}\,,\quad \mathbf{3'}\otimes \mathbf{6}=\mathbf{2}\oplus \mathbf{4'}\oplus \mathbf{6_1}\oplus \mathbf{6_2}\,,\\
\nonumber &&\mathbf{4}\otimes \mathbf{4}=\mathbf{1_s}\oplus \mathbf{3_a}\oplus \mathbf{3'_a}\oplus \mathbf{4_s}\oplus \mathbf{5_s}\,,\quad \mathbf{4}\otimes \mathbf{4'}=\mathbf{4'}\oplus \mathbf{6_1}\oplus \mathbf{6_2}\,,\\
\nonumber && \mathbf{4}\otimes \mathbf{5}=\mathbf{3}\oplus \mathbf{3'}\oplus \mathbf{4}\oplus \mathbf{5_1}\oplus \mathbf{5_2}\,,\quad \mathbf{4}\otimes \mathbf{6}=\mathbf{2}\oplus \mathbf{2'}\oplus \mathbf{4'_1}\oplus \mathbf{4'_2}\oplus \mathbf{6_1}\oplus \mathbf{6_2}\,,\\
\nonumber &&\mathbf{4'}\otimes \mathbf{4'}=\mathbf{1_a}\oplus \mathbf{3_s}\oplus \mathbf{3'_s}\oplus \mathbf{4_s}\oplus \mathbf{5_a}\,,\quad \mathbf{4'}\otimes \mathbf{5}=\mathbf{2}\oplus \mathbf{2'}\oplus \mathbf{4'}\oplus \mathbf{6_1}\oplus \mathbf{6_2}\,,\\
\nonumber &&\mathbf{4'}\otimes \mathbf{6}=\mathbf{3}\oplus \mathbf{3'}\oplus \mathbf{4_1}\oplus \mathbf{4_2}\oplus \mathbf{5_1}\oplus \mathbf{5_2}\,,\\
\nonumber && \mathbf{5}\otimes \mathbf{5}=\mathbf{1_s}\oplus \mathbf{3_a}\oplus \mathbf{3'_a}\oplus \mathbf{4_s}\oplus \mathbf{4_a}\oplus \mathbf{5_{1,s}}\oplus \mathbf{5_{2,s}}\,,\\
\nonumber &&\mathbf{5}\otimes \mathbf{6}=\mathbf{2}\oplus \mathbf{2'}\oplus \mathbf{4'_1}\oplus \mathbf{4'_2}\oplus \mathbf{6_1}\oplus \mathbf{6_2}\oplus \mathbf{6_3}\,,\\
\label{eq:kronecker-Gamma'5}&& \mathbf{6}\otimes \mathbf{6}=\mathbf{1_a}\oplus \mathbf{3_{1,s}}\oplus \mathbf{3_{2,s}}\oplus \mathbf{3'_{1,s}}\oplus \mathbf{3'_{2,s}}\oplus \mathbf{4_s}\oplus \mathbf{4_a}\oplus \mathbf{5_{1,s}}\oplus \mathbf{5_{2,a}}\oplus \mathbf{5_{3,a}}\,,
\end{eqnarray}
where the subscripts $\mathbf{s}$ and $\mathbf{a}$ denote symmetric and antisymmetric combinations respectively. For the product decomposition $\mathbf{3}\otimes \mathbf{6}=\mathbf{2'}\oplus \mathbf{4'}\oplus \mathbf{6_1}\oplus \mathbf{6_2}$, $\mathbf{6_1}$ and $\mathbf{6_2}$ refer to the two sextet representations appearing in the tensor products of $\mathbf{3}$ and $\mathbf{6}$, and similar notations are adopted for other tensor products. We now list the Clebsch-Gordan coefficients which are quite useful in model construction. we use $\alpha_i$ to indicate the elements of the first representation of the product and $\beta_i$ to indicate those of the second representation.

\cgtwo{6cm}{6cm}{
\makecell[t]{\vspace{0.2in}
\fbox{$\mathbf{2} \otimes \mathbf{2} = \mathbf{1_a} \oplus \mathbf{3_s}$} \\ \vspace{0.2in}
$\mathbf{1_a} = \alpha_2 \beta_1-\alpha_1 \beta_2$ \\ \vspace{0.2in}
$\mathbf{3_s} = 
$}
}

\section{\label{app:weight-1-MF}Construct weight 1 and level 5 modular forms with Klein form}

\setcounter{equation}{0}
\renewcommand{\theequation}{\thesection.\arabic{equation}}

The linear space $\mathcal{M}_{k}(\Gamma(5))$ of modular forms of positive integral weight $k$ and level 5 has been explicitly constructed through the Dedekind eta-function and Klein form as follow~\cite{schultz2015notes}
\begin{equation}
\label{eq:MF_Gamma5}\mathcal{M}_{k}(\Gamma(5))=\bigoplus_{a+b=5k,\,a,b\ge0} \mathbb{C} \frac{\eta^{15k}(5\tau)}{\eta^{3k}(\tau)} \mathfrak{k}^a_{\frac{1}{5},\frac{0}{5}}(5\tau)\mathfrak{k}^b_{\frac{2}{5},\frac{0}{5}}(5\tau)\,,
\end{equation}
where the expression of the eta-function $\eta(\tau)$ is~\cite{diamond2005first,Bruinier2008The,lang2012introduction},
\begin{equation}
\label{eq:eta_function}\eta(\tau)=q^{1/24}\prod_{n=1}^\infty \left(1-q^n \right),\qquad q\equiv e^{i 2 \pi\tau}\,.
\end{equation}
The Klein form $\mathfrak{k}_{(r_1,r_2)}(\tau)$ in Eq.~\eqref{eq:MF_Gamma5} is a holomorphic function which has no zeros and poles on the upper half complex plane. The Klein form can be written into an infinite product expansion~\cite{K_lang1981,lang1987elliptic,lang2012introduction,eum2011modularity}:
\begin{equation}
\label{KleinForm}
\mathfrak{k}_{(r_1,r_2)}(\tau)=q^{(r_1-1)/2}_z(1-q_z)\prod_{n=1}^\infty(1-q^nq_z)(1-q^nq_z^{-1})(1-q^n)^{-2}\,,
\end{equation}
where $q_z=e^{2\pi iz}$ with $z=r_1\tau+r_2$.
From Eq.~\eqref{eq:MF_Gamma5} we see that the modular forms of weight $k$ and level 5 spans a linear space of dimension $5k+1$. For the weight $k=1$ modular forms, the modular space $\mathcal{M}_{1}(\Gamma(5))$ has dimension 6 and the basis vectors can be chosen to be
\begin{eqnarray}
\nonumber&& e_1(\tau)=\frac{\eta^{15}(5\tau)}{\eta^3(\tau)}\mathfrak{k}^{5}_{\frac{2}{5}, 0}(5\tau),~~~~~~~~~~\quad \qquad e_2(\tau)=\frac{\eta^{15}(5\tau)}{\eta^3(\tau)}\mathfrak{k}_{\frac{1}{5}, 0}(5\tau)\mathfrak{k}^{4}_{\frac{2}{5}, 0}(5\tau),\\
\nonumber&& e_3(\tau)=\frac{\eta^{15}(5\tau)}{\eta^3(\tau)}\mathfrak{k}^{2}_{\frac{1}{5}, 0}(5\tau)\mathfrak{k}^{3}_{\frac{2}{5}, 0}(5\tau),\quad\qquad e_4(\tau)=\frac{\eta^{15}(5\tau)}{\eta^3(\tau)}\mathfrak{k}^{3}_{\frac{1}{5}, 0}(5\tau)\mathfrak{k}^{2}_{\frac{2}{5}, 0}(5\tau),\\
&& e_5(\tau)=\frac{\eta^{15}(5\tau)}{\eta^3(\tau)}\mathfrak{k}^{4}_{\frac{1}{5}, 0}(5\tau)\mathfrak{k}_{\frac{2}{5}, 0}(5\tau),\quad\qquad e_6(\tau)=\frac{\eta^{15}(5\tau)}{\eta^3(\tau)}\mathfrak{k}^{5}_{\frac{1}{5}, 0}(5\tau)\,.
\end{eqnarray}
The above basis vectors are linearly independent and they span the weight 1 modular space $\mathcal{M}_{1}(\Gamma(5))$. Any modular function of weight 1 and level 5 can be expressed as a linear combination of the basis elements $e_i$ with $i=1, 2, \ldots, 6$. Under the action of the generator $T$, they transform as
\begin{equation}
\begin{aligned}
&e_1(\tau)\rightarrow e_1(\tau),\qquad e_2(\tau)\rightarrow e^{i\frac{2\pi}{5}}e_2(\tau),\qquad e_3(\tau)\rightarrow e^{i\frac{4\pi}{5}}e_3(\tau),\\
&e_4(\tau)\rightarrow e^{i\frac{6\pi}{5}}e_4(\tau),\qquad e_5(\tau)\rightarrow e^{i\frac{8\pi}{5}}e_5(\tau),\qquad e_6(\tau)\rightarrow e_6(\tau)\,.
\end{aligned}
\end{equation}
Furthermore, we find the following transformation properties under another generator $S$,
\begin{eqnarray}
\nonumber e_1(\tau)&\rightarrow&-\tau\frac{\sqrt{5}i}{\sqrt{5\sqrt{5}\phi}}
\Bigg\{\frac{\phi^3}{5}e_1(\tau)+\phi^2e_2(\tau)+2\phi e_3(\tau)+2e_4(\tau)+\frac{1}{\phi}e_5(\tau)+\frac{1}{5\phi^2}e_6(\tau)\Bigg\}\,,\\
\nonumber e_2(\tau)&\rightarrow&-\tau\frac{i}{\sqrt{5\sqrt{5}\phi}}
\Bigg\{\frac{\phi^2}{\sqrt{5}}e_1(\tau)+e_2(\tau)-2e_3(\tau)-2\phi e_4(\tau)-\phi e_5(\tau)-\frac{1}{\sqrt{5}\phi}e_6(\tau)\Bigg\}\,, \\
\nonumber e_3(\tau)&\rightarrow&-\tau\frac{i}{\sqrt{5\sqrt{5}\phi}}
\Bigg\{\frac{\phi}{\sqrt{5}}e_1(\tau)-e_2(\tau)-\phi e_3(\tau)+e_4(\tau)+\phi e_5(\tau)+\frac{1}{\sqrt{5}}e_6(\tau)\Bigg\}\,, \\
\nonumber e_4(\tau)&\rightarrow&-\tau\frac{i}{\sqrt{5\sqrt{5}\phi}}
\Bigg\{\frac{1}{\sqrt{5}}e_1(\tau)-\phi e_2(\tau)+e_3(\tau)+\phi e_4(\tau)-e_5(\tau)-\frac{\phi}{\sqrt{5}}e_6(\tau)\Bigg\}\,, \\
\nonumber e_5(\tau)&\rightarrow&-\tau\frac{i}{\sqrt{5\sqrt{5}\phi}}
\Bigg\{\frac{1}{\sqrt{5}\phi}e_1(\tau)-\phi e_2(\tau)+2\phi e_3(\tau)-2e_4(\tau)-e_5(\tau)+\frac{\phi^2}{\sqrt{5}}e_6(\tau)\Bigg\}\,, \\
e_6(\tau)&\rightarrow&-\tau\frac{\sqrt{5}i}{\sqrt{5\sqrt{5}\phi}}
\Bigg\{\frac{1}{5\phi^2}e_1(\tau)-\frac{1}{\phi}e_2(\tau)+2e_3(\tau)-2\phi e_4(\tau)+\phi^2e_5(\tau)-\frac{\phi^3}{5}e_6(\tau)\Bigg\}\,,
\end{eqnarray}
where $\phi=(1+\sqrt{5})/2$ is the golden ratio. We see that the basis vectors $e_i$ are closed under $S$ and $T$ up to multiplicative factors, and each element is exactly mapped into itself under the action of $S^4$, $(ST)^3$ and $T^5$. Therefore we conclude that $e_i$ really span the whole modular space $\mathcal{M}_{1}(\Gamma(5))$. As shown in~\cite{Feruglio:2017spp,Liu:2019khw}, the modular form space of weight $k$ and level $N$ can always be decomposed into different irreducible representations of $\Gamma'_{N}$. We can perform linear combinations of the six basis vectors $e_{1,2,3,4,5,6}$ to form a six dimensional representation $\mathbf{6}$ of $A'_5$,
\begin{equation}
Y_{\mathbf{6}}(\tau)=\left(Y_1(\tau), Y_2(\tau), Y_3(\tau), Y_4(\tau), Y_5(\tau), Y_6(\tau)\right)^{T}\,,
\end{equation}
with
\begin{eqnarray}
\nonumber &&Y_1(\tau)=e_1(\tau)+2e_6(\tau),\quad Y_2(\tau)=2e_1(\tau)-e_6(\tau),\quad Y_3(\tau)=5e_2(\tau),\\
\label{eq:wt1_MF_Kelin}&&Y_4(\tau)=5\sqrt{2}e_3(\tau),\quad Y_5(\tau)=-5\sqrt{2}e_4(\tau),\quad Y_6(\tau)=5e_5(\tau)\,.
\end{eqnarray}
Under the actions of the modular symmetry generators $S$ and $T$, the modular form $Y_{\mathbf{6}}(\tau)$ transforms as
\begin{equation}
Y_{\mathbf{6}}(-1/\tau)=-\tau\rho_{\mathbf{6}}(S)Y_{\mathbf{6}}(\tau), \quad Y_{\mathbf{6}}(\tau+1)=\rho_{\mathbf{6}}(T)Y_{\mathbf{6}}(\tau)\,,
\end{equation}
where the six dimensional representation matrices $\rho_{\mathbf{6}}(S)$ and $\rho_{\mathbf{6}}(T)$ are given in table~\ref{tab:rep-matrices}. Furthermore, we can read out the $q$-expansion of the modular forms $Y_i$ as
\begin{eqnarray}
\nonumber&&Y_1(\tau)=1+5q+10q^3-5q^4+5q^5+10q^6+5q^9+\ldots\,, \\
\nonumber&&Y_2(\tau)=2+5 q+10 q^2+5 q^4+5 q^5+10 q^6+10q^7-5q^9+\ldots\,, \\
\nonumber&&Y_3(\tau)=5q^{1/5}\left(1+2q+2q^2+q^3+2q^4+2q^5+2q^6+q^7+2q^8+2q^9+\ldots\right)\,,\\
\nonumber&&Y_4(\tau)=5\sqrt{2}q^{2/5}\left(1+q+q^2+q^3+2q^4+q^6+q^7+2 q^8+q^9+\ldots\right)\,,\\
\nonumber&&Y_5(\tau)=-5\sqrt{2}q^{3/5}\left(1+q^2+q^3+q^4-q^5+2 q^6+q^8+q^9+\ldots\right)\,,\\
\label{eq:q_series_wt1}&&Y_6(\tau)=5q^{4/5}\left(1-q+2q^2+2q^6-2q^7+2q^8+q^9+\ldots\right)\,.
\end{eqnarray}

\setcounter{equation}{0}
\renewcommand{\theequation}{\thesection.\arabic{equation}}

\section{\label{sec:proof} Prove the identities in Eq.~\eqref{eq:identity}}

In section~\ref{sec:rational-weight-MF}, we find the identities between the theta constants and Kelin forms in Eq.~\eqref{eq:identity} which is checked numerically. We will prove them by using the famous Jacobi triple product identity. Given the definition of eta function, Klein form and theta constant in Eqs.~(\ref{eq:eta_function}, \ref{KleinForm}, \ref{eq:q-series_F1F2}), the identities in Eq.~\eqref{eq:identity} become the following formulas,
\begin{align}
\label{eq:theta-Klein-1}
&\sum_{n=-\infty}^{\infty} (-1)^n q^{(5n^2+n)/2} =(1-q^2)\prod^{\infty}_{n=1}(1-q^{5n})(1-q^{5n-2})(1-q^{5n+2}) \,,\\[0.1in]
\label{eq:theta-Klein-2}&\sum_{n=-\infty}^{\infty} (-1)^n q^{(5n^2+3n)/2} = (1-q)\prod^{\infty}_{n=1}(1-q^{5n})(1-q^{5n+1})(1-q^{5n-1})\,.
\end{align}
Notice that these formulas relate the infinite triple product with infinite summation, this reminds us of the famous Jacobi triple product identity which is the special case of the more general Macdonald identity~\cite{kohler2011eta,macdonald1971affine} :
\begin{equation}
\label{eq:Jacobi_TriProd_Identity}
\sum_{n=-\infty}^{\infty} x^{n^2} y^n = \prod^{\infty}_{n=1}(1-x^{2n})(1+x^{2n-1}y)(1+x^{2n-1}y^{-1}) \,,
\end{equation}
where $x,y\in \mathbb{C}$ and $|x|<1,\,y\neq 0$.
Taking $x=q^{5/2}$ and $y=-q^{1/2}$ in the above Jacobi triple product identity, we can straightforwardly obtain Eq.~\eqref{eq:theta-Klein-1}, and similarly Eq.~\eqref{eq:theta-Klein-2} follows from $x=q^{5/2}$ and $y=-q^{3/2}$.

\section{\label{sec:higher-weight-MF} Higher weight modular forms of level 5}

\setcounter{equation}{0}
\renewcommand{\theequation}{\thesection.\arabic{equation}}

The weight 4 modular multiplets can be generated from the product of $Y_{\mathbf{6}}$ and the modular forms of weight 3. We find there are 21 linearly independent modular forms which can be taken to be
\begin{equation}
  \begin{split}
Y_{\mathbf{1}}^{(4)}&=(Y_{\mathbf{6}}^{(1)}Y_{\mathbf{6}I}^{(3)})_{\mathbf{1}_{a}}=
\,.
\end{split}\nonumber
\end{equation}
Using the Clebsch-Gordan coefficients listed in Appendix~\ref{sec:A5_group_DC}, we can express each entry of these weight four modular forms as quartic polynomial of $Y_i$. In the same fashion, we can obtain the modular multiplets of weight 5 as follow,
\begin{equation}
  \begin{split}
Y_{\mathbf{2}}^{(5)}&=(Y_{\mathbf{6}}^{(1)}Y_{\mathbf{5}II}^{(4)})_{\mathbf{2}}=
%
\,.
  \end{split}\nonumber
\end{equation}
Finally we report the linearly independent weight 6 modular multiplets,
  \begin{align*}
Y_{\mathbf{1}}^{(6)}&=(Y_{\mathbf{6}}^{(1)}Y_{\mathbf{6}II}^{(5)})_{\mathbf{1}_{a}}=
 Y_2 Y_{\mathbf{6}II,1}^{(5)}-Y_1 Y_{\mathbf{6}II,2}^{(5)}+Y_6 Y_{\mathbf{6}II,3}^{(5)}+Y_5 Y_{\mathbf{6}II,4}^{(5)}-Y_4 Y_{\mathbf{6}II,5}^{(5)}-Y_3 Y_{\mathbf{6}II,6}^{(5)}\,, \\
Y_{\mathbf{3}I}^{(6)}&=(Y_{\mathbf{6}}^{(1)}Y_{\mathbf{6}I}^{(5)})_{\mathbf{3}_{1, s}}=
%
\,.
  \end{align*}
We see that the linear space of modular forms of level 5 and weight 6 has dimension 31 which matches with the dimension formula $5k+1=5\times6+1=31$.

\subsection{Higher rational weight modular forms of level 5}

In the following, we report the higher rational weight modular forms at level 5, and they are organized into irreducible multiplets of $\widetilde{\Gamma}_5$.

\begin{equation}
\nonumber
Y^{(\frac{6}{5})}_{\mathbf{3}'^4}= \frac{1}{2}\left(Y^{(\frac{1}{5})}_{\mathbf{2}^4}Y^{(1)}_{\mathbf{6}^0}\right)_{\mathbf{3}'^4}=
%
\,.
\end{equation}

\end{appendix}

\providecommand{\href}[2]{#2}\begingroup\raggedright\endgroup

\end{document}